\documentclass[10pt, a4paper, twoside, fleqn]{article}

\usepackage[left=1.5cm,top=2.cm,right=1.5 cm,bottom=2.2cm]{geometry}
\usepackage{graphicx}
\usepackage[dvipsnames]{xcolor}

\usepackage{amsfonts}
\usepackage[T1]{fontenc}
\usepackage{lmodern}
\usepackage{amsmath}
\usepackage{dsfont}
\usepackage{color}
\usepackage{epsfig}
\usepackage{comment}
\usepackage{amssymb}
\usepackage{multicol}
\usepackage{mathtools}
\usepackage{float}
\usepackage{cancel}
\usepackage{ulem}

\numberwithin{equation}{section}

\def\gtwid{\mathrel{\raise.3ex\hbox{$>$\kern-.75em\lower1ex\hbox{$\sim$}}}}
\def\ltwid{\mathrel{\raise.3ex\hbox{$<$\kern-.75em\lower1ex\hbox{$\sim$}}}}
\def\square{\kern1pt\vbox{\hrule height 1.2pt\hbox{\vrule width 1.2pt\hskip 3pt
   \vbox{\vskip 6pt}\hskip 3pt\vrule width 0.6pt}\hrule height 0.6pt}\kern1pt}

\begin{document}


\begin{titlepage}

\begin{flushright}
Date: \today
\end{flushright}

\vskip .5cm

\begin{center}
{\bf\LARGE Large-scale geometry of the Universe}

\end{center}

\vskip .5cm

\begin{center}
\bf \large Yassir Awwad$^{\spadesuit}$ and  Tomislav Prokopec$^{\diamondsuit}$
\end{center}

\vskip .5cm

\begin{center}
{
$^{\diamondsuit}$
Institute for Theoretical Physics, Spinoza Institute  \& EMME$\Phi$ \\
Utrecht University, Princetonplein 5,
3584 CC Utrecht, The Netherlands
}
\end{center}

\begin{center}
{\bf Abstract} \\
\end{center}
The large scale geometry of the late Universe can be decomposed
as ${\mathds R}\times {\Sigma}_3$, where ${\mathds R}$
stands for cosmic time and ${\Sigma}_3$ is the three dimensional
spatial manifold.
We conjecture that the geometry of the Universe's spatial
section ${\Sigma}_3$
conforms with the Thurston-Perelman theorem,
according to which the geometry of $\Sigma_3$ is either one of the
eight geometries from the Thurston geometrization conjecture,
or a combination of Thurston geometries smoothly sewn together.
We assume that topology of individual geometries plays
no observational role, {\it i.e.} the size of individual
geometries is much larger than the Hubble radius today.
We  investigate the dynamics of each of the individual
geometries by making use of the simplifying assumption
that our local Hubble patch consists of only one such geometry,
which is approximately homogeneous on very large scales,
but spatial isotropy is generally violated.

Spatial anisotropies grow in time in decelerating universes, but they
decay in accelerating universes. The thus-created {\it anisotropy problem} can be solved
by a period of primordial inflation, akin to how the flatness problem is solved.
Therefore, as regards Universe's large scale geometry,
any of the Thurston's geometries should be considered on {\it a par} with
Friedmann's geometries.

We consider two observational methods that can be used to test our conjecture: one based on
luminosity distance and one on angular diameter
distance measurements,
but leave for the future their detailed forecasting implementations.


\begin{flushleft}
\end{flushleft}

\vskip .5cm

\begin{flushleft}
$^{\spadesuit}$   e-mail: Yassir@Awwad.nl \\
$^{\diamondsuit}$ e-mail: T.Prokopec@uu.nl \\
\end{flushleft}

\vspace*{\fill}

\end{titlepage}


\newpage
\tableofcontents


\newpage
\section{Introduction \& Motivation}

\subsection{The assumption of isotropy}

Much of modern cosmology, in particular the Lambda-CDM model, or Cold Dark Matter with cosmological constant, is based on the \textit{Cosmological Principle}. This Principle is succinctly summarised by Milne~\cite{Milne:1933}, according to which
``Two observers in uniform relative motion have {\it identical} views of the Universe'', {\it i.e.} that each sees the same evolving sequence of world-pictures.
In its modern rendition, the Principle states that statistical properties of
the Universe are the same to all local observers~\footnote{
The principle applies only
to inertial observers in the rest frame of the Cosmic Microwave Backgroud (CMB) photons, which defines the rest frame of the Universe. Any observer moving with respect
to that frame perceives a CMB dipole and a large scale motion of
the Universe's Large Scale Structure
(LSS).}
and, in particular, that it is spatially homogeneous and isotropic at large enough scales.

The Cosmological Principle naturally leads one to also use a spatially homogeneous and isotropic metric -- that is to say, the
Friedmann-Lema\^itre-Robertson-Walker (FLRW) metric -- to describe the space-time background of the Universe. By placing small metric perturbations on top of this background to break these symmetries, one can account for the formation of structures such as galaxies and clusters of galaxies. However, there is no strong {\it a priori} reason to believe that the symmetries of the background metric ought to be exact. In fact, the observational evidence for spatial isotropy and homogeneity is rather weak, if not controversial. \\

Based on the WMAP data,
the assumption of spatial isotropy
was questioned in Ref.~\cite{deOliveira-Costa:2003utu},
where the authors pointed at the anomalous alignment of the
Cosmic Microwave Backround (CMB) quadrupole
and octopole (at the level $\sim 1/60$)
(for a recent review of CMB anomalies, see Ref.~\cite{Abdalla:2022yfr}).
The authors of Ref.~\cite{deOliveira-Costa:2003utu} pointed out that
the Universe with toroidal topology could explain such an alignment
(albeit other features of this model were absent in the data); but
they did not attempt geometric explanations.
Later Refs.~\cite{Planck:2015gmu,Planck:2013okc}
found no evidence for toroidal topology in the Planck satellite data.
This analysis used a limited number of (Euclidean) topologies,
and further investigations are warranted for a more complete understanding of
what the data tell us about
Universe's topology, which include both Euclidean and curved spaces~\cite{COMPACT:2022gbl}.

Land and Magueijo~\cite{Land:2004bs,Land:2005ad,Land:2006bn}
further worked out the ideas in Ref.~\cite{deOliveira-Costa:2003utu},
and pointed out that the Universe
may have a preferred axis, which approximately corresponds to
that of the
CMB dipole, suggesting that, on very large scales, the Universe
violates spatial isotropy.
Given that the WMAP observations of low multipoles were precise enough, the Planck data have not brought deep new insights
into this question, see {\it e.g.} Refs.~\cite{Copi:2013jna}
and~\cite{Planck:2019evm}.~\footnote{The Planck team
is cautious regarding whether statistical anomalies
in the CMB are real or just a statistical fluke:
``The existence of these features is uncontested, but,
given the modest significances at which they deviate from
the standard $\Lambda$CDM cosmological model, and the
{\it a posteriori} nature of their detection, the extent to which they provide evidence for a violation of isotropy in the CMB remains
unclear. It is plausible that they are indeed simply statistical fluctuations.''}

Other CMB anomalies, summarized in~\cite{Planck:2019evm,Planck:2015igc,Schwarz:2015cma}, include
the absence of large-angle correlations in the angular two-point
function~\cite{BennettHalpernHinshaw:2003},
point parity symmetry,
the hemispherical power anomaly~\cite{Eriksen:2003db}, the Cold Spot and other large-scale peaks~\cite{Vielva:2003et,Cruz:2004ce}, with statistical significance for individual anomalies typically up to about 3$\sigma$ (and sometimes more). More recent work \cite{Jones:2023ncn} has studied the joint probability of multiple anomalies happening simultaneously by chance and concludes that this violates statistical isotropy by more than $5\sigma$.

Recent LSS data tend to corroborate some of the the CMB anomalies, and even suggest new ones,
For example, in his recent essay,
Peebles~\cite{Peebles:2022akh} takes a more positive
view on the anomalies, regarding them as tantalizing
hints for new, as-yet-undisclosed physics. In particular,
in section 3, Peebles considers anomalies in
the large scale structure, and remarks: ``The measured
dipoles in the distribution of quasars and in the distributions of radio
galaxies cataloged at several radio frequencies are in about the predicted direction, but the dipole amplitudes are too large, an anomaly,''
for details see
Refs.~\cite{Jain:2003sg,Marinello:2016aay,Aluri:2022hzs,Secrest:2022uvx}.
Peebles also mentions some other (more local) anomalies, including
the local void.
For a more comprehensive overview of the existing anomalies and tensions
in $\Lambda$CDM
(which include the Hubble and $\sigma_8$ tensions)
see Refs.~\cite{Schwarz:2015cma}, \cite{Abdalla:2022yfr}
and~\cite{Perivolaropoulos:2021jda}.

When combined, these observations and remarks present a well-grounded
motivation for considering more general cosmological models that do not make the assumption of spatial isotropy, but are nonetheless capable of accounting what we observe in the night sky.
 \\

Another important question to which we do not have an
unambiguous answer is: ``do we live in a spatially flat or in a spatially curved
universe?'' When CMB data from the Planck mission~\cite{Planck:2018vyg}
are combined with those from
the Atacama Cosmology Telescope (ACT)~\cite{ACTPol:2016kmo,ACT},
and assuming a FLRW geometry,
one obtains a Universe which is consistent with flat spatial sections.
When LSS and Baryon Acoustic Oscillations (BAO) data are included one obtains,
$\Omega_\kappa=-\kappa/H_0^2  = 0.001\pm 0.002$,
where $H_0$ is the expansion rate of the Universe today
and $\kappa$ Gauss' curvature of the spatial sections.
However, Planck's measurements of the primary CMB anisotropies
show evidence for a positively
curved universe ($\kappa>0$),
and depending on the type of analysis used, one obtains
$\Omega_\kappa = -0.044{+0.018\atop -0.015}\; (3.4\sigma)$
when baseline Plik likelihood is used
and $\Omega_\kappa = -0.035  {+0.018\atop -0.013}\; (\gtrsim 2\sigma)$
when CamCode likelihood analysis is used.
At this moment it is unclear whether that is yet another anomaly,
or a calibration problem.
While the Planck collaboration analyses are based on the whole sky data
(with our galaxy and point sources masked out),
the ACT covers a fraction of the sky, possibly indicating a directional dependence in spatial curvature.
A definite answer of this intriguing question will have to wait for
EUCLID~\cite{EUCLID,Euclid:2021icp} and SKAO~\cite{SKAO}, whose observations will break
Planck data degeneracies with regard
to $\Omega_\kappa$ and allow for a highly accurate measurement
of spatial curvature, with an error of the order
$\Delta\Omega_\kappa\sim 10^{-3}$~\cite{Leonard:2016evk,DiDio:2016ykq}.
 \\

In this paper, we explore the consequences of dropping the assumption of spatial isotropy from the outset and considering a broader set of possible {\it spatial geometries}. We are specifically interested in space-times that decompose into one time-like dimension and a three-dimensional spatial section as,
\begin{eqnarray}
\label{eq:LargeScale1} \mathcal{M} &=& \mathds{R} \times \Sigma_3
\label{breakdown: R x Sigma}
\\
\label{eq:LargeScale2}
{\rm d}s^2
&=& -{\rm d}t^2 + a^2(t) \Big ( \gamma_{ij,\Sigma} \ {\rm d}x^i {\rm d}x^j \Big )
\,.
\label{breakdown: metric}
\end{eqnarray}
We retain the assumption that spatial sections expand isotropically in~\eqref{eq:LargeScale2} for now,~\footnote{In
section~\ref{sec:AnisotropicScaleFactors} we drop the assumption of spatial isotropy
of the scale factor, and present a detailed analysis of its dynamics.}
but place no spatial isotropy requirement on the spatial section
($\Sigma_3$, $\gamma_{ij,\Sigma}$) itself. \\

 The breakdown in
equations~(\ref{breakdown: R x Sigma}--\ref{breakdown: metric})
hinges on the assumption that there exists a spacelike three-dimensional
hypersurface on which fluctuations in matter density vanish or
are very small,
and it exists if the {\it Cosmological Principle} holds approximately.
There is a large body of observations which supports
its approximate version;
as discussed above, the data suggest a small violation of spatial
isotropy.

Here we consider
geometry of the Universe on large scales,
and leave the question of topology for a future investigation.
Namely, to each geometry one can
associate different topologies.~\footnote{For example,
if the Universe's spatial sections are flat, then its geometry
is ${\mathds R}^3$, which can be considered as the covering
space of various topologies, the simplest ones being
toroidal (${\mathds T}^3$), cuboidal
(${\mathds T}^2\times {\mathds R}$) and slab
(${\mathds T}\times{\mathds R}^2\simeq S^1\times{\mathds R}^2$).
}
There is a large body of
work devoted to investigating topology
of the Universe, for reviews of existing attempts
see Refs.~\cite{Cornish:1997rp,Luminet:1999qh,Sandhu:2016gbz,Planck:2015gmu,Planck:2013okc,COMPACT:2022gbl,COMPACT:2022nsu}.
Early attempts include the works of Starobinsky~\cite{Starobinsky:1993yx}
and de Oliveira-Costa and Smoot~\cite{deOlivieraCosta:1994eb},
where the authors use the COBE data to
investigate whether the Universe has toroidal topology,
and find no evidence for it. Refs.~\cite{Cornish:2003db,Vaudrevange:2012da} look for circles in the CMB sky that would be an
important sign of topology, but find no compelling evidence
in favor of such circles.

It is interesting to note that topology can affect the amplitude
of CMB fluctuations on large angular scales.
Thus Ref.~\cite{Luminet:2003dx} showed that
dodecahedral space topology in a spatially flat universe
can suppress the amplitude
of CMB fluctuations on the largest angular scales,
thus explaining some of the large scale CMB anomalies.
Similar results were found
in Ref.~\cite{Bernui:2018wef},
where the authors used
${\mathds R}^2\times S^1$ topology
to explain the lack of CMB large-angle correlations.~\footnote{
These findings are to be contrasted with
the Planck 2013~\cite{Planck:2013okc}, where one reads:
``we consider flat spaces with cubic toroidal (T3), equal-sided chimney
(T2) and slab (T1) topologies, three multi-connected spaces of constant positive curvature (dodecahedral, truncated cube and octahedral) and
two compact negative-curvature spaces. These searches yield no detection of the compact topology with the scale below the diameter of the last
scattering surface.'' However, a more recent investigation~\cite{COMPACT:2022nsu}
showed that, depending on topology, the lower constraint on the topology (length) scale may be weaker by a factor that ranges from 2 to at least 6.}
 \\

The search for candidates for $\Sigma_3$ leads us quite naturally to consider classification schemes for three-dimensional manifolds. For such a classification scheme, in this section and the next, we look towards the eight model geometries described in William Thurston's geometrization conjecture~\cite{Thurston}, proven by Grigori Perelman in the early 2000s~\cite{Perelman:2003uq,Perelman:2006un,Perelman:2006up}. We consider the effect of the large-scale anisotropy inherent in these model geometries on the evolution of the Universe (Sections~\ref{sec:BackgroundEvolution} and \ref{sec:AnisotropicScaleFactors}). Additionally, we
study how large-scale anisotropy warps trajectories of light, thus deforming the image of faraway objects and affecting distance measures such as angular diameter distance and luminosity distance (Sections~\ref{sec:DistanceMeasures} and \ref{sec:DistanceMeasurePlots}).

\subsection{Thurston-Perelman's geometrization theorem}

Thurston-Perelman's geometrization theorem~\footnote{This theorem was initially formulated as a conjecture by Thurston and was later proven in Perelman. We therefore prefer to term it as a theorem as opposed to the conjecture moniker it commonly retains in literature.} is a partial classification of three-dimensional manifolds, analogous to the uniformization, theorem that classifies the possible geometries of Riemann surfaces. The main difference lies in the fact that not every 3-manifold can be endowed with a unique geometry, but rather every 3-manifold can be cut into pieces that can. Many formulations of the theorem exist, we will present the wording of Thurston's original publication.

\vspace{1.5em}
\hspace{0.05 \linewidth}\begin{minipage}[c]{0.8 \linewidth}
\noindent\textbf{Thurston-Perelman's geometrization theorem} \cite{Thurston}, \cite{Perelman:2003uq,Perelman:2006un,Perelman:2006up} \\
\noindent\textit{The interior of every compact 3-manifold has a canonical decomposition into pieces which have geometric structures.}
\end{minipage}
\vspace{1.5em}

Central to this theorem are the concepts of a geometry and a geometric structure. Briefly put, a \emph{Geometry} is a pair $(X,\text{Isom(}X\text{)})$ consisting of a simply connected, complete and homogeneous Riemannian manifold $X$ and its isometry group. A geometry $(Y,\text{Isom(}Y\text{)})$ is said to have a \emph{Geometric Structure} based on $X$ if there is a subgroup $A \subset \text{Isom(}X\text{)}$ such that $Y$ is isometric to $X/A$ and $\text{Isom(}Y\text{)}$ is homeomorphic to $\text{Isom(}X\text{)}/A$.

To give an example, the manifold $\mathds{H}^2 \times S^1$ has a geometry based on $\mathds{H}^2 \times \mathds{R}$, since one can obtain the former from the latter taking a quotient with the subgroup $\mathds{1}_{\mathds{H}^2} \times \mathds{Z}$ of $\text{Isom(}\mathds{H}^2 \times \mathds{R}\text{)}$. \\

This notion can be used to define an ordering within the set of geometries, where geometry $A$ is said to be be of lower order of $B$ if $\text{Isom(}A\text{)}$ is properly contained in $\text{Isom(}B\text{)}$. A natural follow-up question is to ask whether there are maximal geometries with respect to this odering: are there geometries $(X,\text{Isom(}X\text{)})$ for which $\text{Isom(}X\text{)}$ is not properly contained in the isometry group of any other manifold? Thurston provides an answer to this exact question in his paper, which we will paraphrase here.

\vspace{1.5em}
\hspace{0.05 \linewidth}\begin{minipage}[c]{0.8 \linewidth}
\textit{Any maximal, simply connected, three-dimensional geometry $X$ that admits a compact quotient is equivalent to one of the eight geometries below.}
\begin{multicols}{4}
\begin{itemize}
    \item $\mathds{R}^3$
    \item $\widetilde{\text{U}(\mathds{H}^2)}$
    \item $\mathds{H}^3$
    \item $\mathds{H}^2 \times \mathds{R}$
    \item $S^3$
    \item $S^2 \times \mathds{R}$
    \item $\text{Nil}$
    \item $\text{Solv}$
    \,.
\end{itemize}
\end{multicols}
\end{minipage}
\vspace{1.5em}

\noindent These eight maximal geometries can be said to form the building blocks of all compact 3-manifolds. This means that if we want to investigate space-time manifolds that decompose as in equation \eqref{eq:LargeScale1}, then we can make a good start by modeling $\Sigma_3$ as one of the eight geometries of the above classification. \\

The fact that this classification applies only to compact manifolds or that a given manifold may consist of multiple copies of these geometries should not worry us too much. Due to the fact that inflation exponentially enlarges spatial sections of the Universe, it is highly likely that any complex geometric structure of the very early Universe is hidden beyond the Hubble radius, if we assume cosmic inflation occurred. Indeed, we argue in Section \ref{sec:LengthScales} that the curvature radius of any of these geometries is larger than the diameter of the observable Universe. It is therefore reasonable to assume that the observable patch of the Universe corresponds to only \textit{one} of the Thurston geometries and, similarly, to make the additional simplifying assumption that many subtleties arising due to topology are not easily observable.~\footnote{Nevertheless, from the phenomenological point of view, it is worth investigating the signatures of topology of the Universe and confront them with the data. In fact, there is a large body of work mentioned above and doing precisely that, for reviews see Refs.~\cite{Cornish:1997rp,Luminet:1999qh,Sandhu:2016gbz,Planck:2015gmu,Planck:2013okc,COMPACT:2022nsu}.
All of these works assume that the geometry of the spatial section is flat, {\it i.e.} $\Sigma_3 = {\mathds R}^3$. The question of how topology may affect the analyses performed in this work is not something we will
address in this work. Needless to say, this would be a natural question to investigate.}

We will concentrate on the consequences of introducing large-scale spatial anisotropy in the spatial part of the metric. We will make the simplifying assumption that the spatial section of the (observable) Universe corresponds to a single Thurston geometry and put aside any further topological considerations. An investigation into the possible effects of topology we will leave for future work.


\section{Thurston Space-Times}
\label{sec:ThurstonSpacetimes}

In this section we will present explicit coordinate representations of space-times based on Thurston geometries. Following the approach outlined in the previous section, we will decompose space-time at large scales as
\begin{align}
\label{eq:ThustonCosmo1} \mathcal{M} &= \mathds{R} \times \Sigma_3 \\
\label{eq:ThustonCosmo2} {\rm d}s^2        &= -{\rm d}t^2 + a^2(t) \bigg ( \gamma_{ij,\text{Thurston}} \ {\rm d}x^i {\rm d}x^j \bigg ),
\end{align}
where $\gamma_{ij,\text{Thurston}}$ is specific to each of the eight geometries of this theorem. \\

\noindent The spatial part contains a real-valued curvature parameter $\kappa$ that distinguishes between positive ($\kappa > 0$), zero ($\kappa = 0$) and negative ($\kappa < 0$) curvature. This parameter also defines a length scale, the radius of curvature, which can be written as $L = 1/\sqrt{\kappa}$ for geometries with positive curvature, $L = 1 /\sqrt{-\kappa}$ for those with negative curvature, and in the zero curvature case $L \rightarrow \infty$ as $\kappa$ approaches zero. We will use the curvature parameter $\kappa$ and curvature radius $L$ interchangeably in this text.

\vspace{.5em}
\subsection{The FLRW geometries, $\mathds{R}^3$, $\mathds{H}^3$ and $\mathbf{S}^3$ }

The first three Thurston geometries, $\mathds{R}^3$, $\mathds{H}^3$ and $\mathbf{S}^3$, are the only isotropic geometries in Thurston's classification. They are, of course, exactly the spatial slices of the familiar FLRW space-time. We can parameterise all three simultaneously in hyperspherical coordinates as follows.
\begin{align}
\label{eq:FLRWmetric0}
{\rm d}s^2 &= -{\rm d}t^2 + a^2(t)
\bigg ( {\rm d}\chi^2 + S^2_\kappa(\chi) {\rm d}\mathbf{\Omega}^2 \bigg )
= -{\rm d}t^2 + a^2(t) \bigg ( {\rm d}\chi^2 + S^2_\kappa(\chi) \big({\rm d}\theta^2 + \sin^2(\theta) {\rm d}\phi^2 \big) \bigg )
\,, \\
\label{eq:SKappa} S_\kappa(\chi) &=    \begin{cases}
            \sin  (\chi \sqrt {\kappa }) / \sqrt{ \kappa } & \text{ if } \kappa > 0 \\
            \chi                                           & \text{ if } \kappa = 0 \\
            \sinh (\chi \sqrt{-\kappa}) / \sqrt{-\kappa} & \text{ if } \kappa < 0
            \end{cases} \\
            &=    \begin{cases}
            L \sin  (\chi/L) & \text{ if } \kappa > 0 \\
            \chi             & \text{ if } \kappa = 0 \\
            L \sinh (\chi/L) & \text{ if } \kappa < 0\,,
            \end{cases}
\end{align}
where $\chi \in [0,\infty)$, $\theta \in [0,\pi)$ and $\phi \in [0, 2\pi)$. The coordinate $\chi$ measures comoving distance along a radial geodesic and the angles $\theta$ and $\phi$ are the \emph{polar} and \emph{azimuthal} angles, respectively.

\vspace{.5em}
\subsection{$\mathds{R}\times\mathds{H}^2$ and $\mathds{R}\times\mathbf{S}^2$}

The next two geometries, $\mathds{R}\times\mathds{H}^2$ and $\mathds{R}\times\mathbf{S}^2$, are the first anisotropic spaces of Thurston's classification schema. We can think of the resulting space-time as an FLRW space-time of two dimensions, together with a third dimension that is flat. The resulting space is still homogeneous, but is anisotropic due to this special direction. We will therefore present them using hyperspherical coordinates of one dimension lower, with a free coordinate $z$.
\begin{equation}
\label{eq:FLRW2Dmetric0}
{\rm d}s^2 = -{\rm d}t^2 + a^2(t)
\bigg ( {\rm d}z^2 + {\rm d}\chi^2 + S_\kappa^2(\chi) {\rm d}\phi^2 \bigg )
\,.
\end{equation}
Here  $\chi \in [0,\infty)$ and $\phi \in [0, 2\pi)$ as before; but rather than including a polar angle $\theta$, we instead have a real coordinate $z \in \mathds{R}$ orthogonal to the $(\chi,\phi)$ plane. The parameter $\kappa$ again distinguishes between positive and negative curvature through $S_\kappa$, defined in \eqref{eq:SKappa}. \\

\vspace{.5em}
\subsection{$\widetilde{\text{U}(\mathds{H}^2)}$}

The sixth Thurston geometry, $\widetilde{\text{U}(\mathds{H}^2)}$, is the universal cover of the unit tangent bundle of the hyperbolic plane. To derive its metric we will mostly follow the derivation of Fagundes
in~\cite{Fagundes:1991uy} and begin with the following metric of $\mathds{H}^2$.
\begin{equation}
\label{eq:ChirstoffelUH2}
{\rm d}{\Sigma^2_{\mathds{H}^2}} = {\rm d}x^2 + \cosh^2(x) {\rm d}y^2
\,.
\end{equation}
\noindent A unit tangent vector $\hat{u}_p \in \text{U}_p(\mathds{H}^2)$ at any point $p = (x,y) \in \mathds{H}^2$ must now satisfy $\hat{u}_p \cdot \hat{u}_p = 1$. This means means we can write $\hat{u}_p = \big(\cos(\phi),\ \tfrac{\sin(\phi)}{\cosh(x)}\big)$, with $0 \leq \phi < 2\pi$. For a small displacement $dp^i$ we can calculate the total differential,
\begin{equation}
{\rm D} u^i = \left( \frac{\partial u^i}{\partial p^k} + \Gamma^i_{jk} u^j \right) {\rm d}p^k + \frac{\partial u^i}{\partial \phi} {\rm d}\phi,
\end{equation}
where $i,\ j,\ k \in \{1,2\}$, $p^1 = x$ and $p^2 = y$. The nonzero Christoffel symbols obtained from \eqref{eq:ChirstoffelUH2} are $\Gamma^1_{22}
= - \sinh(x) \cosh(x)$ and $\Gamma^2_{12} = \Gamma^2_{21} = \tanh(x)$. Therefore we get
\begin{align}
{\rm D} \hat{u}^1 &= -\sinh(x)\sin(\phi) {\rm d}y - \sin(\phi) {\rm d}\phi \\
{\rm D} \hat{u}^2 &= +\tanh(x)\cos(\phi) {\rm d}y + \frac{\cos(\phi)}{\cosh(x)}
 {\rm d}\phi.
\end{align}
The length of ${\rm D}\hat{u}$ is then given by $\big({\rm D} \hat{u}^1\big)^2 + \big({\rm D} \hat{u}^2\big)^2 \cosh^2(x)$, so that the metric on $\text{U}(\mathds{H}^2)$ can be written as:
\begin{align}
\label{eq:UH2MetricA}
{\rm d}\Sigma^2_{\text{U}(\mathds{H}^2)} &= {\rm d}x^2 + \cosh^2(x) {\rm d}y^2
+ \big({\rm D}\hat{u}^1\big)^2 + ({\rm D} \hat{u}^2)^2 \cosh^2(x) \\
\label{eq:UH2MetricB}             &= {\rm d}x^2 + \cosh^2(x) {\rm d}y^2 + \big({\rm d}\phi + \sinh(x) {\rm d}y \big)^2.
\end{align}
Note that the topology of $\text{U}(\mathds{H}^2)$ is homeomorphic to the Cartesian product of the 2-plane with the circle. This means that this space is path-connected, but not simply-connected: there are loops that wind around $\phi$ that cannot be shrunk to a point. Taking the universal cover of $\text{U}(\mathds{H}^2)$ means that we must `unroll' the circle $\mathbf{S}$ to a line by promoting the angle $\phi$ to a real variable $z$. Since the metric in~(\ref{eq:UH2MetricA}--\ref{eq:UH2MetricB}) does not contain a length scale, we will introduce it by setting $x \rightarrow x \sqrt{-\kappa} =  x/L$ in the argument of the hyperbolic functions. \\

\noindent The (identity component of the) $\widetilde{\text{U}(\mathds{H}^2)}$ space-time can then be presented as $\mathds{R}^{1,3}$ with the following metric: 
\begin{align}
\label{eq:UH2Metric}
{\rm d}s^2 &= -{\rm d}t^2 + a^2(t) \left({\rm d}x^2 + \cosh^2(x\sqrt{-\kappa}){\rm d}y^2 + \big({\rm d}z+ \sinh(x\sqrt{-\kappa}) {\rm d}y \big)^2 \right) \\
                            &= -{\rm d}t^2 + a^2(t) \left({\rm d}x^2 + \cosh^2(x/L) {\rm d}y^2 + \big({\rm d}z+ \sinh(x/L ) {\rm d}y \big)^2 \right).
\end{align}

\subsubsection{A sidenote on $\widetilde{\text{SL}(2,\mathds{R})}$}

In literature, $\widetilde{\text{U}(\mathds{H}^2)}$ is often used interchangeably with $\widetilde{\text{SL}(2,\mathds{R})}$ in the context of the geometrization theorem. This identification is sensible in a topology or differential geometry context, as $\text{SL}(2,\mathds{R})$ acts naturally on $\mathds{H}^2$ by way of a M\"obius transformation. This action can be extended to $\text{U}(\mathds{H}^2)$ through the tangent map, which induces a diffeomorphism between the two manifolds. However, this diffeomorphism is \textit{not} an isometry. To see this, recall the definition of $\text{SL}(2,\mathds{R})$,
\begin{equation}
\text{SL}(2,\mathds{R}) = \left\{ \begin{pmatrix} a & b \\ c & d \end{pmatrix} \hspace{.5em}\bigg|\hspace{.5em} a,b,c,d \in \mathds{R} \hspace{.5em}\&\hspace{.5em} ad - bc = 1 \right\},
\end{equation}
and parameterise any 2-dimensional real matrix $g$ as
\begin{equation}
g=\begin{pmatrix} a & b \\ c & d \end{pmatrix} = \begin{pmatrix} X_1 + X_3 & X_4 + X_2 \\ X_4 - X_2 & X_1 - X_3 \end{pmatrix}
\end{equation}
with $X_i \in \mathds{R}$. For $g$ to be in $\text{SL}(2,\mathds{R})$ it must hold that $X_1^2 + X_2^2 - X_3^2 - X_4^2 = \det(g) = 1$. In other words, $\text{SL}(2,\mathds{R})$ can be constructed as a unit (hyper)sphere in $\mathds{R}^{(2,2)}$. The metric on $\text{SL}(2,\mathds{R})$ can then be induced from the ambient space $\mathds{R}^{(2,2)}$ by using the Iwasawa decomposition and we can take similar steps to the derivation above to pass to the universal cover and restore a length scale $L$. We then get the following metric,
\begin{align}
{\rm d}s &= -{\rm d}t^2 + a^2(t) \left( {\rm d}x^2 + {\rm d}y^2 - {\rm d}z^2 + 2 \sinh(2x\sqrt{-\kappa}){\rm d}y {\rm d}z \right) \\
         &= -{\rm d}t^2 + a^2(t) \left( {\rm d}x^2 + {\rm d}y^2 - {\rm d}z^2 + 2 \sinh(2x/L            ){\rm d}y {\rm d}z \right).
\end{align}
Note that ${\rm d}z^2$ has a minus sign in this expression, as the spatial section has inherited a $(+,+,-)$ signature from the metric of $\mathds{R}^{(2,2)}$. This shows that $\text{SL}(2,\mathds{R})$ is not locally Euclidean and so cannot be isometric to $\text{U}(\mathds{H}^2)$. For the purposes of this paper we will therefore refrain from identifying these two spaces.

\vspace{.5em}
\subsection{Nil and Solv}

The last two geometries, Nil and Solv, are hyperbolic geometries ($\kappa < 0$) that are the geometries of a Lie group. Nil can be described as the geometry of the Heisenberg group, while Solv can be described as the geometry of the identity component of the 2-dimensional Poincar\'e group. There are standard ways of presenting these manifolds as $\mathds{R}^3$ endowed with a special metric. \\

\noindent In the case of Nil, the space-time can be presented as,
\begin{align}
\label{eq:Nilmetric0}   
{\rm d}s^2 &= -{\rm d}t^2 + a^2(t) \Bigg ({\rm d}x^2 + \Big (1-\kappa x^2 \Big) {\rm d}y^2 + {\rm d}z^2 - 2 x \sqrt{-\kappa} \ {\rm d}y {\rm d}z \Bigg ) \\
     &= -{\rm d}t^2 + a^2(t) \Bigg ({\rm d}x^2 + \Big (1+x^2/L^2 \Big) {\rm d}y^2 + {\rm d}z^2 - 2 x/L \ {\rm d}y {\rm d}z \Bigg ).
\end{align}
The space-time based on Solv can be presented as,
\begin{align}
\label{eq:Solvmetric0}  
{\rm d}s^2 &= -{\rm d}t^2 + a^2(t) \Bigg ( {\rm e}^{2z\sqrt{-\kappa}} {\rm d}x^2 + {\rm e}^{-2z\sqrt{-\kappa}} {\rm d}y^2 + {\rm d}z^2 \Bigg ) \\
     &= -{\rm d}t^2 + a^2(t) \Bigg ( {\rm e}^{2z/L} {\rm d}x^2 + {\rm e}^{-2z/L} {\rm d}y^2 + {\rm d}z^2 \Bigg )\,,
\end{align}
with $x,y,z\in\mathbb{R}$, $\kappa<0$ and $L>0$. Note that the Solv metric is diagonal,
while that of Nil is not.

Finally, we note that the $\widetilde{\text{U}(\mathds{H}^2)}$~(\ref{eq:UH2Metric})
and Nil~(\ref{eq:Nilmetric0}) geometries contain off-diagonal terms,
signifying a twist and deformation of the $yz$-plane, see also Figures~\ref{fig:UH2Geodesics1}--\ref{fig:NilGeodesics3} from the appendix.
To get an idea of the size of the deformation, one can diagonalize these geometries
by performing $x$-dependent rotations (which are {\it not} coordinate transformations).
The eigenvalues in these local diagonal frames are
{\it e.g.} for Nil: $1$ in the $x$-direction and $1-\kappa x^2 \pm  x\sqrt{-\kappa}\sqrt{1-\kappa x^2}$
in the $yz$-plane,
implying that (at short distances) the contraction/dilation along the principal axes in the $yz$-plane grows
as $\sim\pm x\sqrt{-\kappa} =  \pm x/L$.


\section{Background Evolution}
\label{sec:BackgroundEvolution}

In this section we will derive the evolution of the cosmological background for each of the space-times. It will turn out that the evolution of all eight Thurston space-times is very similar to the evolution of the first three space-times, which we recall are the ordinary FLRW cases. This is because the metrics in equations \eqref{eq:FLRWmetric0}, \eqref{eq:FLRW2Dmetric0}, \eqref{eq:UH2Metric}, \eqref{eq:Nilmetric0} and \eqref{eq:Solvmetric0} presented in the previous section all admit a very similar Einstein tensor:
\begin{equation}
G^{\mu}_{\hspace{1ex}\nu} = - \text{diag}\left( 3\frac{\dot{a}}{a}, \hspace{1ex} \frac{\dot{a}+2 \ddot{a} a}{a^2}, \hspace{1ex} \frac{\dot{a}+2 \ddot{a} a}{a^2}, \hspace{1ex} \frac{\dot{a}+2 \ddot{a} a}{a^2}\right) + \frac{\kappa}{a^2} \text{diag}\left(K^{(0)}, \hspace{1ex}  K^{(1)}, \hspace{1ex}  K^{(2)}, \hspace{1ex}  K^{(3)}  \right)
\,.
\end{equation}
Here $K^{(0)}$, $K^{(1)}$, $K^{(2)}$ and $K^{(3)}$ are a set of four parameters~\footnote{We have used round brackets around the indices to indicate that these are a set of parameters and not a vector.} specific to each Thurston geometry that determine how strongly terms in the energy-momentum tensor are coupled to the curvature parameter $\kappa$. This makes the calculation fairly straightforward, as we can leave these parameters implicit to solve for all geometries simultaneously. \\

\noindent The Nil and $\widetilde{\text{U}(\mathds{H}^2)}$ geometries are slightly more complicated in that their Einstein tensor has an additional nonzero off-diagonal term.
\begin{align}
G^{3}_{\hspace{1ex}2,\text{Nil}}                         &=        x\sqrt{-\kappa}  \frac{\kappa}{a} \\
G^{3}_{\hspace{1ex}2,\widetilde{\text{U}(\mathds{H}^2)}} &= 2\sinh(x\sqrt{-\kappa}) \frac{\kappa}{a} .
\end{align}
\noindent For all geometries, the Ricci scalar takes the form
\begin{equation}
\label{eq:RicciscalarGEN} R = 6 \, \frac{ a \ddot{a}+\dot{a}^2}{a^2} - 2 K^{(0)} \frac{\kappa}{a^2}.
\end{equation}
This expression is devoid of any coordinates, which confirms that the Thurston space-times are indeed homogeneous -- despite this not being immediately manifest from their metric representation. This means that we are free to choose the origin of our coordinate system. We will exploit this in later sections to simplify calculations.

\subsection{General solution}

The most general fluid solution compatible with this Einstein tensor is:
\begin{equation}
T^{\mu}_{\hspace{1ex}\nu} = \rho \; u^\mu u_\nu + p \; \delta^\mu_{\hspace{1ex}\nu}
 + \pi^\mu_{\hspace{1ex}\nu}
\,,
\label{Tmn with pi}
\end{equation}
where $ \rho=\rho(t)$ and $p=p(t)$ denote the fluid energy density and pressure, respectively
$u^\mu=u^\mu(t)$ is the velocity vector of the fluid and $\pi^{\hspace{1ex}\mu}_\nu$ is the shear tensor, which is symmetric,  $\pi^{\mu}_{\hspace{1ex}\nu} = \pi^{\hspace{1ex}\mu}_\nu$, traceless, $\pi^\mu_{\hspace{1ex}\mu} = 0$, and transverse, $\pi^\mu_{\hspace{1ex}\nu} u^\nu = 0$.  Clearly the {\it Ansatz}~(\ref{Tmn with pi})
goes beyond the perfect fluid {\it Ansatz} of the standard Friedmann geometries, for which
the shear tensor vanishes. In section~\ref{sec:AnisotropicScaleFactors} we consider an alternative approach where we set
$\pi^\mu_{\hspace{1ex}\nu} = 0$, but allow for anisotropic expansion rates. \\

\noindent We now consider the Einstein equation in the rest frame of this fluid, where $u^\mu = (1,0,0,0)$. This yields a slight modification of the familiar Friedmann equations,
\begin{align}
\label{eq:FriedmannIGEN}      \hspace{0em} ^0_{\hspace{1ex}0}&\hspace{1em}\text{equation:}  \hspace{-5em} &
H^2 &= \frac{8 \pi G}{3} \rho + \frac{\Lambda}{3} + \frac{\kappa K^{(0)}}{3 a^2} \\
\label{eq:FriedmannIIGENUgly} \hspace{0em} ^{\hat{i}}_{\hspace{1ex}\hat{i}}&\hspace{1em}\text{equation:} \hspace{-6em} &
\frac{\ddot{a}}{a} &= - \frac{4 \pi G}{3} (\rho + 3 p) + \frac{\Lambda}{3} + \frac{\kappa}{a^2} \left(\frac{K^{(\hat{i})}}{2} - \frac{K^{(0)}}{6} \right) - 4 \pi G \pi^{\hat{i}}_{\hspace{1ex}\hat{i}}.
\end{align}
There is no summation over repeated indices carrying hats in the equations in this section.
We have omitted the $\hat{i}_{\hspace{1ex}\hat{j}}$ equation for $\hat{i} \neq \hat{j}$; for all geometries except Nil and $\widetilde{\text{U}(\mathds{H}^2)}$ this reads $\pi^{\hat{i}}_{\hspace{1ex}\hat{j}}=0$
for $\hat{i} \neq \hat{j}$, except for $\pi^{3}_{\hspace{1ex}2}$, in which case one gets,
\begin{align}
\label{shear tensor: Nil} \pi^{3}_{\hspace{1ex}2,\text{Nil}}                         &=        x\sqrt{-\kappa}  \frac{\kappa}{8 \pi G a^2} \\
\label{shear tensor: UH2} \pi^{3}_{\hspace{1ex}2,\widetilde{\text{U}(\mathds{H}^2)}} &= 2\sinh(x\sqrt{-\kappa}) \frac{\kappa}{8 \pi G a^2}
\,.
\end{align}
Therefore these off-diagonal stresses are suppressed when compared with
the diagonal components $\pi^i_{\;i}$
by a factor $\sim x\sqrt{-\kappa}=x/L$, which is much smaller than unity
on distances small in comparison to the curvature radius, {\it i.e.} $|x|\ll L$. This suggests that
their contribution to the dynamics of the Universe (in the local Hubble patch) is much smaller than that of the diagonal components,
and can therefore be -- to the leading approximation -- neglected.

The right-hand side of \eqref{eq:FriedmannIIGENUgly} has terms containing the (fixed) spatial index $\hat{i}$, however the left-hand side does not. This means that all index-carrying terms on the right-hand side must be the same, independent of the choice of index $\hat{i}$. Therefore for all $\hat{i}$, $\hat{j}$ it must hold that:
\begin{align}
8 \pi G a^2 \pi^{\hat{i}}_{\hspace{1ex}\hat{i}} - \kappa K^{(\hat{i})} = 8 \pi G a^2 \pi^{\hat{j}}_{\hspace{1ex}\hat{j}} - \kappa K^{(\hat{j})}
\end{align}
Since $\pi$ is traceless, we can solve for its diagonal elements,
\begin{equation}
\label{eq:ShearTensor} \pi^{\hat{i}}_{\hspace{1ex}\hat{i}} = \frac{\kappa}{24 \pi G a^2} \left( 2 K^{(\hat{i})} - \sum_{\hat{j} \neq \hat{i}} K^{(\hat{j})}  \right).
\end{equation}
This can be used to rewrite the second Friedmann-like equation \eqref{eq:FriedmannIIGENUgly} into a more complete form,
\begin{align}
\label{eq:FriedmannIIGEN} \frac{\ddot{a}}{a}  &= - \frac{4 \pi G}{3} (\rho + 3 p) + \frac{\Lambda}{3} + \frac{\kappa}{6 a^2} \left(- K^{(0)} + K^{(1)} + K^{(2)} + K^{(3)} \right).
\end{align}
When we plug in the values for the $K$-parameters, it will turn out that the last term in this equation, $- K^{(0)} + K^{(1)} + K^{(2)} + K^{(3)}$, vanishes for all of the Thurston space-times and so \eqref{eq:FriedmannIIGEN} reduces to the ordinary form of the second Friedmann equation. We obtain the evolution of $\rho$ directly by differentiating \eqref{eq:FriedmannIGEN} with respect to time
\begin{equation}
2 H \left ( \frac{\ddot{a}}{a} - H^2 \right) =   \frac{8 \pi G}{3} \dot{\rho} - 2 H \frac{K^{(0)} \kappa}{3a^2}. \\
\end{equation}
Using\eqref{eq:FriedmannIGEN} and \eqref{eq:FriedmannIIGEN}, we can rewrite this as,
\begin{equation}
\label{eq:EnergyEvolutionGEN} \dot{\rho} + 3 H (\rho + p ) = \frac{\kappa H}{8 \pi G a^2} \,\left(- K^{(0)} + K^{(1)} + K^{(2)} + K^{(3)} \right) = 0
\,. \\
\end{equation}

\subsection{Friedmann Equations}

Let's now consolidate the equations from the previous section into the following set:
\begin{align}
\label{eq:FriedmannI}      &\textbf{Friedmann I:}          &\hspace{-2em} H^2                \ &= \  \frac{8 \pi G}{3} \rho + \frac{\Lambda}{3} + \frac{\kappa K^{(0)}}{3 a^2} & & \\
\label{eq:FriedmannII}     &\textbf{Friedmann II:}         &\hspace{-2em} \frac{\ddot{a}}{a} \ &= \ -\frac{4 \pi G}{3} (\rho + 3 p) + \frac{\Lambda}{3} & &\\
\label{eq:EnergyEvolution} &\textbf{Energy Evolution:}     &\hspace{-2em}                  0 \ &= \  \dot{\rho}+ 3 H (\rho + p ) & & \\
\label{eq:ShearConstraints}&\textbf{Shear Constraints I:}  & &\hspace{-2em} \pi^{\hat{i}}_{\hspace{1ex}\hat{i}} \ = \ \frac{\kappa}{24 \pi G a^2} \left( 2 K^{(\hat{i})} - \sum_{\hat{j} \neq \hat{i}} K^{(\hat{j})}  \right) & &\\
\label{eq:SpecialShear}    &\textbf{Shear Constraints II:} &\hspace{2em}    \pi^{3}_{\hspace{1ex}2,\text{Nil}}                \ &= \ x\sqrt{-\kappa}  \frac{\kappa}{8 \pi G a^2}, &\hspace{-2em} \pi^{3}_{\hspace{1ex}2,\widetilde{\text{U}(\mathds{H}^2)}} = 2\sinh(x\sqrt{-\kappa}) \frac{\kappa}{8 \pi G a^2} & &
\end{align}
The first three of these equations tell us that the last five (anisotropic) Thurston space-times admit Friedmann equations that are of the same form as the first three (isotropic) cases. Importantly, the space-times based on Thurston geometries admit the usual matter, radiation, and dark energy or cosmological constant contributions as constituents of the universe that we are used to from ordinary FLRW space-time. The only difference is that the strength of the curvature contribution to the energy density is determined by the parameter $K^{(0)}$, which varies between geometries and is given in Table~\ref{table:thurstonparameters}. Aside from this difference, many of the standard analyses apply as \textit{e.g.} we can solve the \textit{flatness problem} by the usual means we are familiar with from FLRW Cosmology. \\

\begin{table}[!ht]
\centering
\begin{tabular}{|c|cccc|ccc|c|}
\hline
\textbf{Space-Time}                             & $K^{(0)}$ & $K^{(1)}$  & $K^{(2)}$  & $K^{(3)}$ & $8 \pi Ga^2 \pi^1_{\hspace{1ex}1}/\kappa$ & $8 \pi Ga^2 \pi^2_{\hspace{1ex}2}/\kappa$ & $8 \pi Ga^2 \pi^3_{\hspace{1ex}3}/\kappa$  & $8 \pi Ga^2 \pi^3_{\hspace{1ex}2}/\kappa$ \\ \hline
\textbf{FLRW}                                   & -3        & -1         & -1         & -1        &  0                                        &  0                                        &  0                                         &  0                                        \\
$\mathbf{\mathds{R} \times \mathds{H}^2 / S^2}$ & -1        &  0         &  0         & -1        &  1/3                                      &  1/3                                      & -2/3                                       &  0                                        \\
$\widetilde{\text{U}(\mathds{H}^2)}$            & -5/4      &  1/4       &  1/4       & -7/4      &  2/3                                      &  2/3                                      & -4/3                                       &  $-2 \sinh(x \sqrt{-\kappa})$             \\
\textbf{Nil }                                   & -1/4      &  1/4       &  1/4       & -3/4      &  1/3                                      &  1/3                                      & -2/3                                       &  $x\sqrt{-\kappa}$                        \\
\textbf{Solv}                                   & -1        & -1         & -1         &  1        & -2/3                                      & -2/3                                      &  4/3                                       &  0                                        \\ \hline
\end{tabular}
\caption{Several parameters for the Thurston space-times.}
\label{table:thurstonparameters}
\end{table}

However, the last two equations tell us that this comes at the cost of picking up one or two  additional constraints from the Einstein field equations on the shear tensor $\pi$ that must be satisfied in order for these standard analyses to apply (see also Table~\ref{table:thurstonparameters}).
One might imagine solving this on the matter side of the Einstein field equations by introducing some exotic, anisotropic fluid that has the correct $\propto a^{-2}$ scaling properties. However, this approach creates more problems than it solves. Firstly, one would have to come up with a candidate fluid whose energy density scales appropriately as $\propto a^{-2}$ to make up this exotic fluid. Secondly, one would have to propose a mechanism by which this fluid exerts pressure anisotropically and explain why this is fine-tuned to exactly satisfy the Shear Constraints~\eqref{eq:ShearConstraints}--\eqref{eq:SpecialShear}. Thirdly, since we now have a second energy contribution that scales as $\propto a^{-2}$, curvature being the first, we are introducing an additional contribution to the evolution of $a$ in equations~\eqref{eq:FriedmannI}--\eqref{eq:FriedmannII}; this means must extend the \textit{flatness problem} and find not only an explanation for why the primordial curvature density is very close to zero, but also why the same is the case for the primordial energy density this exotic fluid. \\

Instead, we will opt to solve this on the geometry side of the Einstein field equations in Section~\ref{sec:AnisotropicScaleFactors} by introducing anisotropies in the scale factor.
This alternative approach allows us to solve the Friedmann equations using a perfect fluid {\it Ansatz} at the cost of picking up anisotropies in expansion of the space-time, which grow (decay) in decelerating
(accelerating) space-times, suggesting inflation
as the natural solution both for
the flatness and anisotropy problems. With this in mind,
one can assume anisotropies to be small throughout the expansion history of the Universe.\\

The Nil and $\widetilde{\text{U}(\mathds{H}^2)}$ space-times remain an exception, as their $\pi^3_{\hspace{1ex}2}$ terms cannot be absorbed into anisotropies in the scale factor. However, this term is additionally suppressed by $x/L$ (see Table~\ref{table:thurstonparameters}), which is much less than unity on sub-Hubble scales. Since this will be observationally small compared to other curvature terms in the energy-momentum tensor, we will ignore this term in the dynamics of the Universe.

\subsection{Length scales}
\label{sec:LengthScales}

As a last topic in this section, we will calculate several length scales; these will be used to generate the plots in Section \ref{sec:DistanceMeasurePlots}. In particular, we are interested in the inherent curvature radius $L$ of the geometry and in $\eta_*$, the conformal time until the last scattering surface. \\

\noindent To this end, we introduce  $\rho_\Lambda = \Lambda / (8 \pi G)$
and $\rho_\kappa = K^{(0)} \kappa / (8 \pi G a^2)$ to put the cosmological constant and curvature on the same footing as the other matter contents of the Universe. We can use this to rewrite \eqref{eq:FriedmannI} in terms of energy densities,
\begin{equation}
\label{eq:TimeEvolution1}
 H^2 = \frac{8 \pi G}{3} \big( \rho_\text{matter} + \rho_\Lambda + \rho_\kappa)
\,.
\end{equation}
By setting $\rho_i(t) = \rho_i(0) a(t)^{-2\epsilon_i}$ ($i=$matter, $\Lambda, \kappa$),
where $\epsilon_i$ denotes the slow roll parameter associated with the fluid $i$
and is defined as $\epsilon_i=-\dot H/H^2$ if that fluid would dominate the energy content of the Universe,
we can re-express Eq.~(\ref{eq:TimeEvolution1})
 in terms of fractions of the present critical energy density $\Omega_{i,0}=\rho_{i,0}/\rho_{\rm crit,0}$ as,
\begin{equation}
\label{eq:TimeEvolution2}
 H^2(t) 
 = \frac{8 \pi G}{3} \sum_i \frac{\rho_{i,0}}{a(t)^{2\epsilon_i}}
 = H_0^2 \sum_i \frac{\Omega_{i,0}}{a(t)^{2\epsilon_i}}
  = H_0^2 \sum_i \Omega_{i,0}\big(1+z(t)\big)^{2\epsilon_i}
 \,,
\end{equation}
where $z(t)$ is the redshift,
 $H_0^2 = \frac{8 \pi G}{3} \rho_{\text{crit},0}$ denotes the expansion rate today
and $\rho_{\text{crit},0}$ is the total energy density supporting it. Assuming the Universe has a matter, radiation, curvature and cosmological constant (dark energy) component, we can derive an integral expression for the (conformal) time between two events,
\begin{align}
\label{eq:TimeIntegral}
\eta       &= \int^{\eta_2}_{\eta_1} {\rm d} \eta' \
   = \int^{t_2}_{t_1} \frac{{\rm d}t}{a} \
            = \int^{a_2}_{a_1} \frac{{\rm d}a}{a^2H(a)} \
            = \frac{1}{H_0} \int^{a_2}_{a_1}\frac{ {\rm d}a}
            {\big(\sum_i \Omega_{i,0}a^{4-2\epsilon_i}\big)^{1/2}}
\,.
\end{align}
This is an elliptic integral that does not admit an easy analytical solution in terms of elementary
functions~\cite{Coquereaux:2014sva,Coquereaux:1981ya}.
Fortunately, we will not need a full analytical  solution for~(\ref{eq:TimeIntegral})
 and we can instead compute the integral numerically by plugging in the relevant observed values. \\

\noindent Next, we calculate the effective curvature length scale from equations \eqref{eq:FriedmannI} and \eqref{eq:TimeEvolution2},
\begin{equation}
\Omega_\kappa = \frac{K^{(0)} \kappa}{3 H^2 a^2} \ \ \Leftrightarrow \ \ \kappa = \frac{3 H^2 a^2 \Omega_\kappa}K^{(0)} \ \ \Leftrightarrow \ \ L =  \frac{1}{a H}\sqrt{\left|\frac{K^{(0)}}{3 \Omega_\kappa}\right|}.
\end{equation}
To get a sense of the largest-scale of observable effects in the next chapter, we will calculate $\eta_*$, the conformal time to the surface of last scattering at $z_*\simeq1091$, $a_* \simeq 1/1092$ in terms of the inherent  curvature length scale $L$ of the underlying 3-manifolds,~\footnote{$L$ coincides with the physical curvature radius at present. For earlier times, the physical curvature radius can be expressed as the invariant spatial distance $a(t) L$, but this will not be used in this paper.}
\begin{equation}
\eta_* = L \sqrt{\left|\frac{3\Omega_{\kappa,0}}{K^{(0)}}\right|} \int^{1}_{a_*} \frac{1}{\sqrt{\Omega_{\Lambda,0} \ a^4 + \Omega_{\kappa,0} \ a^2 + \Omega_{\rm m,0} \ a + \Omega_{R,0} }} {\rm d}a.
\label{eta star}
\end{equation}
We now evaluate the integral using numerical values from the 2018 Planck Results \cite{Planck:2018vyg} in Table~\ref{table:Planck2018data}.

\begin{table}[!ht]
\centering
\begin{tabular}{|c|c|l|c|c|}
\hline\rule{0pt}{3ex} & $\Omega_{\Lambda,0}  $ & \multicolumn{1}{|c|}{$\Omega_{\kappa,0}$}  & $\Omega_{R,0} = \Omega_{\rm m,0}/(1+z_{eq}) $ & $\Omega_{\rm m,0}  $ \\ [1ex] \hline \rule{0pt}{3ex}%
Pure CMB constraints  & $0.6834 \pm 0.0084   $ & $-0.0440^{+0.018}_{-0.015}               $ & $(9.29 \pm 0.25)\cdot 10^{-5}             $ & $0.3166 \pm 0.0084 $ \\ [1ex] \hline \rule{0pt}{3ex}%
CMB + Lensing         & $0.6847 \pm 0.0073   $ & $-0.0406 \pm 0.0065                      $ & $(9.31 \pm 0.18)\cdot 10^{-5}             $ & $0.3153 \pm 0.0073 $ \\ [1ex] \hline \rule{0pt}{3ex}%
CMB + Lensing + BAO   & $0.6889 \pm 0.0056   $ & $+0.0007 \pm 0.0019                      $ & $(9.18 \pm 0.17)\cdot 10^{-5}             $ & $0.3111 \pm 0.0056 $ \\ [1ex] \hline
\end{tabular}
\caption{Cosmological parameters from~\cite{Planck:2018vyg}.}
\label{table:Planck2018data}
\end{table}

It is worth reflecting that these constraints were obtained from observations assuming an isotropic model; we have no reason to believe they would stay the same if we assumed an anisotropic model instead. However, the proper statistical analysis for the Thurston geometries is beyond the scope of this paper, so we will work with the constraints given in Table~\ref{table:Planck2018data} for the moment.

We will use the average values for $\Omega_{\Lambda,0}$, $\Omega_{R,0}$ and $\Omega_{\rm m,0}$ and we will take the two standard deviations (2$\sigma$)
 bounds on $\Omega_{\kappa,0}$ from CMB + Lensing + BAO. We arrive at $\Omega_{\kappa,0} = 0.0007+2\times 0.0019 = 0.0045$ for negatively curved geometries and $\Omega_{\kappa,0} = 0.0007-2\times 0.0019 = -0.0031$ for positively curved geometries.~\footnote{It is worth noting that the constraints on $\Omega_\kappa$ are significantly less strong on the first two rows of
 Table~\ref{table:Planck2018data}, for which $\Omega_\kappa$ is strictly negative ($\kappa > 0$) within the 2$\sigma$ bounds. As only the $\mathbf{\mathds{R} \times S^2}$ geometry is anisotropic and positively curved, this would significantly limit the scope of the paper, see Table~\ref{table:LengthScales2},
 unless one would include broader confidence limits ($>3\sigma$).
 Hence we have opted for the constraints that allow for both positively and negatively curved spaces.

If one were to instead allow for both strongly negatively and positively curved space-times, the assumption that proper distance $\lambda$ is small compared to the curvature radius $L$ would start to break down for distances comparable to the Hubble scale. This would necessitate the inclusion of increasingly higher-order contributions,
until the power expansion breaks down entirely. Since the plots in Section~\ref{sec:DistanceMeasurePlots} are generated through a power series in $\eta_*/L$, one would expect contributions at higher order to become increasingly important, which has a potential to change these plots dramatically. Lastly, we would similarly expect that topological considerations will become markedly more important in this regime, as \textit{e.g.} boundary conditions of the Thurston patches will no longer be hidden behind the Hubble horizon.} The resulting values are shown in Table~\ref{table:LengthScales1}, with Table~\ref{table:LengthScales2} contrasting the values from CMB constraints only.

\begin{table}[!ht]
\centering
\begin{minipage}[t]{0.45\linewidth}
\centering
\begin{tabular}{|c|c|c|c|c|}
\hline
\textbf{Space-Time}                         & $\Omega_\kappa$ & $\eta_* H_0$ & $L H_0$  & $\eta_*/L$ \\ \hline
$\mathbf{\mathds{R}^3}$                     &  0              & 3.133        & $\infty$   & 0        \\
$\mathbf{\mathds{H}^3}$                     &  0.0045         & 3.136        & 17.96      & 0.175    \\
$\mathbf{  S^3}$                            & -0.0031         & 3.138        & 14.91      & 0.210    \\
$\mathbf{\mathds{R} \times \mathds{H}^2}$   &  0.0045         & 3.136        & 10.37      & 0.302    \\
$\mathbf{\mathds{R} \times   S^2}$          & -0.0031         & 3.128        &  8.61      & 0.363    \\
$\widetilde{\text{U}(\mathds{H}^2)}$        &  0.0045         & 3.136        & 11.59      & 0.271    \\
\textbf{Nil }                               &  0.0045         & 3.136        &  5.18      & 0.605    \\
\textbf{Solv}                               &  0.0045         & 3.136        & 10.37      & 0.302    \\ \hline
\end{tabular}
\caption{Length Scales conforming to the 2$\sigma$ observational bounds, $-0.0031\leq \Omega_\kappa\leq 0.0045$ from CMB + Lensing + BAO.}
\label{table:LengthScales1}
\end{minipage}\hspace{0.05\linewidth}
\begin{minipage}[t]{0.47\linewidth}
\centering
\begin{tabular}{|c|c|c|c|c|}
\hline
\textbf{Space-Time}                         & $\Omega_\kappa$ & $\eta_* H_0$ & $L H_0$  & $\eta_*/L$ \\ \hline
$\mathbf{\mathds{R}^3}$                     & 0             & 3.111        & $\infty$  & 0        \\
$\mathbf{\mathds{H}^3}$                     & &  &  &  \\
$\mathbf{  S^3}$                            & -0.074        & 3.125        & 8.45      & 0.37    \\
$\mathbf{\mathds{R} \times \mathds{H}^2}$   & & & & \\
$\mathbf{\mathds{R} \times   S^2}$          & -0.074        & 3.125        & 4.88      & 0.64    \\
$\widetilde{\text{U}(\mathds{H}^2)}$        & & & & \\
\textbf{Nil }                               & & & & \\
\textbf{Solv}                               & & & & \\ \hline
\end{tabular}
\caption{The same scales based on the 2$\sigma$ bounds from CMB, $-0.074 \leq \Omega_\kappa\leq -0.08$. Since $\kappa$ is strictly positive for these bounds, hyperbolic spaces are left bank.}
\label{table:LengthScales2}
\end{minipage}
\end{table}

\noindent These tables tell us two important things. Firstly, it tells us that $L$ larger than $\eta_*$ (by about half an order of magnitude for most geometries), which means that we can expand complicated expressions in powers of $\eta_* / L$ to get approximate results. This will be very useful in Section \ref{sec:DistanceMeasurePlots}, as we are not able to find exact expressions for all geometries.

Secondly, and perhaps more importantly, the curvature radius $L$ is larger than the Hubble radius $H_0^{-1}$ for all Thurston geometries. This allows us to substantiate the following assumption: If spatial sections of our Universe are well-described by a patchwork of Thurston geometries, then the Hubble sphere -- and moreso Earth's low-redshift surroundings -- are very likely to belong only a \emph{single} of these patches. As a result, we argue that any detectable effects of anisotropy are similarly likely to be sourced by a \emph{single} geometry.\\

Under this assumption we can start to make predictions about how our Universe would look if we took any one of the eight Thurston-Perelman geometries as our local spatial section and our Universe were indeed curved and anisotropic. In the next section we will discuss two ways in which these effects may become manifest.


\section{Distance Measures}
\label{sec:DistanceMeasures}

Since our chosen constraints put the curvature radius as significantly larger than the size of the observable universe, we likely stand no chance of devising a local experiment that is sensitive enough to detect curvature anisotropy. Rather, we must look for the effects of anisotropy over very long distances or in very large-scale phenomena. We propose to use photon trajectories for this purpose, as they are bent by the presence of curvature and thus directly affected by the anisotropies in the metric. The farther a photon travels, the more apparent any deflection in its trajectory will be, so we expect any effects to be magnified for higher redshifts.

Anisotropic curvature will behave like a large lens that deforms the image of distant objects. The magnitude and shape of this deformation will depend on the position of the object in the night sky in a way that is unique to each of our geometries. This pattern of deformations can serve as a fingerprint by which we can identify a geometry. \\

In this section, we will derive expressions for these deformations by considering the angular diameter distance $d_A$, which is affected by the deformations directly, and the luminosity distance $d_L$, which is affected by the net change in apparent luminosity. In the next section, we will present several figures to visualise this effect for each geometry. \\

Plotting these distance measures will require us to derive null geodesics for each of the Thurston space-times. Fortunately, this becomes a very tractable problem when we transform Eq.~(\ref{breakdown: metric})
 to conformal time:
\begin{align}
0 = {\rm d}s^2  = -{\rm d}t^2 + a^2(t)  {\rm d}\Sigma_3^2
         = a^2(t) \Big(\! -{\rm d}\eta^2 + {\rm d}\Sigma_3^2 \Big)
         \,,\qquad
         \big({\rm d}\Sigma_3^2 =  \gamma_{ij,\Sigma} \ {\rm d}x^i {\rm d}x^j \big)
\,.
\label{lightlike distances}
\end{align}
This equation has the implicit solution ${\rm d}\eta^2 = {\rm d}\Sigma_3^2$,~\footnote{
Given that ${\rm d}\Sigma_3^2$ depends on spatial variables only,
Eq.~(\ref{lightlike distances}) implies that conformal time is a good affine parameter along geodesics, and
can be used to parametrize geodesic distances on $\Sigma_3$.
Rigorously speaking this ceases to be the case when anisotropic expansion rates are considered,
in which case $d\Sigma_3^2$ becomes dynamical (see
section~\ref{sec:AnisotropicScaleFactors} for more details).
}
 which reduces the problem to a 3-dimensional one. That is to say, if we can find some spatial geodesic $x^i(\lambda)$ parameterised by proper distance $\lambda$ on $\Sigma_3$, then $(\eta, x^i(\eta))$ is a geodesic in the full Thurston space-time parameterised by conformal time $\eta$. Finding these spatial geodesics is further simplified by the fact that the manifolds in Thurston's theorem are homogeneous and, as explained in Section \ref{sec:BackgroundEvolution}, we are at liberty to place the observer at the origin of the coordinate system and to consider only radial geodesics. We will leave the derivation of such geodesics to the appendices and assume for the rest of this section that they are known.

Rigorously speaking the analysis in this section applies to rigid geometries. To capture the dynamical aspects of various Thurston geometries discussed in detail in section~\ref{sec:AnisotropicScaleFactors}, one would have to suitably modify the analysis of this section. However, we will show later in Section \ref{sec:AnisotropicScaleFactors} that geometries evolve rather slowly, so that the results derived in this section will apply to most of the situations of interest.

\subsection{Distance measures in an isotropic universe}

In FLRW space-time, calculating both distance measures is straightforward due to spherical symmetry on spatial slices.

\subsubsection{Angular diameter distance}
\label{sec:ADDFLRW}

Angular diameter distance $d_A$ can be defined as the ratio of an object's physical size $h$ at the time of emission, to the object's apparent angular size $\delta\omega$ as viewed from the observer. In essence, it is the answer to the question, `in a flat universe, how far away would an object of a known size need to be in order to appear as large as it does?'

\begin{figure}[!ht]
\centering
\includegraphics[width=0.95 \textwidth]{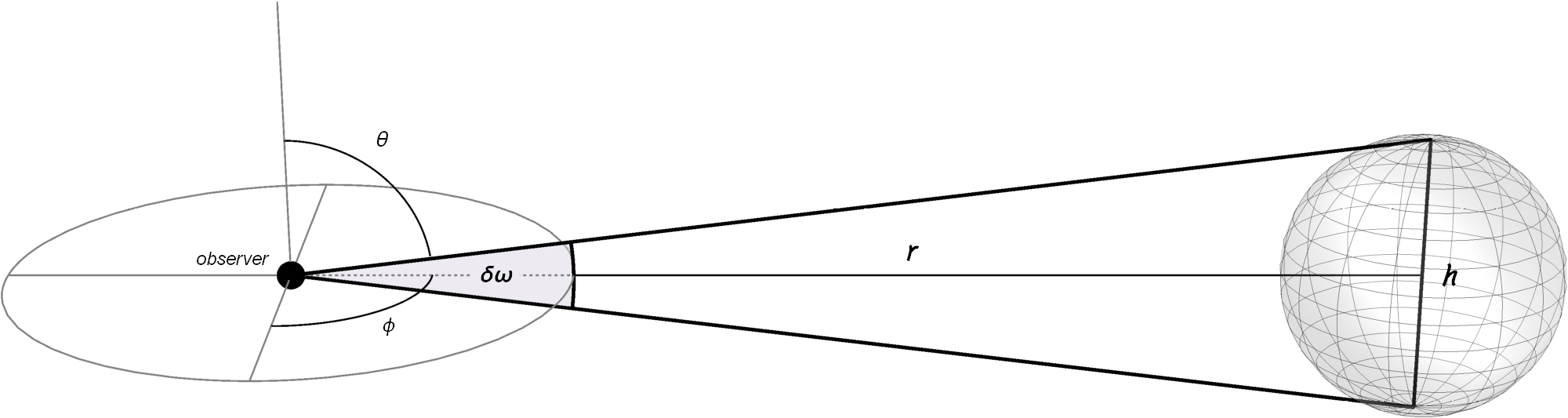}
\caption{Angular Diameter Distance in FLRW space-time}
\label{fig:ADDFLRW}
\end{figure}

\noindent Due to spatial isotropy, we may choose a spherical coordinate system so that the object, which we will assume to be spherical itself,~\footnote{The assumption of sphericity is convenient, but not necessary, for the derivation of Eq.~(\ref{eq:ADD}). More generally, the source size can
be characterised by an arc, see Figure~\ref{fig:ADDFLRW}, and we will adopt a more general approach in the anisotropic case.} lies along the equator at coordinate-distance $r$. Again exploiting isotropy, we can decide to measure the angular diameter along an arc that lies in the $(r,\theta)$-plane, as in Figure~\ref{fig:ADDFLRW}. Assuming that the angular size $\delta\omega$ of the object is small, we may write
\begin{equation}
\label{eq:ADD}
 d_{A,\text{ FLRW}} = \frac{h}{\delta\omega} = \frac{a \, r \, \tan(\delta\omega)}{\delta\omega} \approx \frac{a \, r \, \delta\omega}{\delta\omega} = a r = a S_\kappa(\chi)
\, ,
\end{equation}
where the $a$ in this equation and in the rest of this section is taken at the time of emission of the photon,
and for the scale factor today we take, $a_0=a(t_0)=1$.

\subsubsection{Luminosity distance}

Luminosity distance $d_L$ can be defined through a relationship between the intrinsic luminosity $\mathcal{L}$ (in Js$^{\text{-}1}$) of a distant source and the observed
flux $f$ (in Jm$^{\text{-}2}$s$^{\text{-}1}$) as measured by an observer on Earth,
\begin{equation}
\label{eq:LDdef1} f = \frac{\mathcal{L}}{4 \pi d_L^2} \hspace{2em} \longleftrightarrow \hspace{2em} d_L = \sqrt{\frac{\mathcal{L}}{4 \pi f}}
\,.
\end{equation}
In essence, it answers a question similar to the angular diameter distance: `in a flat universe, how far away would an object of known luminosity need to be to appear as bright as it does?' \\

\noindent In FLRW space-time, the calculation is again fairly straightforward. Since the space-time is homogenous and isotropic, light emitted from a source spreads evenly in all directions. The flux $f$ measured by an observer at proper distance $\chi$ is simply the luminosity $\mathcal{L}$ of the source divided by the area $4 \pi S^2_\kappa(\chi)$ of a hypersphere with the same radius.
We also need to take into account the expansion of the Universe, which will multiply the flux by a factor of $a$ due to red-shifting of photons and another factor of $a$ due to the expansion slowing down the rate of incoming photons,
\begin{equation}
\label{eq:LDdef2} f = \frac{\mathcal{L} a^2}{A_{\text{hypersphere}}}
= \frac{\mathcal{L} a^2}{4 \pi S_\kappa^2(\chi)}
\,.
\end{equation}
It follows directly from \eqref{eq:LDdef1} and \eqref{eq:LDdef2} that,
\begin{equation}
\label{eq:LDFLRW}
d_{L,\text{ FLRW}} = \frac{S_\kappa(\chi)}{a}
\,.
\end{equation}

\subsubsection{Etherington's reciprocity theorem}

By comparing equations \eqref{eq:ADD} and \eqref{eq:LDFLRW} we see that there is a relationship between angular diameter distance and luminosity distance known as Etherington's reciprocity theorem. Sometimes also referred to as the distance duality relation, it states simply that
\begin{equation}
\label{eq:Etherington} d_L = (1+z)^2 d_A = \frac{d_A}{a^2}
\,.
\end{equation}

\subsection{Distance measures in an anisotropic universe}

In an anisotropic setting, spherical symmetry is broken. This means that the angular diameter distance and luminosity distance will become an explicit function of the direction in which an observer is looking. Furthermore, if axial symmetry is also broken along this direction, the angular diameter distance will,
in addition, depend on the orientation of the arc along which the observer measures the object's angular size. In a more general setting, these two distance measures become a quantity assigned to a particular choice of arc or solid angle, and a proper distance.

\subsubsection{Angular diameter distance}
\label{sec:ADDGeneral}

In what follows we first exploit homogeneity of the Thurston-Perelman geometries to put the observer at the origin of the coordinate system. We then introduce some additional notation to parameterize arcs on the unit sphere around the observer. For a given direction $\hat{P} = (P_x,P_y,P_z)$ on the unit sphere in which we point a telescope, we can define two orthogonal unit vectors,
\begin{align}
\label{coordinate frame1}   \hat\theta  &= \left(\frac{P_x P_z}{\sqrt{1-P_z^2}},\frac{P_y P_z}{\sqrt{1-P_z^2}},-\sqrt{1-P_z^2}\right) \\
\label{coordinate frame2}   \hat\phi    &= \left(-\frac{P_y}{\sqrt{1-P_z^2}},\frac{P_x}{\sqrt{1-P_z^2}},0\right)
\end{align}
so that the triple ($\hat{P}$, $\hat\theta$, $\hat\phi$) is an orthonormal basis on $\mathds{R}^3$. We can now define any arc $\mathcal{A}$ through $\hat{P}$ by picking two angles, $\zeta$ and $\delta\omega$, and then writing,
\begin{equation}
\mathcal{A}({\hat{P}},\zeta,\delta\omega)  :=\bigl\{\mathcal{G}({\hat{P}};\zeta,s)\,|\, \zeta = {\rm fixed}\; \& \; s\in[-\delta\omega/2,\delta\omega/2)\bigr\}
\,.
\label{eq:ArcDefinition}
\end{equation}
where $\mathcal{G}$ characterises points on the unit sphere~\footnote{%
Even though the coordinate frame in~(\ref{coordinate frame1}) and (\ref{coordinate frame2}) is singular when $P_z = \pm 1$,
it is more convenient for our purposes than a nonsingular frame one would obtain by {\it e.g.}
a Gram-Schmidt procedure. The results of this section are not affected by the choice of approach.} through,
\begin{equation}
\mathcal{G}({\hat{P}};\zeta,s) = \cos(s) \hat{P} + \sin(s) \left(\cos(\zeta) \hat\theta + \sin(\zeta) \hat\phi\right).
\label{eq:ArcParameterization}
\end{equation}
Here $\delta\omega \in (0,2\pi)$ determines the angular size of the arc and the parameter $\zeta$ specifies the orientation of the arc around the vector $\hat{P}$. For instance, if $\hat{P}$ lies along the equator, then setting $\zeta = 0$ means the arc lies orthogonal to the equator, while if we set $\zeta = \pi/2$ then it lies parallel to the equator. \\

\begin{figure}[H]
\centering
\includegraphics[width=0.95 \textwidth]{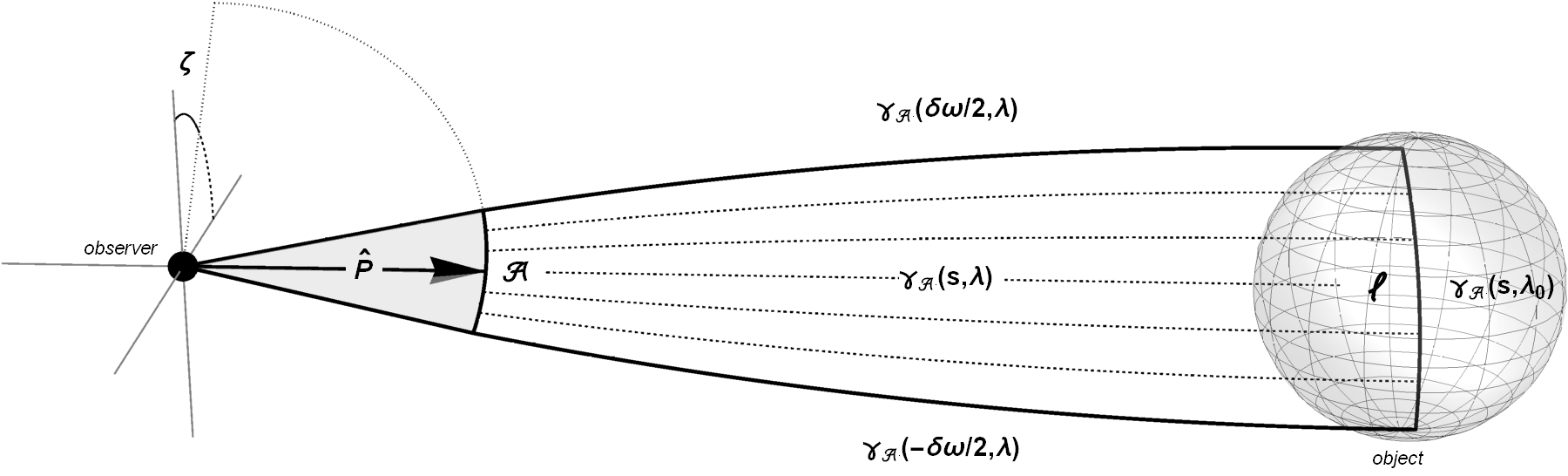}
\caption{Angular diameter distance in an anisotropic space-time.}
\label{fig:ADDGeneral}
\end{figure}

Now suppose that we have chosen $\hat P$ and $\mathcal{A}$ so that our telescope points at a distant object so that $\delta\omega$ coincides with the object's apparent (angular) size along the arc $\mathcal{A}$. In order to derive the angular diameter distance, we then want to know how the angular size $\delta\omega$ relates to the object's proper size $h$ along this direction. In order to do this, we need to trace the rays of incoming light back to their source along this arc. However, since we can no longer assume isotropy, we cannot simply extend the initial directions specified by $\mathcal{A}$ to a set of straight lines. Instead, we must account for the fact that the curvature of space may curve photon trajectories, as illustrated in Figure \ref{fig:ADDGeneral}. \\

\noindent This means we must solve the geodesic equation for light-like (radial) geodesics with an arbitrary initial direction, which we do explicitly in Appendices~A-D. For now, we assume that we know a family of geodesics $\{\gamma(\hat{P},\lambda)\}$ characterised by the initial direction $\hat{P}$ and dependent on geodesic proper distance $\lambda$ along the spatial 3-manifold $\Sigma_3$.~\footnote{Or, equivalently, conformal time, see \eqref{lightlike distances}.} The geodesics within this set is additionally assumed to satisfy,
\begin{equation}
\gamma(\hat{P},0) = 0\,, \hspace{2em}
\frac{{\rm d}\gamma(\hat{P},\lambda)}{{\rm d}\lambda}\Big|_{\lambda=0} = \hat{P}.
\end{equation}
For each choice of $\hat{P}$ and $\mathcal{A}$, we can define a 1-parameter sub-family $\gamma_{\mathcal{A}}$ as,
\begin{equation}
\gamma_{\mathcal{A}}(s,\lambda) = \gamma\left(\mathcal{G}(\hat{P};\zeta,s),\lambda\right)\,,
\end{equation}
where $\zeta$ is the (fixed) orientation of the arc and the angle $s \in [-\delta\omega/2,\delta\omega/2]$ parameterizes the internal angle between the initial direction of a given geodesic and the midpoint of the arc $\hat{P}$. By construction, the set of initial directions of this sub-family corresponds exactly to the arc $\mathcal{A}$, that is to say,
\begin{equation}
\left\{ \frac{{\rm d}\gamma_\mathcal{A}(s,\lambda)}{{\rm d}\lambda}\, \Bigg|\, s \in [-\delta\omega/2,\delta\omega/2] \; \& \; \lambda = 0 \right\} = \mathcal{A} \,.
\end{equation}
If we now fix $\lambda$ to the proper distance $\lambda_0$ of a faraway object, then the angle $s$ traces a distant arc between opposite sides of this faraway object as we let $s$ vary from $-\delta\omega/2$ to $\delta\omega/2$. Under the assumption that $\delta\omega$ is small compared to unity and $\lambda_0$ is small compared to $L$, the arc length $\ell$ of this distant arc multiplied by $a$ coincides with the proper size $h$ of the object. This means we can write,~\footnote{Rigorously speaking, the derivation presented in
this section applies to spaces which expand with a uniform, isotropic expansion rate. To generalize
to spaces that expand anisotropically, discussed in section~\ref{sec:AnisotropicScaleFactors}, one would
have to replace the uniform scalar factor $a(t)$ by the scale factor that characterises the expansion in
direction $\hat P$.}
\begin{equation}
\label{eq:ADDh}
h \simeq a\ell = a\int^{\delta\omega/2}_{-\delta\omega/2} {\rm d}s \sqrt{\gamma_{ij,\Sigma} \,
\frac{{\rm d}\gamma^i\left(\mathcal{G}(\hat{P};\zeta,s),\lambda_0\right)}{{\rm d}s} \,
\frac{{\rm d}\gamma^j\left(\mathcal{G}(\hat{P};\zeta,s),\lambda_0\right)}{{\rm d}s}}
\,,
\end{equation}
where $a$ in this equation, is again understood to be the scale factor at the time of emission of the photons. Hence, for any of the Thurston space-times, the angular diameter distance of an object visible in the direction $\hat{P}$ that sits at proper distance $\lambda_0$ measured along an arc $\mathcal{A}$ of apparent size $\delta\omega$ and orientation $\zeta$ can be expressed as,
\begin{equation}
\label{eq:ADDGeneral}
d_{A}(\hat{P},\lambda_0,\delta\omega,\zeta) := a \frac{\ell}{\delta\omega}
 = \frac{a}{\delta\omega}\int^{\delta\omega/2}_{-\delta\omega/2} {\rm d}s \sqrt{
\gamma_{ij,\Sigma} \,
\frac{{\rm d}}{{\rm d}s}\gamma^i\left(\mathcal{G}(\hat{P};\zeta,s),\lambda_0\right) \,
\frac{{\rm d}}{{\rm d}s}\gamma^j\left(\mathcal{G}(\hat{P};\zeta,s),\lambda_0\right)
}
\,. \\
\end{equation}
In the regime where $L >> \lambda_0 >> h$, the arc is small and far away enough that $\delta\omega$ is small compared to unity, but the curvature effects are not so strong that small deviations from the initial angle will lead to extreme differences at distance $\lambda_0$. In this regime, the integrand in the previous equation is approximately the same for all $s \in [-\delta\omega/2,\delta\omega/2]$ and so can be approximated by a constant. This means that we can approximate the expressions for $\ell$ and $d_A$ to a high degree of accuracy by,
\begin{equation}
\label{eq:ADDhapprox}
\ell \simeq
\delta\omega
\sqrt{\gamma_{ij,\Sigma} \,
\frac{{\rm d}}{{\rm d}s}\gamma^i\left(\mathcal{G}(\hat{P};\zeta,s),\lambda_0\right)  \,
\frac{{\rm d}}{{\rm d}s}\gamma^j\left(\mathcal{G}(\hat{P};\zeta,s),\lambda_0\right)
}\Bigg|_{s=0},
\end{equation}
such that
\begin{equation}
\label{eq:ADDGeneralapprox}
\boxed{
d_{A}(\hat{P},\lambda_0,\zeta) \simeq
a
\sqrt{\gamma_{ij,\Sigma} \,
\frac{{\rm d}}{{\rm d}s}\gamma^i\left(\mathcal{G}(\hat{P};\zeta,0),\lambda_0\right) \,
\frac{{\rm d}}{{\rm d}s}\gamma^j\left(\mathcal{G}(\hat{P};\zeta,0),\lambda_0\right)
}\Bigg|_{s=0}
}
\,,
\end{equation}
and the expression is no longer dependent on $\delta\omega$. This is the anisotropic equivalent of approximating $h \approx a r \delta \omega$ in \eqref{eq:ADD} for sufficiently small $\delta\omega$.
In this regime, $\delta\omega$ drops out of the expression for the angular diameter distance, and we are left with a more manageable expression that is solely dependent on direction $\hat{P}$, distance $\lambda_0$ and orientation $\zeta$ of the arc. \\

\noindent Lastly, we check the isotropic limit of this derivation. In FLRW space-time, equation \eqref{eq:ADDGeneral} reduces to a familiar form
given in Eq.~(\ref{eq:ADD}),
\begin{equation}
d_{A,\text{ FLRW}} = \frac{h}{\delta\omega} \simeq \frac{a}{ \delta\omega}\int^{\delta\omega/2}_{-\delta\omega/2} {\rm d}s \sqrt{ r^2 \Big(\sin^2(s) + \cos^2(s)\Big)} = \frac{a r \delta\omega}{\delta\omega} = aS_\kappa(\chi)
\,. \\
\end{equation}

\hspace{1em}

\subsubsection{Luminosity distance}

\begin{figure}[H]
\centering
\includegraphics[width=0.95 \textwidth]{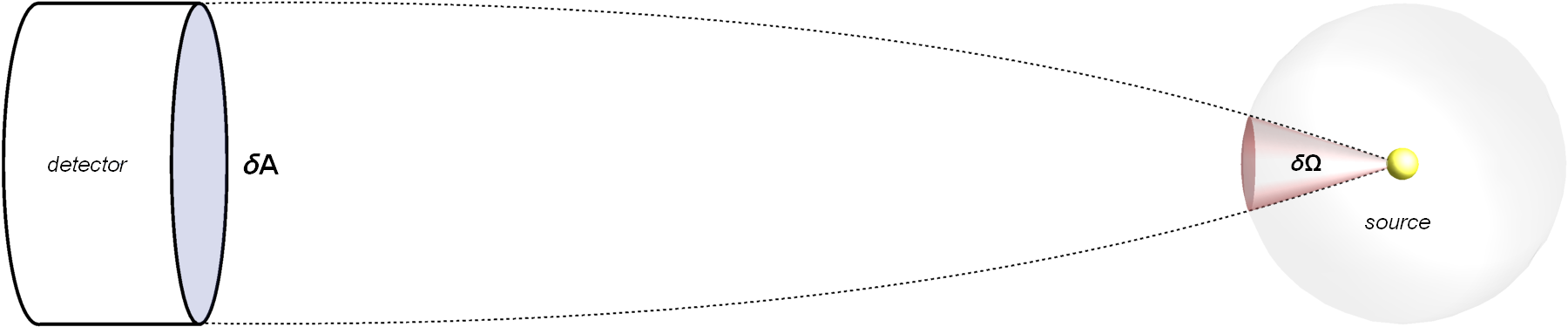}
\caption{Luminosity distance in an anisotropic space-time}
\label{fig:LDGeom}
\end{figure}

\begin{figure}[H]
\centering
\includegraphics[width=0.95 \textwidth]{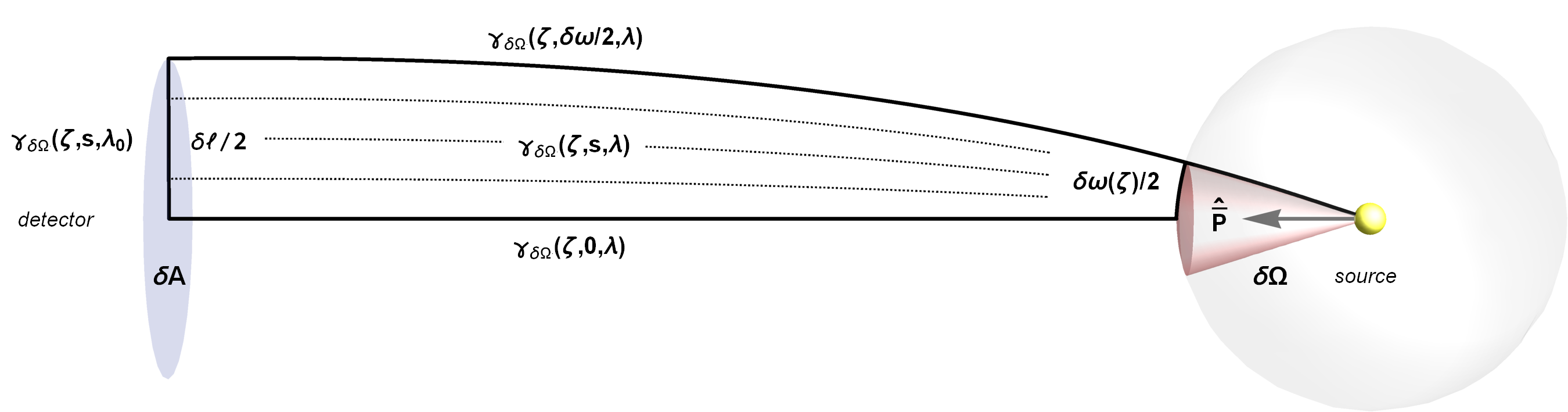}
\caption{Luminosity distance in an anisotropic space-time.}
\label{fig:LDGeneral}
\end{figure}

To study the general case for luminosity distance, consider a beam of light emitted over a small solid angle $\delta\Omega$ from a source, which we will approximate as point-like. If the source radiates equally in every direction, then the power through this solid angle can be written as
\begin{equation}
\mathcal{L}_\text{Beam} = \mathcal{L}_\text{Source} \times\frac{\delta\Omega}{4 \pi}
\,.
\end{equation}
Now suppose that this beam of light terminates on a photosensitive plate of a detector ({\it e.g.} a telescope) at some proper distance $\lambda_0$ as in Figure~\ref{fig:LDGeom}. Conservation of energy implies that the power flowing through this plate must be equal to the power flowing through the initial solid angle $\delta\Omega$, up to powers of $a$ to compensate for the expansion of the universe.~\footnote{Note that we make the additional implicit assumption here $\lambda_0$ is small compared to $L$ so that no extreme lensing effects occur that might cause beams emitted in disparate directions to intersect.
Such intersections are typical for positively curved spaces ($\kappa>0$), but usually do not occur
in negatively curved spaces ($\kappa<0$).
}
\noindent The flux through this plate can thus be written as
\begin{equation}
f_\text{Detector} = \frac{\mathcal{L}_\text{Beam}  a^2}{\delta A}  = \frac{\mathcal{L}_\text{Source}}{4 \pi} \times \frac{\delta\Omega a^2}{\delta A}
\,,
\end{equation}
where $\delta A$ is the beam's cross-sectional area at distance $\lambda$. Comparing this to the definition of $d_L$ in equation \eqref{eq:LDdef1}, we can write a general (geometric) expression for the luminosity distance as
\begin{equation}
\label{eq:LDgeneral} d_L = \frac1a\sqrt{\frac{\delta A}{\delta \Omega}}
\,.
\end{equation}
Following the approach in Section \ref{sec:ADDGeneral}, we use homogeneity of spatial sections to place the source at the origin and pick $\hat{\overline{P}}$ to be the initial direction of an emitted photon. This situation is sketched in Figure \ref{fig:LDGeneral}.

\begin{figure}[H]
\centering
\includegraphics[width=0.95 \textwidth]{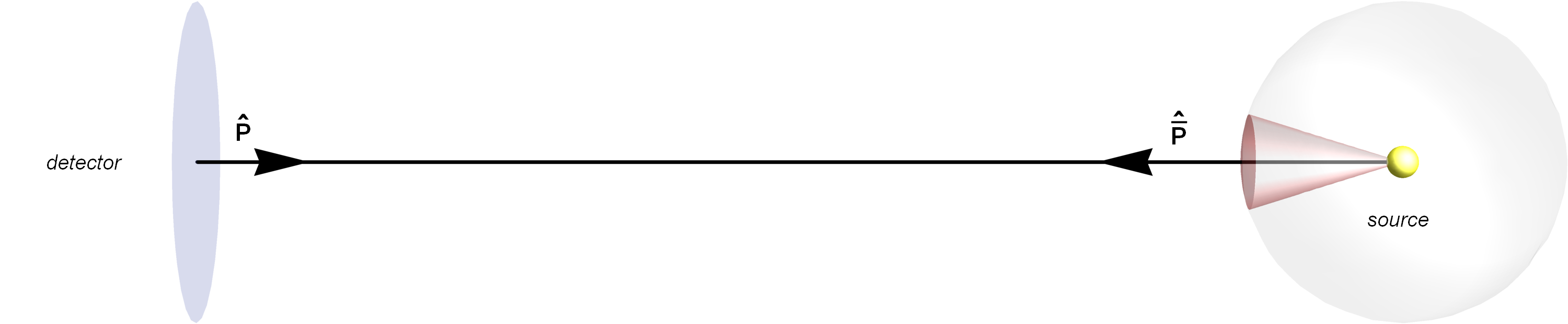}
\caption{Relationship between $\hat P$ and $\hat{\overline{P}}$.}
\label{fig:LDInitialDirections}
\end{figure}

\noindent We can describe this solid angle $\delta\Omega$ by specifying the angular distance $\delta\omega(\zeta)/2$ from the vector $\hat{\overline{P}}$ to the edge of $\delta\Omega$ along every direction $\zeta$.~\footnote{ Note that we have implicitly assumed here that $\delta\Omega$ is convex and so can be conveniently parameterised this manner. This assumption is not necessary to the derivation, as we will eventually arrive at an expression that is independent of $\delta\Omega$. However, making this assumption here means we avoid additional complexity surrounding the parameterization of $\delta\Omega$, which would detract from the clarity of the derivation presented. The assumption is justified in the small-$\delta\Omega$ regime, as the way the small solid angle grows or shrinks as it propagates from source to detector is important at first order and effects of its shape (barring pathological examples) will contribute at higher order.} This means we can write,
\begin{equation}
\delta\Omega(\hat{\overline{P}},\delta\omega(\zeta))
:=\bigl\{\mathcal{G}(\hat{\overline{P}};\zeta,s)\,|\, \zeta \in [0,2\pi)\;
\& \; s\in[0,\delta\omega(\zeta)/2]\bigr\}
\,,
\label{arc 2}
\end{equation}
where $\mathcal{G}$ is the same as in \eqref{eq:ArcParameterization}. As in the previous subsection, for each choice of $\hat{\overline{P}}$ and $\delta\Omega$ described in this way, we can now find a 2-parameter family of geodesics,
\begin{equation}
\gamma_{\delta\Omega}(\zeta,s,\lambda) = \gamma\left(\mathcal{G}(\hat{\overline{P}};\zeta,s),\lambda\right),
\end{equation}
characterized by $\zeta\in[0,2\pi)$ and $s\in[0,\delta\omega(\zeta)/2]$, that terminates on the plate of the detector at $\lambda = \lambda_0$. As before, the set of initial directions of this sub-family corresponds exactly to the solid angle $\delta\Omega$
\begin{equation}
\left\{ \frac{{\rm d}\gamma_{\delta\Omega}(\zeta,s,\lambda)}{{\rm d}\lambda}\, \Bigg|\, s \in [0,\delta\omega(\zeta)/2] \; \& \; \zeta \in [0,2\pi) \; \& \; \lambda = 0 \right\} = \delta\Omega \,.
\end{equation}
From here, it is not difficult to write out the areas $\delta\Omega$ and $\delta A$ in integral form:
\begin{align}
\delta A      &= \frac{1}{2} \int_0^{2\pi} \left(\frac{\ell(\zeta)}{2}\right)^2 {\rm d}\zeta
             = \frac{1}{2} \int_0^{2\pi} \left(\int^{\delta\omega(\zeta)/2}_{0} {\rm d}s \sqrt{\gamma_{ij,\Sigma} \,
\frac{{\rm d}\gamma^i\left(\mathcal{G}(\hat{\overline{P}};\zeta,s),\lambda_0\right)}{{\rm d}s} \,
\frac{{\rm d}\gamma^j\left(\mathcal{G}(\hat{\overline{P}};\zeta,s),\lambda_0\right)}{{\rm d}s}}\right)^2 {\rm d}\zeta
\label{lum distance: delta A}
\\
\delta \Omega &= \frac{1}{2} \int_0^{2\pi} \left(\frac{\delta\omega(\zeta)}{2}\right)^2  {\rm d}\zeta.
\label{lum distance: delta Omega}
\end{align}
If we divide the first integral by the second, we get the ratio we are looking for. However, there is one subtlety to be addressed. We have derived an expression for the luminosity distance dependent on $\hat{\overline{P}}$, which is the direction in which the luminous source emitted the light that reaches the observer, as expressed in the local inertial frame of the source. Observationally, we are more interested in an expression dependent on  $\hat{P}$, see figure \ref{fig:LDInitialDirections}, which is the direction in which an observer views a luminous source as measured in the local inertial frame of the observer. So what remains to be done is to find some way of connecting these vectors and translating them from one frame to another.

\noindent Homogeneity of spatial slices and reversibility of solutions to the geodesic equation make this (conceptually) relatively straightforward. In the frame of the observer, light is incoming along some direction $\hat{P}$ after having traveled some geodesic distance $\lambda_0$. Since geodesic distance is invariant, we can construct the reverse geodesic $\gamma(\hat{P},\lambda)$ from the observer back to the source and find the position of the source at $\gamma(\hat{P},\lambda_0)$. We can then find the initial direction of the photons emitted from the source by taking the derivative of this geodesic at $\lambda = \lambda_0$ and reversing the sign (see figure~\ref{fig:LDInitialDirections}):
\begin{equation}
\hat{\overline{P}} = - \frac{{\rm d} \gamma(\hat{P},\lambda)}{{\rm d}\lambda}\Big|_{\lambda=\lambda_0}\,.
\end{equation}
However, this is $\hat{\overline{P}}_{o}$ expressed in the coordinates of the observer, not those of the source. To obtain $\hat{\overline{P}}_{s}$ in the coordinates of the source, we must construct the transformation $T$ from the frame of the observer to the frame of the source and then apply the tangent map of this transformation, $dT$, to the equation above,~\footnote{ We will work out the Solv geometry case explicitly as an example, as the diagonal metric makes the calculation quick. Suppose that we have two local inertial frames, $O$ and $O'$, and the second frame has its origin at $(a,b,c)$ with respect to the coordinates of the first, then we can express the metric of both frames in terms of the coordinates of the first,
\begin{align}
{\rm d}s^2  &= {\rm e}^{2z/L} {\rm d}x^2 +{\rm e}^{-2z/L}{\rm d}y^2  + {\rm d}z^2 \\
{\rm d}s'^2 &= {\rm e}^{2z'/L}{\rm d}x'^2  +{\rm e}^{-2z'/L} {\rm d}y'^2 + {\rm d}z'^2  \\
      &= {\rm e}^{2(z+c)/L}{\rm d}x'^2   + {\rm e}^{-2(z+c)/L}{\rm d}y'^2 + {\rm d}z'^2
\end{align}
Since both frames must agree about the geometry of the spatial section, we can equate ${\rm d}s^2$ and
${\rm d}s'^2$ and then easily relate the two frames by the transformation,
\begin{align}
T(x,y,z)     &= (x {\rm e}^{c/L} + a, y{\rm e}^{-c/L} + b, z + c) = (x',y',z')
\,.
\end{align}
Therefore, for the Solv geometry this means that can relate the direction of emission of light from the source, $\hat{\overline{P}}_{s}$, to the direction at which the observer receives this light, $\hat{P}_{o}$ by
\begin{equation}
\hat{\overline{P}}_{s}[\hat{P}_{o}] = - \left( \frac{{\rm d} \gamma^1(\hat{P}_{o},\lambda)}{{\rm d}\lambda}{\rm e}^{c/L},\, \frac{{\rm d} \gamma^2(\hat{P}_{o},\lambda)}{{\rm d}\lambda}{\rm e}^{-c/L},\, \frac{{\rm d} \gamma^3(\hat{P}_{o},\lambda)}{{\rm d}\lambda}\right)_{\lambda=\lambda_0}\,.
\end{equation}
Similar analyses can be carried out for the other Thurston geometries, but we refrain from explicating them here
for the sake of brevity. Derivations for other geometries are available upon request.
}
\begin{equation}
\boxed{
\hat{\overline{P}}_{s}[\hat{P}_{o}] := {\rm d}T\left(- \frac{{\rm d} \gamma(\hat{P}_{o},\lambda)}{{\rm d}\lambda}\Big|_{\lambda=\lambda_0}\right)
} \,.
\label{lum distance: change of initial direction}
\end{equation}
 Hence, to obtain the luminosity distance from the point of view of the detector, we need to plug equations \eqref{lum distance: delta A} and \eqref{lum distance: delta Omega} into \eqref{eq:LDgeneral} and replace $\hat{\overline{P}}$ by $\hat{\overline{P}}_s(\hat{P}_o)$ as given in \eqref{lum distance: change of initial direction}. Putting this together, we can write the following expression for the luminosity distance of a source visible in the direction $\hat{P}_o$ that sits at proper distance $\lambda_0$ and emits light at the detector through a solid angle $\delta\Omega$ as
\begin{equation}
\label{eq:LDGeneralb}
\left(d_L(\hat{P_o},\lambda_0,\delta\Omega)\right)^2 \ := \ \frac{1}{a^2}\frac{\delta A}{\delta\Omega}
= \frac{1}{a^2}\frac{\int_0^{2\pi}\left( \int^{\delta\omega(\zeta)/2}_0 {\rm d}s
\sqrt{\gamma_{ij,\Sigma}  \,
\frac{{\rm d}}{{\rm d}s}
\gamma^i\left(\mathcal{G}\left(\hat{\overline{P}}_s[\hat{P}_o];\zeta,s\right),\lambda_0\right)
 \, \frac{{\rm d}}{{\rm d}s}
\gamma^j\left(\mathcal{G}\left(\hat{\overline{P}}_s[\hat{P}_o];\zeta,s\right),\lambda_0\right)
}\right)^2 {\rm d}\zeta}
  {\int_0^{2\pi} \left(\frac{\delta\omega(\zeta)}{2}\right)^2 {\rm d}\zeta}
 \,.
\end{equation}
This expression is not immediately useful, though, as we do not a-priori know what the solid angle $\delta\Omega$ and looks like. One would have to define the shape of the detector plate by setting $\ell(\zeta)$ and then finding $\delta\omega(\zeta)$ so that the family $\gamma_{\delta\Omega}$ maps $\delta\Omega$ exactly onto $\delta A$. This is, however, a hard problem to solve in general. \\

A more tractable approach can be taken by restricting to the regime where $L \gg \lambda_0 \gg \sqrt{\delta A}$ like we did with angular diameter distance. In this regime, the detector is small and far away enough from the source that $\delta\Omega$ is small compared to unity, but the curvature effects are not so strong that small deviations from the initial angle will lead to extreme differences at distance $\lambda_0$. We can then make use of the approximation in \eqref{eq:ADDhapprox} to make the following simplification,
\begin{align}
\delta                              A & \simeq \frac{1}{8} \int_0^{2\pi} \ell^2(\zeta) {\rm d}\zeta \\
\delta                         \Omega & \simeq \frac{1}{8} \int_0^{2\pi} \frac{\ell^2(\zeta)}{
\Big[\gamma_{ij,\Sigma}  \,
\frac{{\rm d}}{{\rm d}s}
\gamma^i\left(\mathcal{G}\left(\hat{\overline{P}}_s[\hat{P}_o];\zeta,s\right),\lambda_0\right)
\frac{{\rm d}}{{\rm d}s}
\gamma^j\left(\mathcal{G}\left(\hat{\overline{P}}_s[\hat{P}_o];\zeta,s\right),\lambda_0\right)\Big]
\Big|_{s=0} }  \, {\rm d}\zeta \,.
\end{align}
This allows us to express the luminosity distance directly as a function of the shape of the detector without making any reference to $\delta\omega$.

Within this same regime, the incoming photon flux is approximately constant across the detector plate. Hence we can make the additional simplifying assumption that the detector is disk-shaped, {\it i.e.} $\ell(\zeta) = \ell$, as it is the area of the detector plate that is important at first order and the effects of the shape will contribute at higher orders. This means that the dependence on $\ell$ also drops out of the expression entirely and $d_L$ becomes a purely geometrical quantity depending only on $\hat{P}_o$ and $\lambda_0$,
\begin{equation}
\label{eq:LDGeneralapprox}
\boxed{
\left(d_L(\hat{P}_o,\lambda_0)\right)^2  \simeq \frac{2\pi}{a^2} \left(\int_0^{2\pi} \frac{1}{
\Big[\gamma_{ij,\Sigma}  \,
\frac{{\rm d}}{{\rm d}s}
\gamma^i\left(\mathcal{G}\left(\hat{\overline{P}}_s[\hat{P}_o];\zeta,s\right),\lambda_0\right)
\frac{{\rm d}}{{\rm d}s}
\gamma^j\left(\mathcal{G}\left(\hat{\overline{P}}_s[\hat{P}_o];\zeta,s\right),\lambda_0\right)\Big]\Big|_{s=0}
} {\rm d}\zeta
\right)^{-1}
}\,.
\end{equation}

\noindent Lastly, we check the isotropic limit of our derivation. In FLRW space-time, equation \eqref{eq:LDGeneralb} reduces to a familiar form,
\begin{equation}
(d_{L,\text{ FLRW}})^2 = \frac{1}{a^2}\frac{\delta A}{\delta\Omega}
= \frac{1}{a^2}\frac{\int_0^{2\pi}\left( \int^{\delta\omega/2}_0 {\rm d}s \sqrt{r^2 \left(\sin^2 (s)+ \cos^2(s)\right)}\right)^2 {\rm d}\zeta}{\int_0^{2\pi} \left(\frac{\delta\omega(\zeta)}{2}\right)^2 {\rm d}\zeta} = \frac{1}{a^2}\frac{2 \pi r^2 \left(\tfrac{\delta\omega}{2}\right)^2}{2 \pi \left(\tfrac{\delta\omega}{2}\right)^2}
 = \frac{S^2_\kappa(\chi)}{a^2}
\,,\quad
\end{equation}
which agrees with Eq.~(\ref{eq:LDFLRW}).

\subsubsection{The anisotropic reciprocity theorem}

The statement of Etherington's theorem can be amended to hold in a more general anisotropic context. With the notational machinery we have developed in the previous section, this is not a difficult task. We will work explicitly in the regime when $L \gg \lambda \gg h$ and $L\gg \lambda \gg \sqrt{\delta A}$, so that we can start from equation \eqref{eq:LDGeneralapprox}. Next, we recognise term in the denominator of the integrand in this equation as the term on the right-hand side of \eqref{eq:ADDGeneralapprox} and we write

\begin{align}
\left(d_L(\hat{P},\lambda)\right)^{-2} &= \frac{a^2}{2 \pi}
\int_0^{2\pi} \frac{1}{
\left[\gamma_{ij,\Sigma}  \,
\frac{{\rm d}}{{\rm d}s}
\gamma^i\left(\mathcal{G}(\hat{\overline{P}}_s[P];\zeta,s),\lambda\right)
\frac{{\rm d}}{{\rm d}s}
\gamma^j\left(\mathcal{G}(\hat{\overline{P}}_s[P];\zeta,s),\lambda\right)\right]
\Big|_{s=0}} {\rm d}\zeta \\
&= \frac{a^2}{2 \pi} \int_0^{2\pi} \frac{a^2}{\left(d_A(\hat{\overline{P}}_s[P], \lambda, \zeta)\right)^2}{\rm d}\zeta
\label{anisotropic Etherington's theorem}
\end{align}
so that,
\begin{equation}
\boxed{
\frac{1}{d_L^{2}(\hat{P},\lambda)}  = \frac{a^4}{2 \pi}\int_0^{2\pi} \frac{1}{d_A^2\big(\hat{\overline{P}}_s[P], \lambda, \zeta\big)} {\rm d}\zeta
\,
}\,.
\label{anisotropic Etherington's theorem}
\end{equation}
In the isotropic case where $d_A$ does not depend on $\zeta$, this of course reduces to the familiar isotropic form of Etherington's theorem in~\eqref{eq:Etherington}.
Its generalization to anisotropic spaces
in Eq.~(\ref{anisotropic Etherington's theorem})
is one of the principal results of this paper.


\section{Distance Measures Visualised}
\label{sec:DistanceMeasurePlots}

With the machinery we have  developed in the previous section, we are only a small step away from visualizing the effect of large-scale curvature on angular diameter distance and luminosity distance. To show the maximal potential extent of these effects, we have opted to generate plots for the largest redshift that we can conceivably optically measure, $z_* \simeq 1091$, or the redshift at the time of recombination. \\

The plots in this section~\footnote{\textit{Mathematica} notebooks are available upon request.} were generated by taking the expressions for $d_A$ and $d_L$, \eqref{eq:ADDGeneralapprox} and \eqref{eq:LDGeneralapprox} from the previous section and setting $\lambda_0$ to $\eta_*$ as shown in Table~\ref{table:LengthScales1}. Where necessary for computational speed, expressions were expanded to order $L^{-6}$, or, equivalently, $\kappa^3$. The resulting length scales are plotted on the $(\phi,\theta)$-plane relative to the flat scenario. We have used {\it red} to signify a distance that is shorter than the spatially flat case and {\it vice versa} for {\it blue}.

\subsection{Angular diameter distance}

The figures below show the angular diameter distance relative to a flat geometry for a large triangle whose far side lies at redshift of $z_* \simeq 1091$. Two bars are drawn for each point on the figure representing the principal axes, meaning that one represents the direction in which the angular diameter distance is {\it smallest} and the other represents the direction in which the angular diameter distance is {\it largest}.

The length of the bar represents the (absolute) magnitude of the effect and the color gradient represents magnitude and sign: the bar is colored red if the angular distance is shorter than in the flat case and {\it vice versa} for blue. Thus a long red bar indicates that the angular diameter is much smaller than flat (relative to other points on the figure) when measured along that direction, while a small blue bar indicates that the distance is a little bit larger than the flat case when measured along that direction. These graphs can also be read as indicating the axes along which objects would be maximally observationally deformed by anisotropic curvature compared to a flat scenario.

\noindent Note that every figure is plotted on a separate color gradient, so the scale varies between plots. The Nil geometry shows the strongest effect on any one single arc, followed by the other anisotropic geometries and then the isotropic geometries.
\vskip -0.1cm
\begin{figure}[H]
\begin{minipage}[l]{0.5 \linewidth}
\includegraphics[width=1 \textwidth]{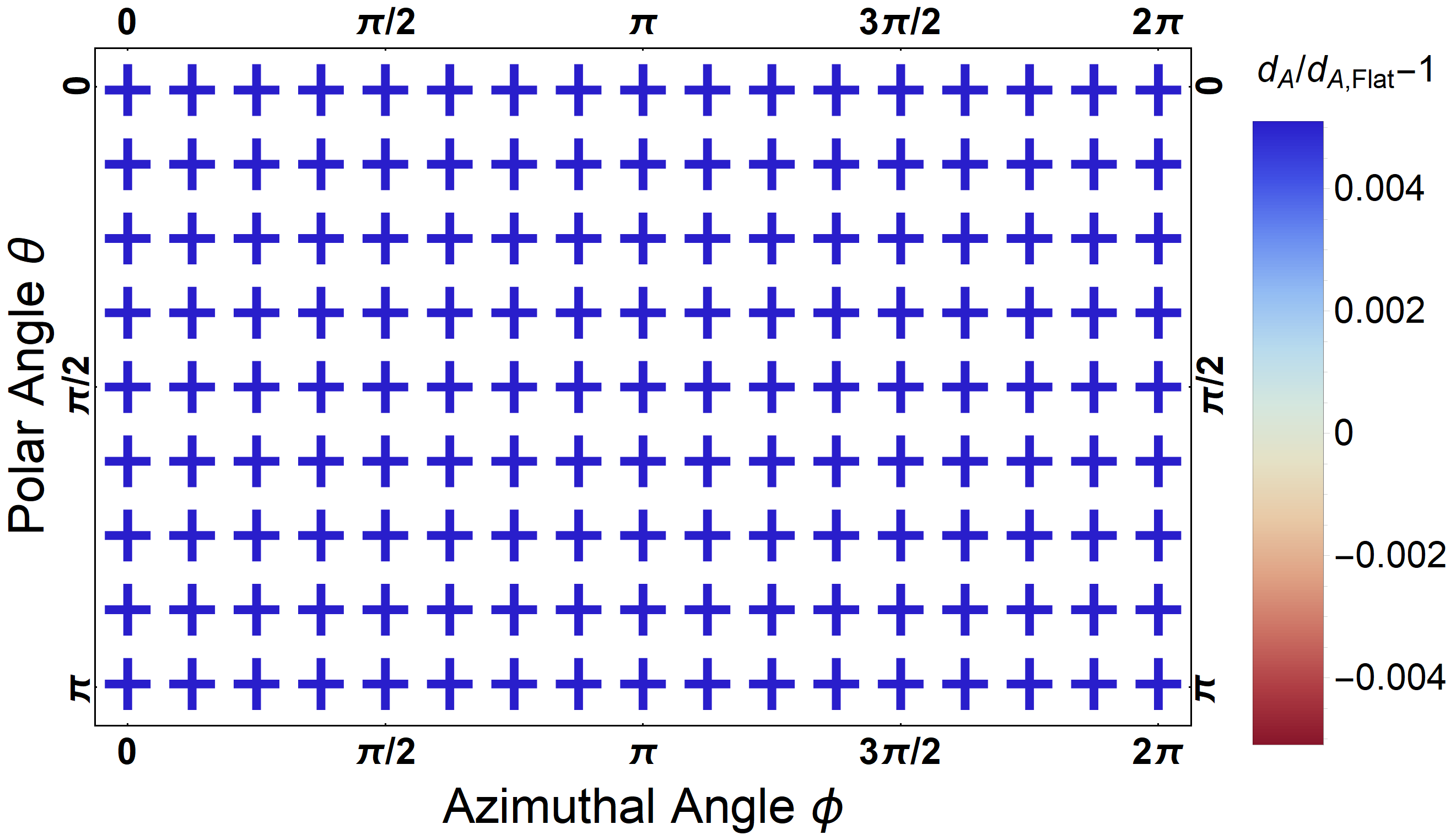}
\caption{$d_A$ for the $\mathds{H}^3$ geometry.}
\label{Angdiadist-1-H3}
\end{minipage}
\begin{minipage}[l]{0.5 \linewidth}
\includegraphics[width=1 \textwidth]{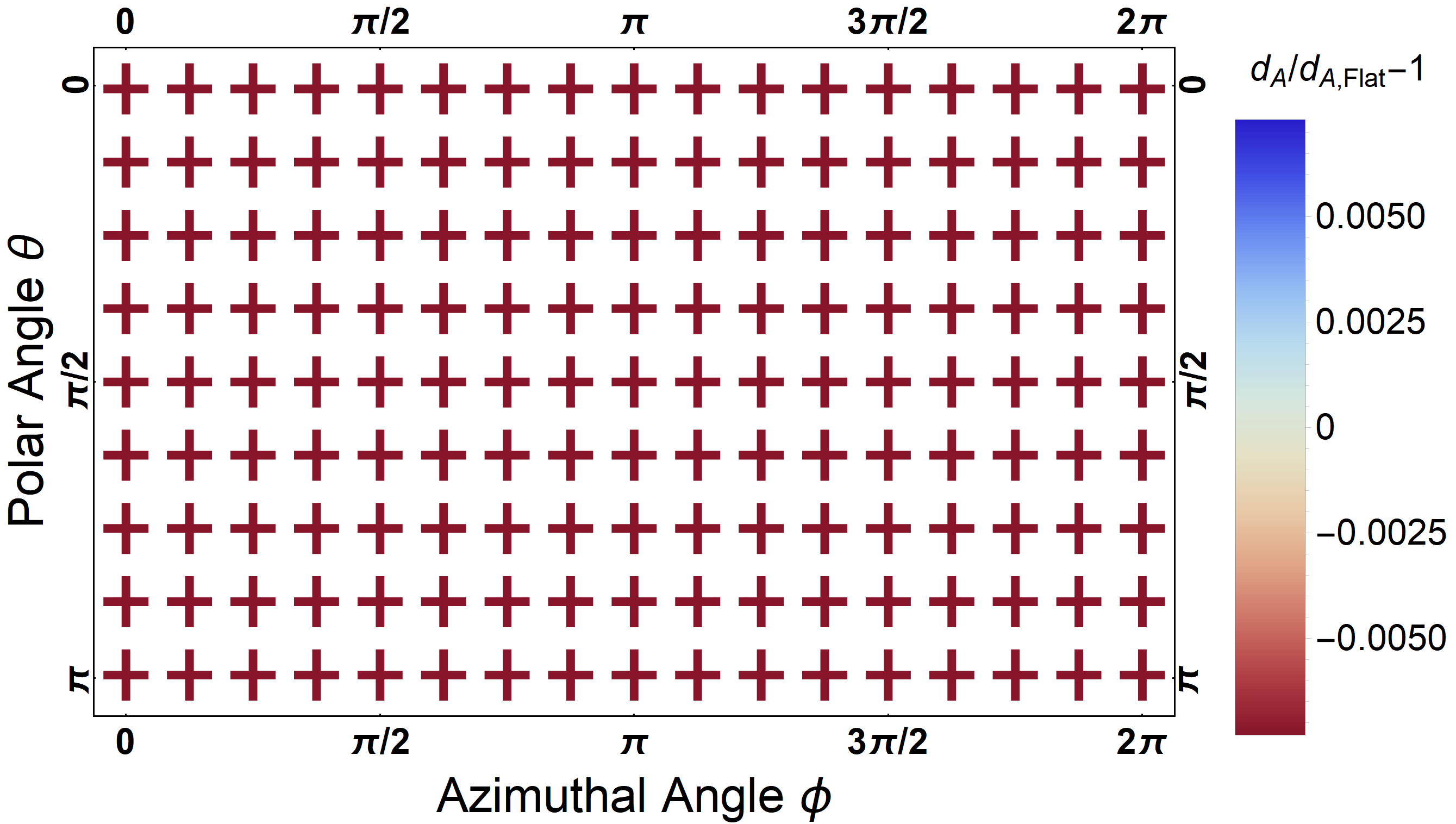}
\caption{$d_A$ for the $S^3$ geometry.}
\label{Angdiadist-2-S3}
\end{minipage}
\end{figure}
\vskip -0.3cm
\noindent As one would expect in the isotropic case, the angular diameter distance shows no dependence on the the direction or orientation of the arc for the $\mathds{H}^3$ and $S^3$ geometries. In the hyperbolic case, a given solid angle in Earth's night sky corresponds to a greater physical area at a given distance from the origin than in a flat model. This area will be populated by proportionally more objects, each of which will appear smaller
and thus seem to be further away (blue), see figure~\ref{Angdiadist-1-H3}.
The opposite is true of the spherical case, objects will appear larger and therefore appear closer (red), because a
smaller volume of space gets projected onto the same solid angle, see figure~\ref{Angdiadist-2-S3}.
\vskip -0.2cm
\begin{figure}[H]
\begin{minipage}[l]{0.5 \linewidth}
\includegraphics[width=1 \textwidth]{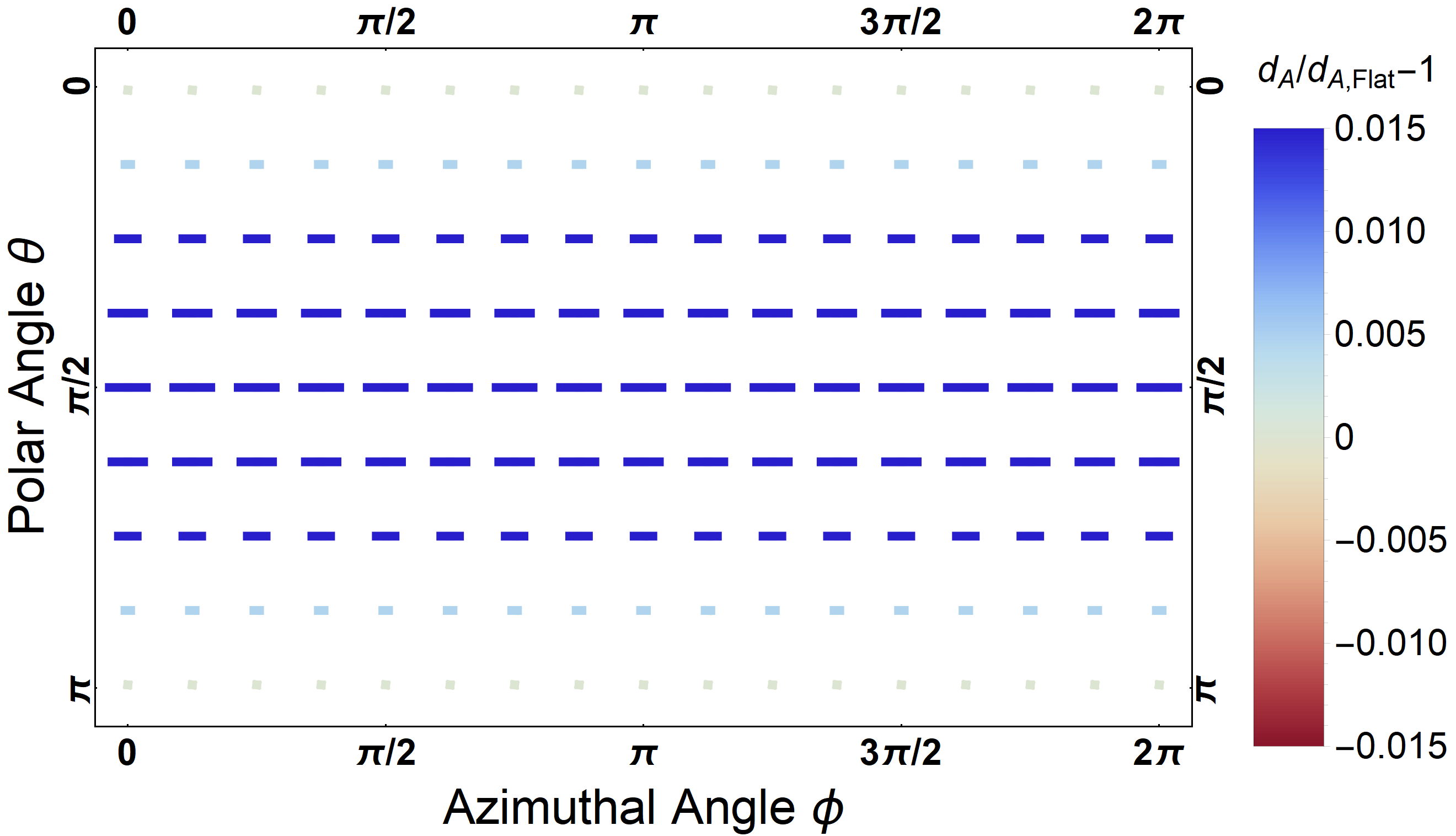}
\caption{$d_A$ for the $\mathds{R} \times \mathds{H}^2$ geometry.}
\label{Angdiadist-3-RH2}
\end{minipage}
\begin{minipage}[l]{0.5 \linewidth}
\includegraphics[width=1 \textwidth]{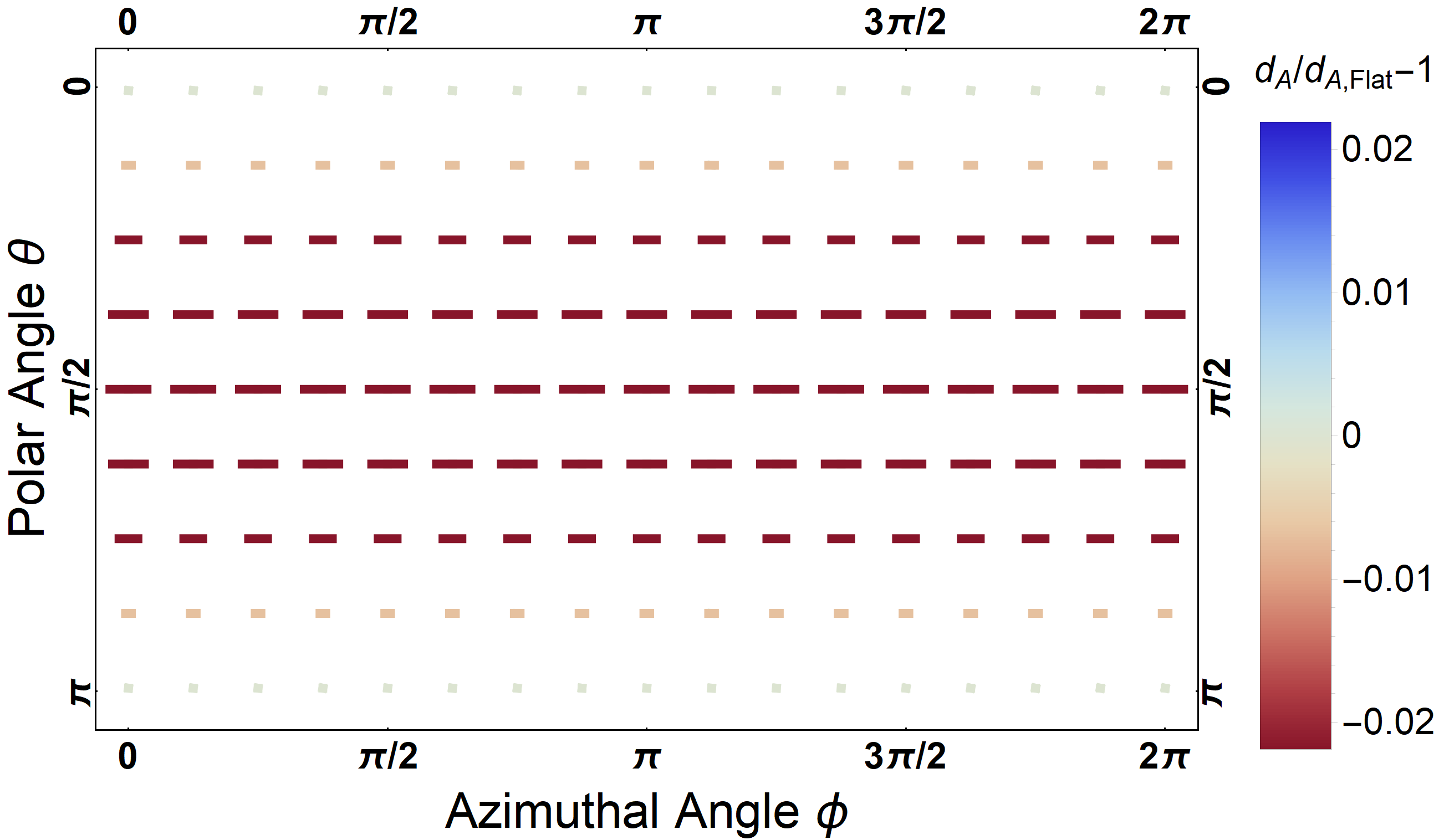}
\caption{$d_L$ for the $\mathds{R} \times S^2$ geometry.}
\label{Angdiadist-4-RS2}
\end{minipage}
\end{figure}
\vskip -0.1cm
\noindent The $\mathds{R} \times \mathds{H}^2$ and $\mathds{R} \times S^2$ geometries show the first signs of anisotropy.  We see again that the hyperbolic geometry tends to make objects appear further and spherical geometry makes them appear closer. While axial symmetry around the $z$-axis is preserved, we see that arcs in different directions and orientations are affected differently.

Arcs along the $\phi$-directions are maximally stretched or compressed, while arcs along the $\theta$ direction are unaffected. The effect is greatest at the equator, which indeed is the plane in which $\mathds{H}^2$ and $S^2$ lie, and vanishes as we turn as we rotate the arc to the poles. This pattern exactly matches the set-up of the spatial sections, which are flat in the $z$-direction and curved along the $(x,y)$-plane.
\vskip -0.1cm
\begin{figure}[H]
\begin{minipage}[l]{0.5 \linewidth}
\includegraphics[width=1 \textwidth]{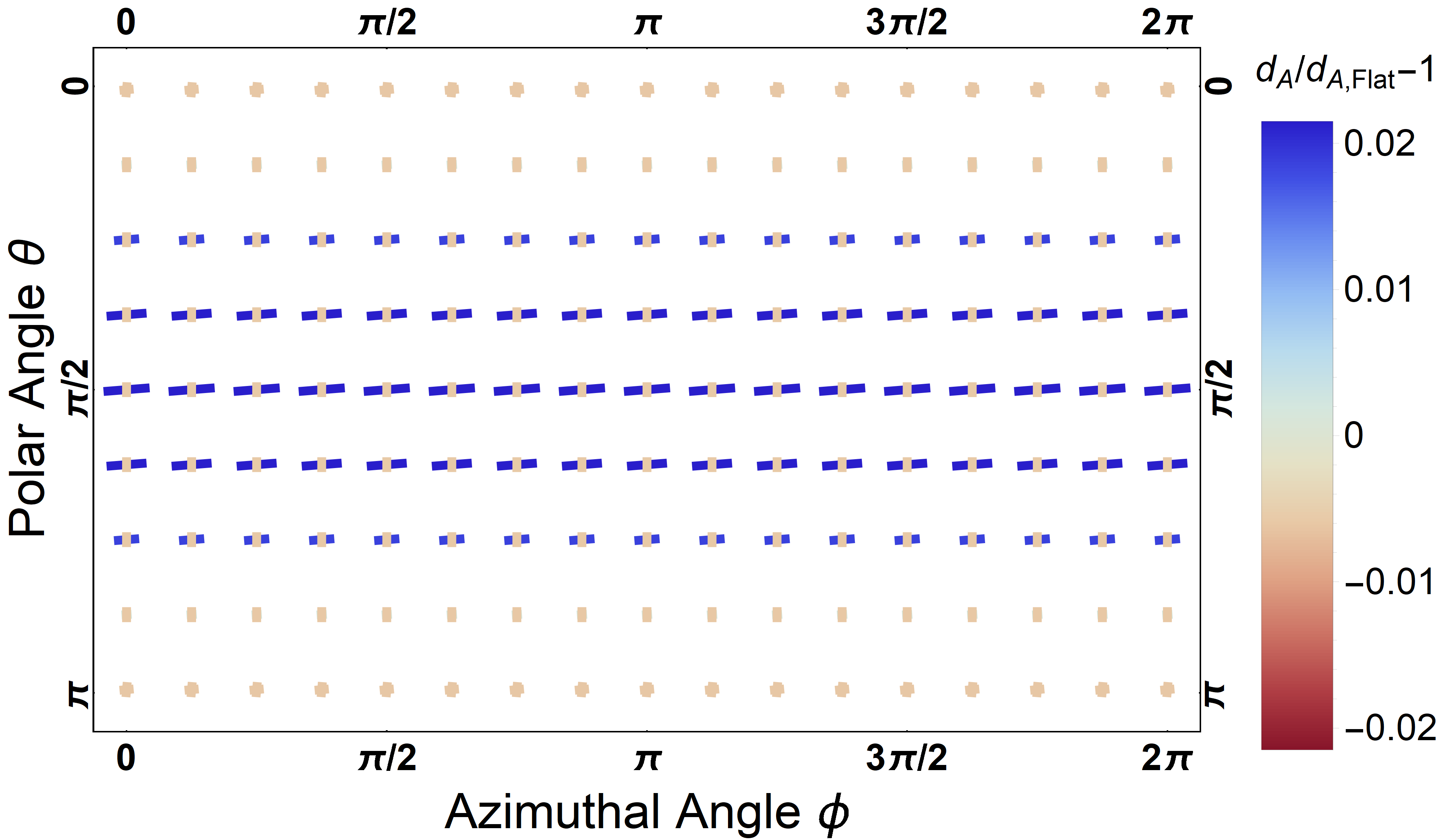}
\caption{$d_A$ for the $\widetilde{\text{U}(\mathds{H}^2})$ geometry.}
\label{Angdiadist-5-UH2}
\end{minipage}
\begin{minipage}[l]{0.5 \linewidth}
\includegraphics[width=1 \textwidth]{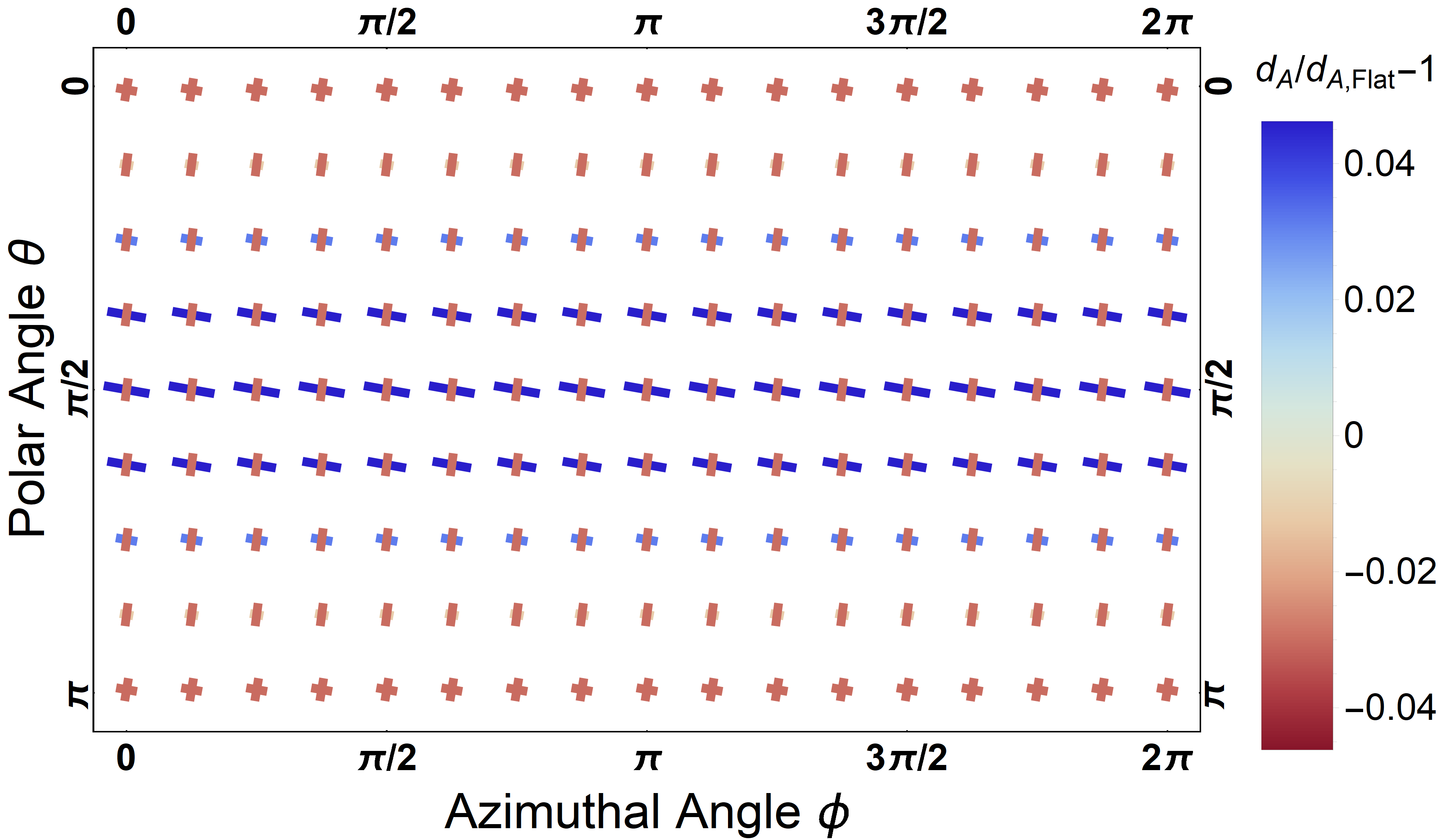}
\caption{$d_A$ for the Nil geometry.}
\label{Angdiadist-6-Nil}
\end{minipage}
\end{figure}
\vskip -0.2cm
The figure for the $\widetilde{\text{U}(\mathds{H}^2)}$ geometry bears some resemblance to that of the $\mathds{R} \times \mathds{H}^2$ geometry
in that arcs along the equator are contracted. This is not entirely unsurprising, as both geometries use the hyperbolic 2-plane $\mathds{H}^2$ as part of their definition. There are some major differences, however.

Most notably, although rotational symmetry around the $z$-axis is preserved, the arcs that are maximally contracted no longer lie along the $\phi$-direction but are tilted slightly out of the $(x,y)$-plane. This means that we lose mirror symmetry along the cardinal directions, which was present for the previous geometries. Additionally, we see that there is some slight stretching happening just off from the $\theta$-direction, and objects also become stretched in the $\phi$-direction towards the poles. \\

Nil looks very similar to a more extreme version of $\widetilde{\text{U}(\mathds{H}^2)}$ in that the overall behaviour is similar (although mirrored), but the magnitudes are more severe. The scale of the graph has increased by more than a factor of $2$, we see that the degree of stretching is more significant compared to the degree of compression and the orientation of the arcs most affected by the anisotropy is tilted out of the cardinal planes to a greater degree.

\begin{figure}[H]
\begin{minipage}[l]{0.5 \linewidth}
\includegraphics[width=1 \textwidth]{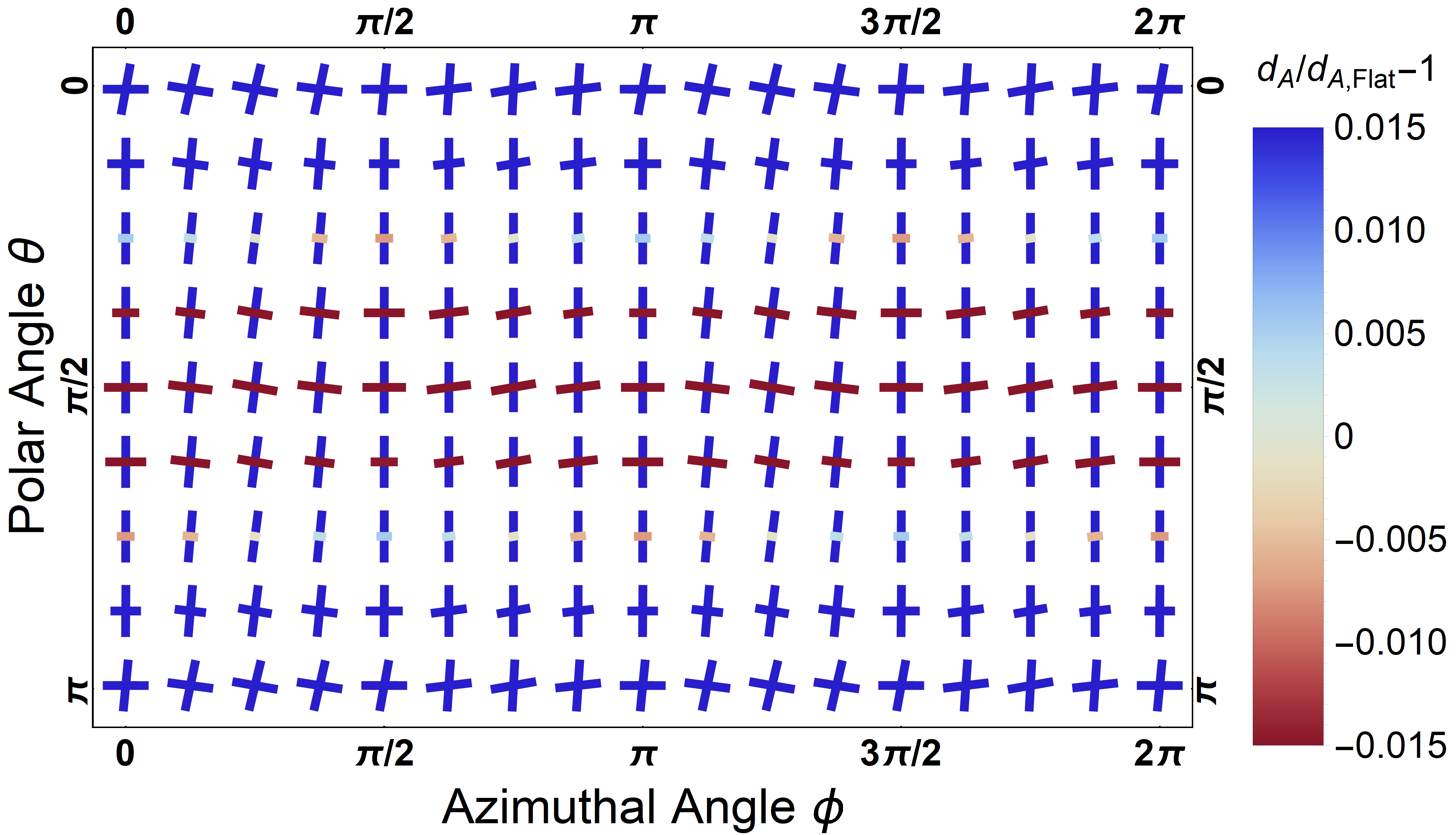}
\caption{$d_A$ for the Solv geometry.}
\label{Angdiadist-7-Solv}
\end{minipage}
\begin{minipage}[l]{0.05 \linewidth}
\end{minipage}
\begin{minipage}[l]{0.45 \linewidth}
The Solv behaves like an inverse of $\widetilde{\text{U}(\mathds{H}^2)}$ geometry, objects are stretched rather than compressed along approximately the $\phi$-direction at the equator and {\it vice versa} for the $\theta$-direction. However, there are also significant differences, as the graph for Solv shows a richer behavior. \\
We see that the angular diameter distance loses its axial symmetry entirely, as we see the two bars are oriented differently at each point on the plot. In addition, we see that the chiefly vertically-oriented arcs are always a deep blue, while the more horizontally-oriented arcs go from a deep blue at the poles to a deep red at the equator. This means that objects are very compressed at the poles, whereas they get deformed significantly at the equator -- being elongated in one direction and flattened in another.
\end{minipage}
\end{figure}

\subsection{Luminosity distance}

\noindent The figures below show the luminosity distance relative to a flat geometry at redshift of $z_* \simeq 1091$. Since this measure is now not dependent on a choice of arc, we can use a heat map to show the behavior in greater resolution. Red again indicates that the distance is shorter than the flat case and {\it vice versa} for blue. Alternatively, these figures can be read as indicating the relative change of apparent luminosity due to anisotropic curvature when compared to a flat geometry. \\

\noindent As before, every figure has its own color gradient, and the scale of the gradient gets larger for each subsequent plot. The growth of the scale in this plots is typically slower than for the angular diameter distance, as the image of an object may be very compressed in one direction but slightly stretched in another direction; this averages out to a moderate decrease in luminosity distance. We see that the anisotropic geometries generate `hot' (brighter) and `cold' (darker) spots at the poles relative to the equator.

\begin{figure}[H]
\begin{minipage}[l]{0.5 \linewidth}
\includegraphics[width=1 \textwidth]{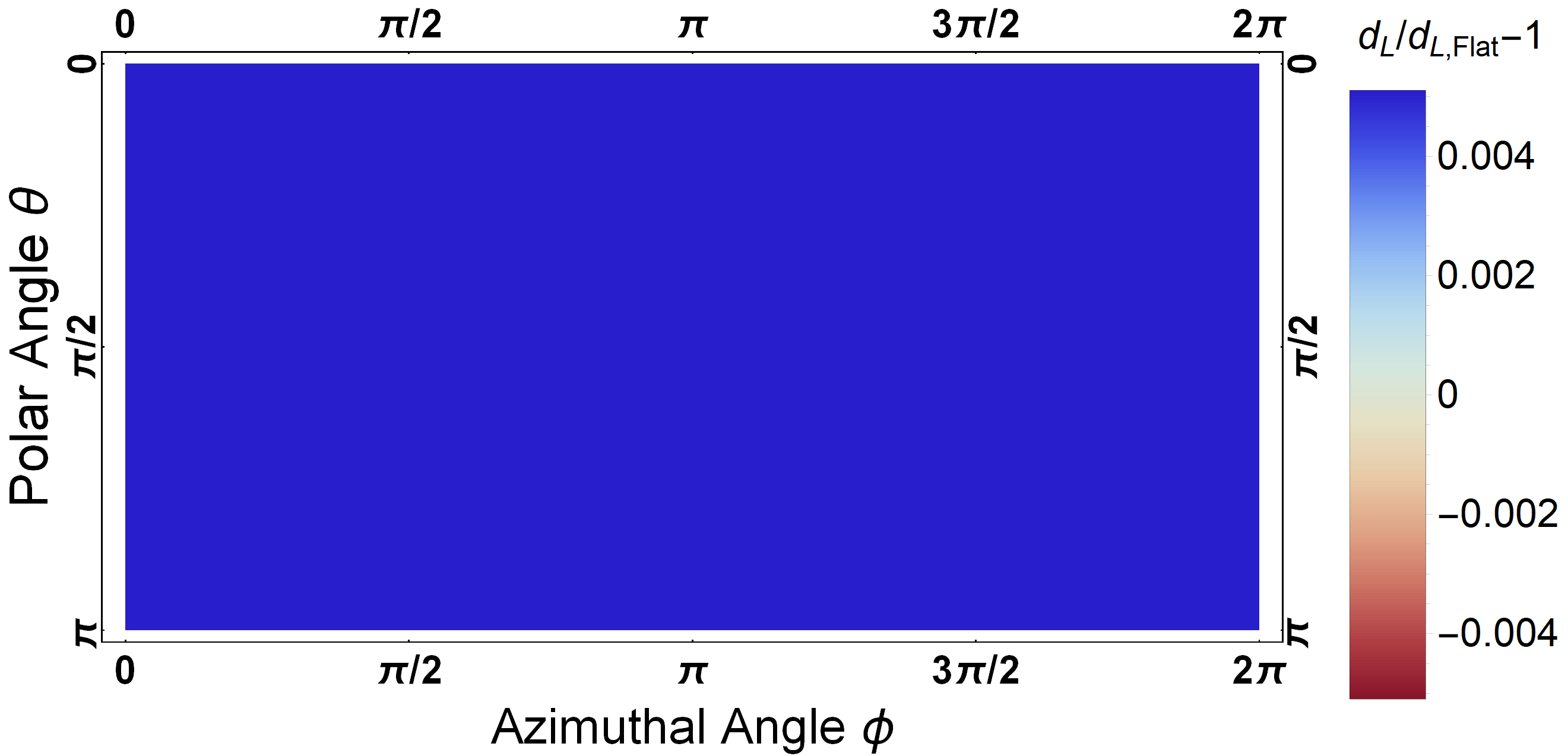}
\caption{$d_L$ for the $\mathds{H}^3$ geometry.}
\label{Lumdist-1-H3-Individual-Scale}
\end{minipage}
\begin{minipage}[l]{0.5 \linewidth}
\includegraphics[width=1 \textwidth]{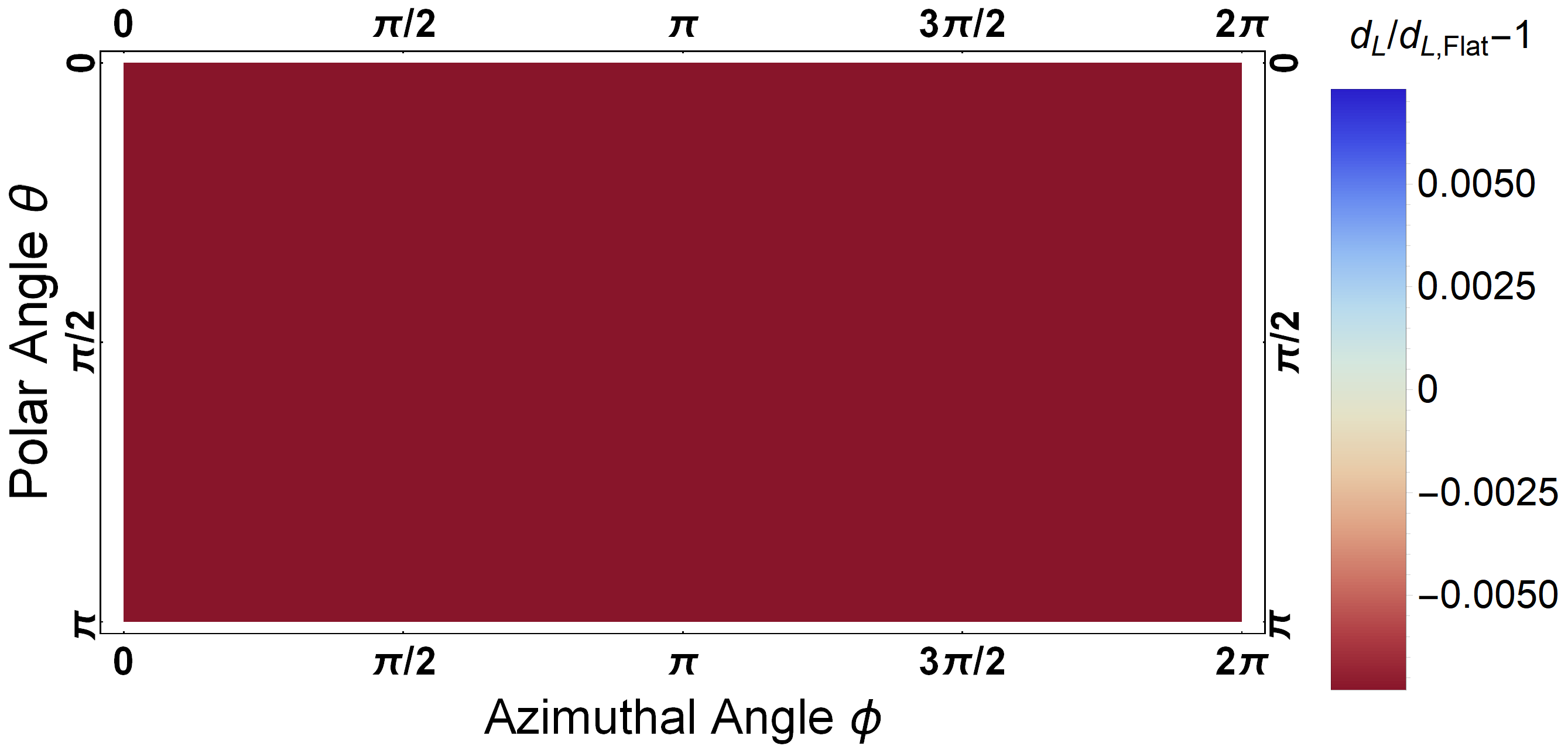}
\caption{$d_L$ for the $S^3$ geometry.}
\label{Lumdist-2-S3-Individual-Scale}
\end{minipage}
\end{figure}

\noindent The heat maps for the isotropic geometries are again exactly what we expect. They are constant with respect to $\theta$ and $\phi$, with luminosity distances being longer
(blue) in the hyperbolic geometric and shorter (red) in the spherical one.

\begin{figure}[H]
\begin{minipage}[l]{0.5 \linewidth}
\includegraphics[width=1 \textwidth]{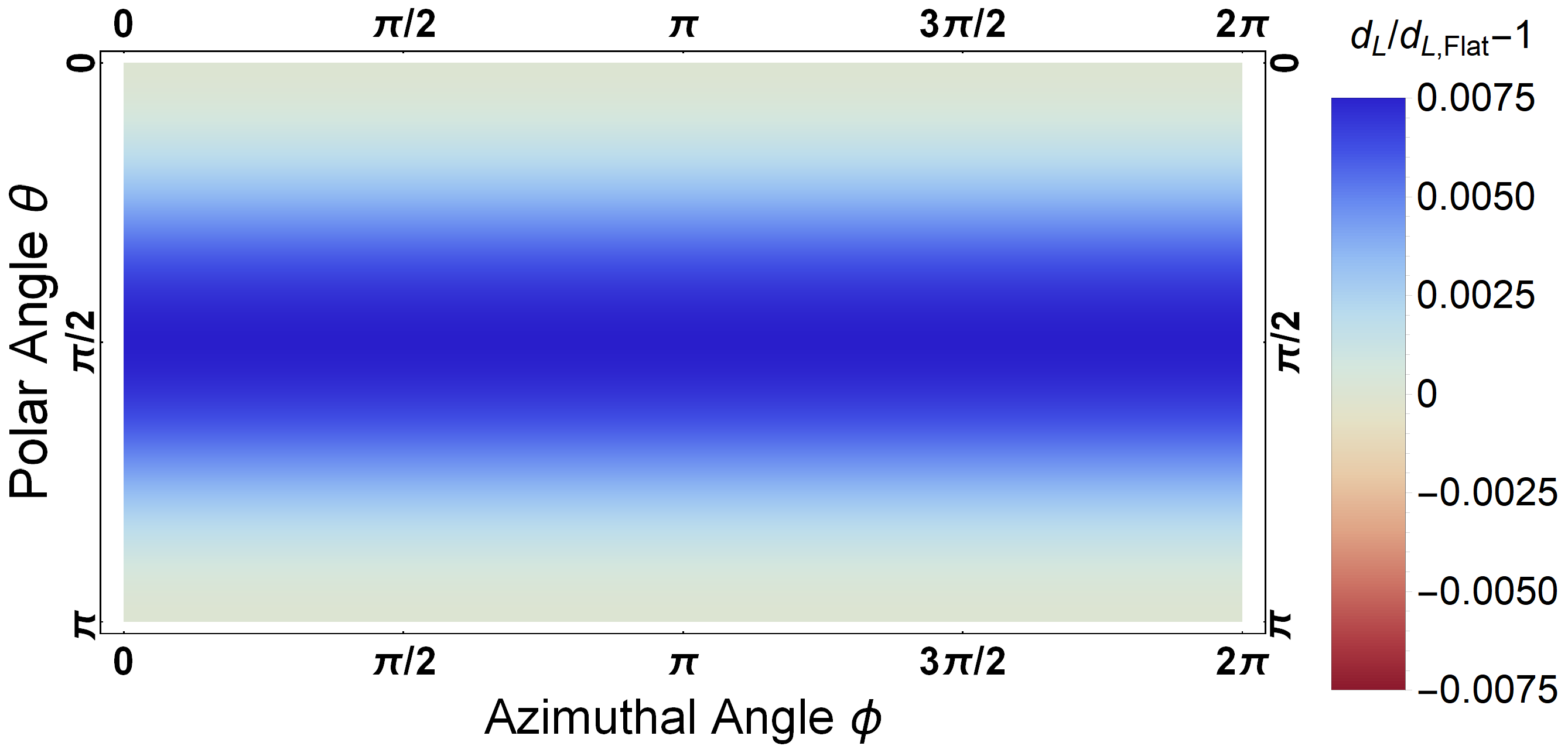}
\caption{$d_L$ for the $\mathds{R} \times \mathds{H}^2$ geometry.}
\label{Lumdist-3-RH2-Individual-Scale}
\end{minipage}
\begin{minipage}[l]{0.05 \linewidth}
\end{minipage}
\begin{minipage}[l]{0.45 \linewidth}
\includegraphics[width=1 \textwidth]{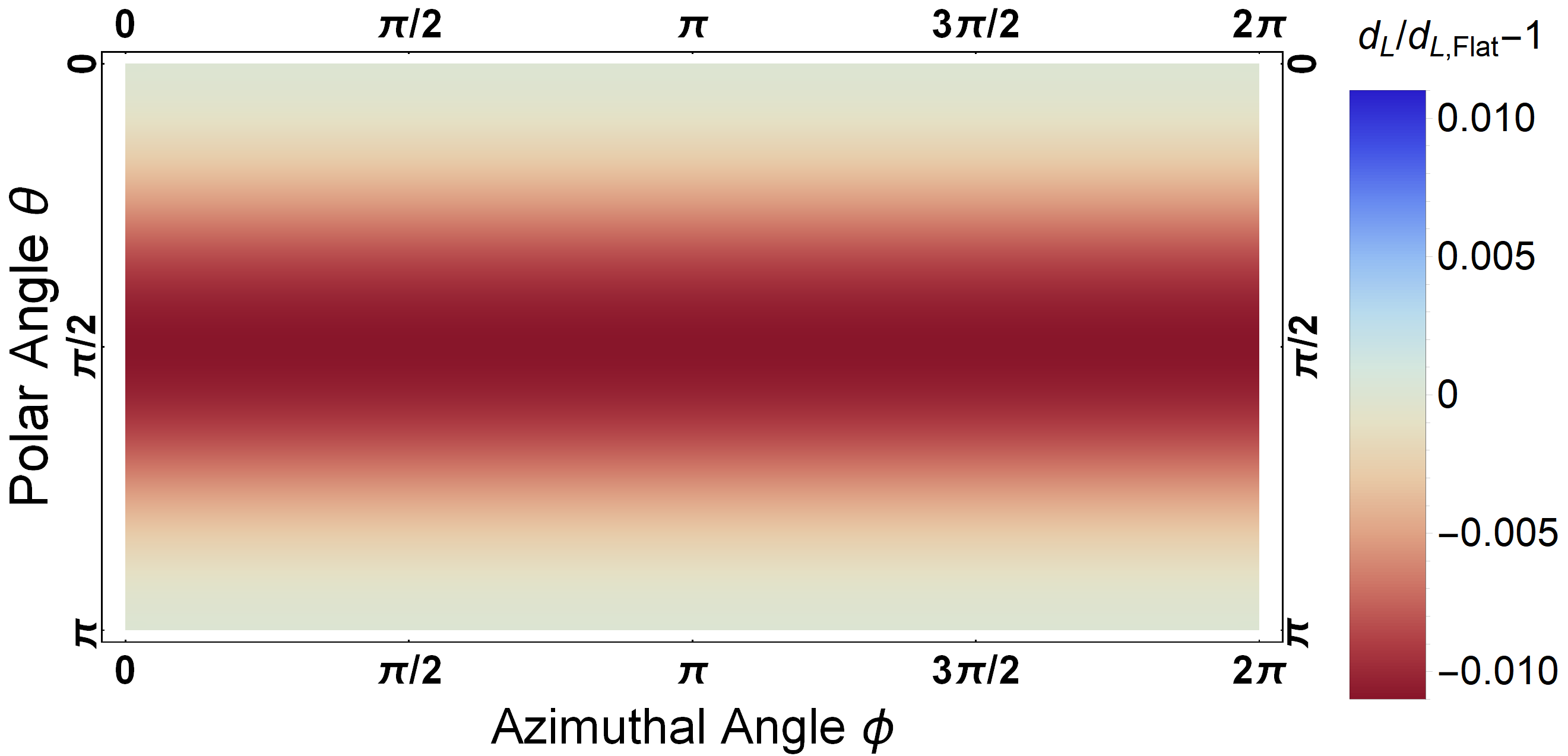}
\caption{$d_L$ for the $\mathds{R} \times S^2$ geometry.}
\label{Lumdist-4-RS2-Individual-Scale}
\end{minipage}
\end{figure}

\noindent The heat maps for the $\mathds{R} \times \mathds{H}^2$ and $\mathds{R} \times S^2$ geometries display the same behaviour as their barred counterparts above. Since objects get deformed in the $\phi$-direction -- compressed (blue) for $\mathds{H}^2$ and elongated (red) for $S^2$ -- the net apparent luminosity changes accordingly. This creates the pattern that we see, the distances are maximally elongated or shortened along the equator of the geometry and the effect vanished towards the poles. Note also that the scale on these luminosity plots differs by about a factor of two from the scale in the angular diameter distance plots, reflecting the fact that the anisotropy only affects one direction.

\begin{figure}[H]
\begin{minipage}[l]{0.5 \linewidth}
\includegraphics[width=1 \textwidth]{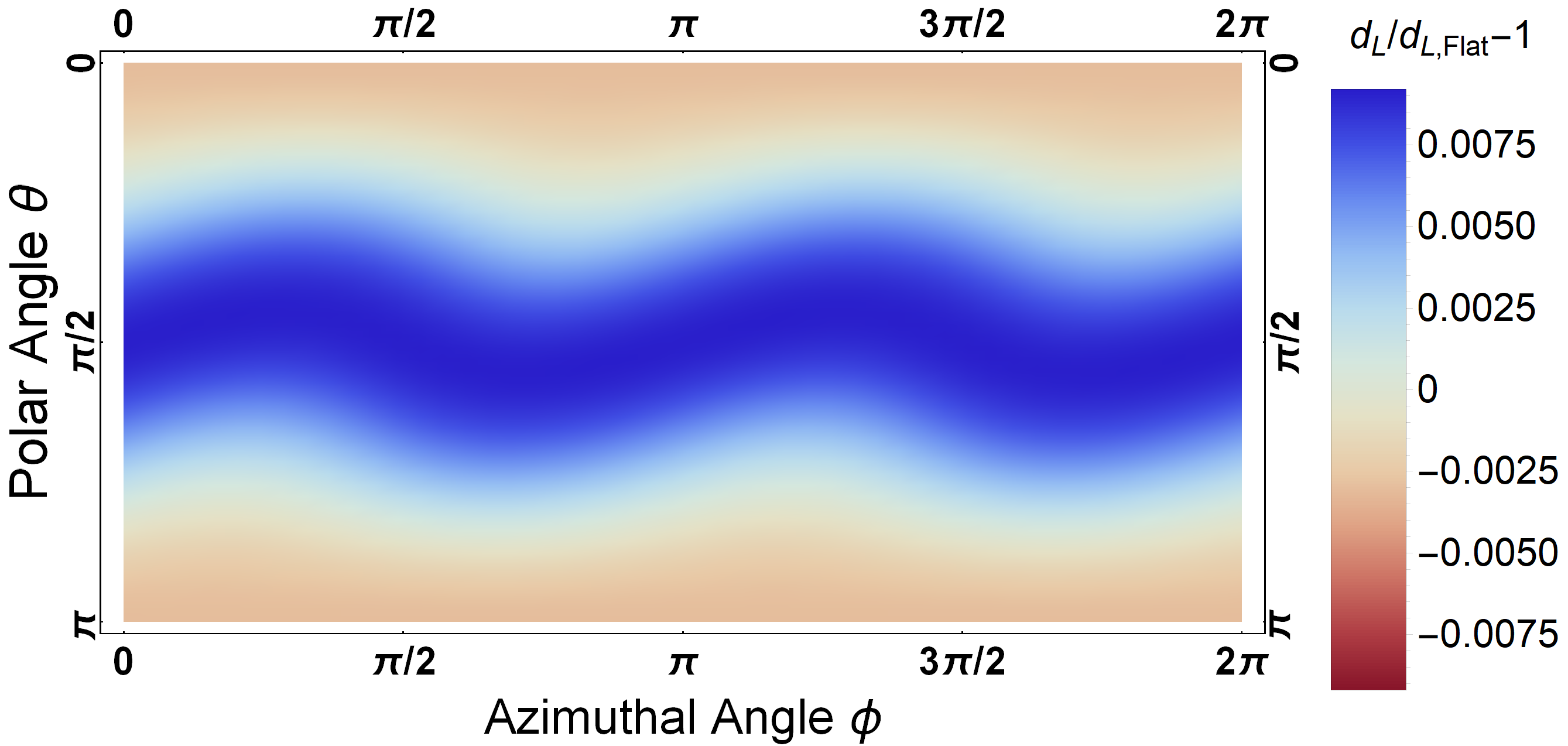}
\caption{$d_L$ for the $\widetilde{\text{U}(\mathds{H}^2)}$ geometry.}
\label{Lumdist-5-UH2-Individual-Scale}
\end{minipage}
\begin{minipage}[l]{0.5 \linewidth}
\includegraphics[width=1 \textwidth]{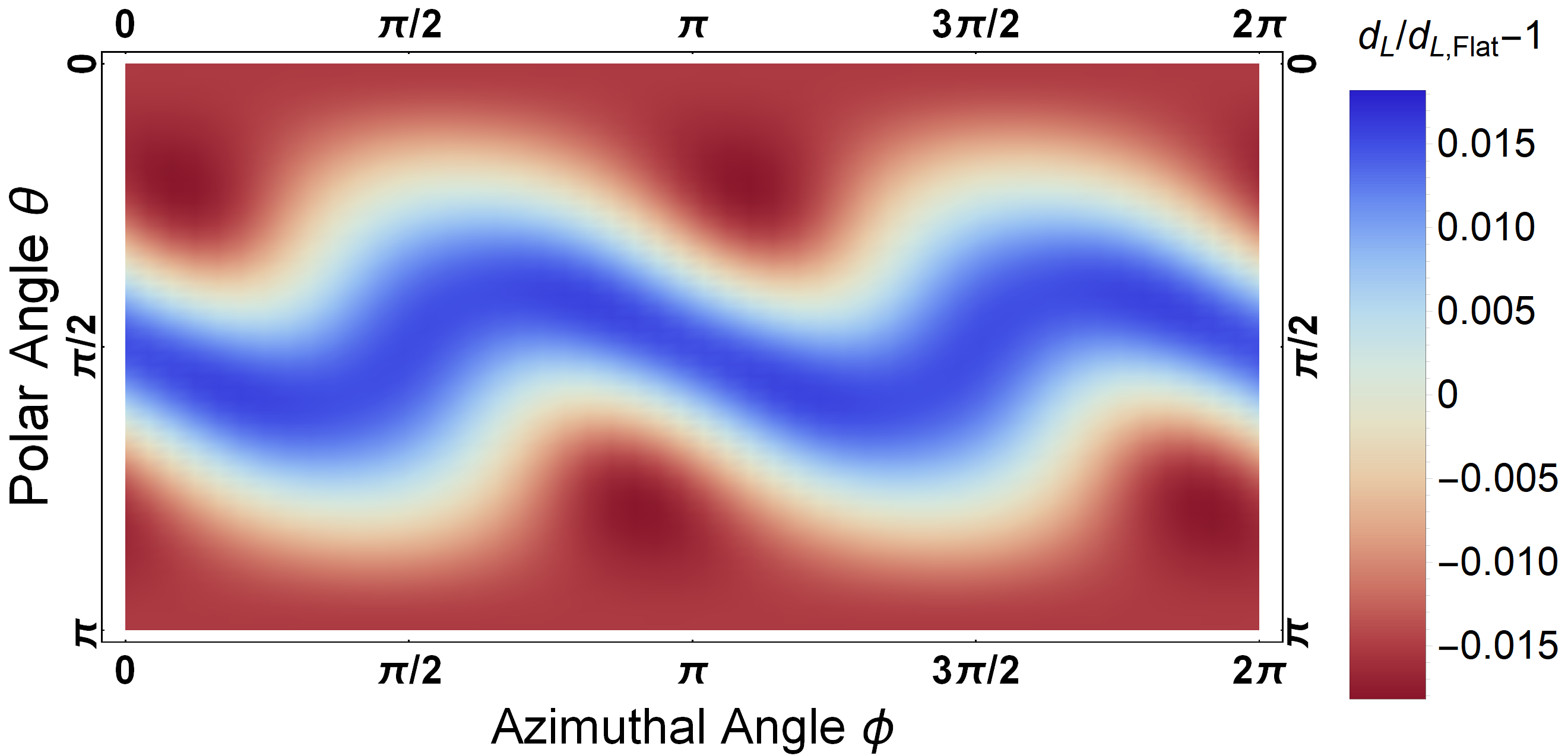}
\caption{$d_L$ for the Nil geometry.}
\label{Lumdist-6-Nil-Individual-Scale}
\end{minipage}
\end{figure}

\noindent The heatmap for $\widetilde{\text{U}(\mathds{H}^2)}$ and Nil geometries approximately match what we would expect from their barred plots, an overall decrease in the apparent luminosity along the equator and in increase towards the poles. However, these heatmaps do not show the neat, axially symmetric pattern that we observed for angular diameter distance, but is instead deformed into a wave-like pattern. This deformation is caused by the transformation between $\hat P$ and $\hat{\overline{P}}$ discussed in the previous section, and the direct result of geodesic trajectories being curved (twisted) by the geometry.

\begin{figure}[H]
\begin{minipage}[l]{0.49 \linewidth}
\includegraphics[width=1 \textwidth]{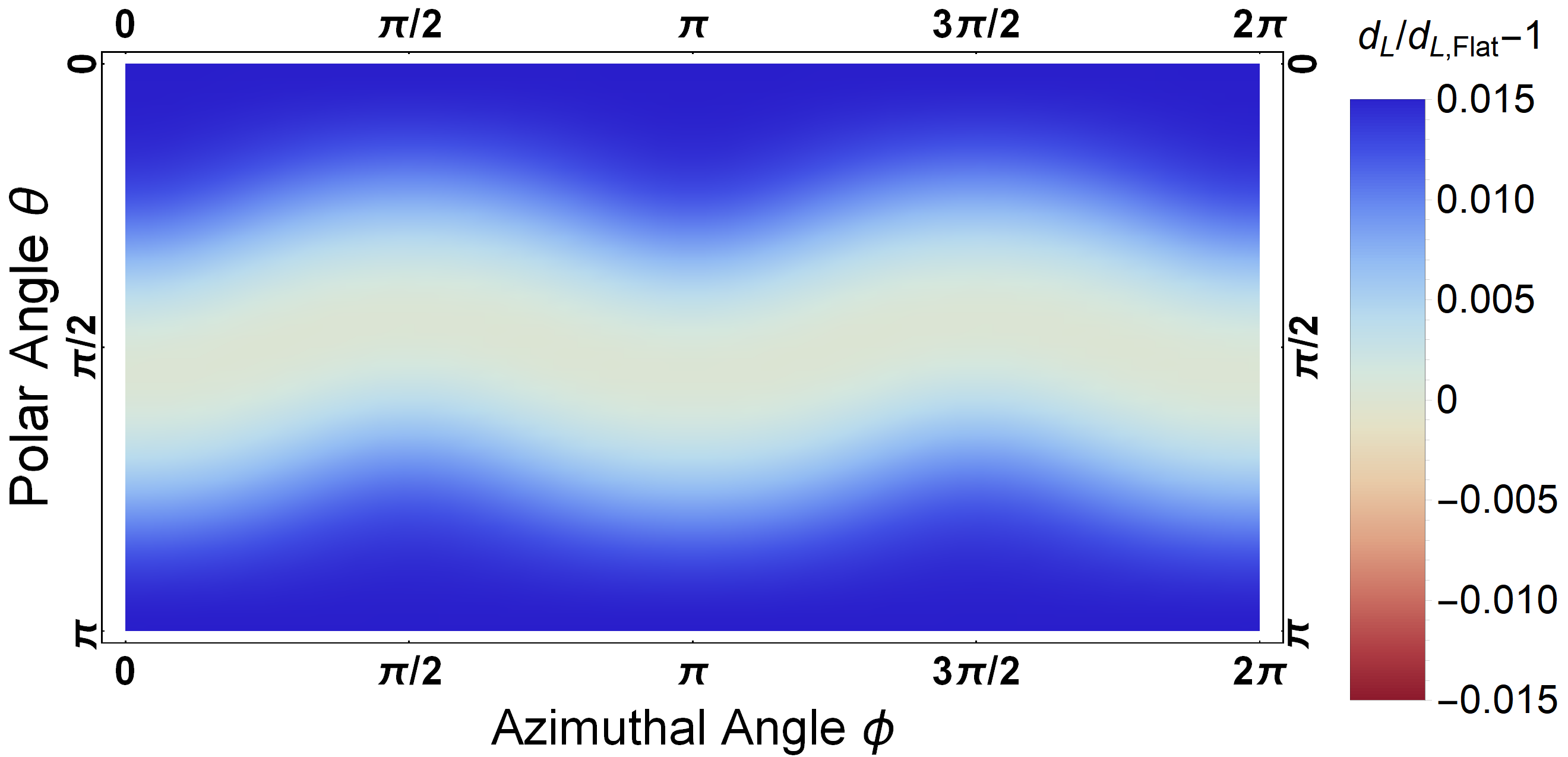}
\caption{$d_L$ for the Solv geometry}
\label{Lumdist-7-Solv-Individual-Scale}
\end{minipage}
\hskip 0.2cm
\begin{minipage}[l]{0.50 \linewidth}
Something similar happens for the Solv geometry, as the equatorial band gets deformed into a wavelike pattern. Note that the scale for the luminosity distance plot of this geometry is similar to that of its angular diameter distance plot, owing to the fact that objects get compressed maximally in both the $\phi$ and $\theta$ directions close to the poles.
\end{minipage}
\end{figure}

\subsubsection{Spherical harmonics}

\begin{figure}[H]
\begin{minipage}[l]{0.5 \linewidth}
\includegraphics[width=1 \textwidth]{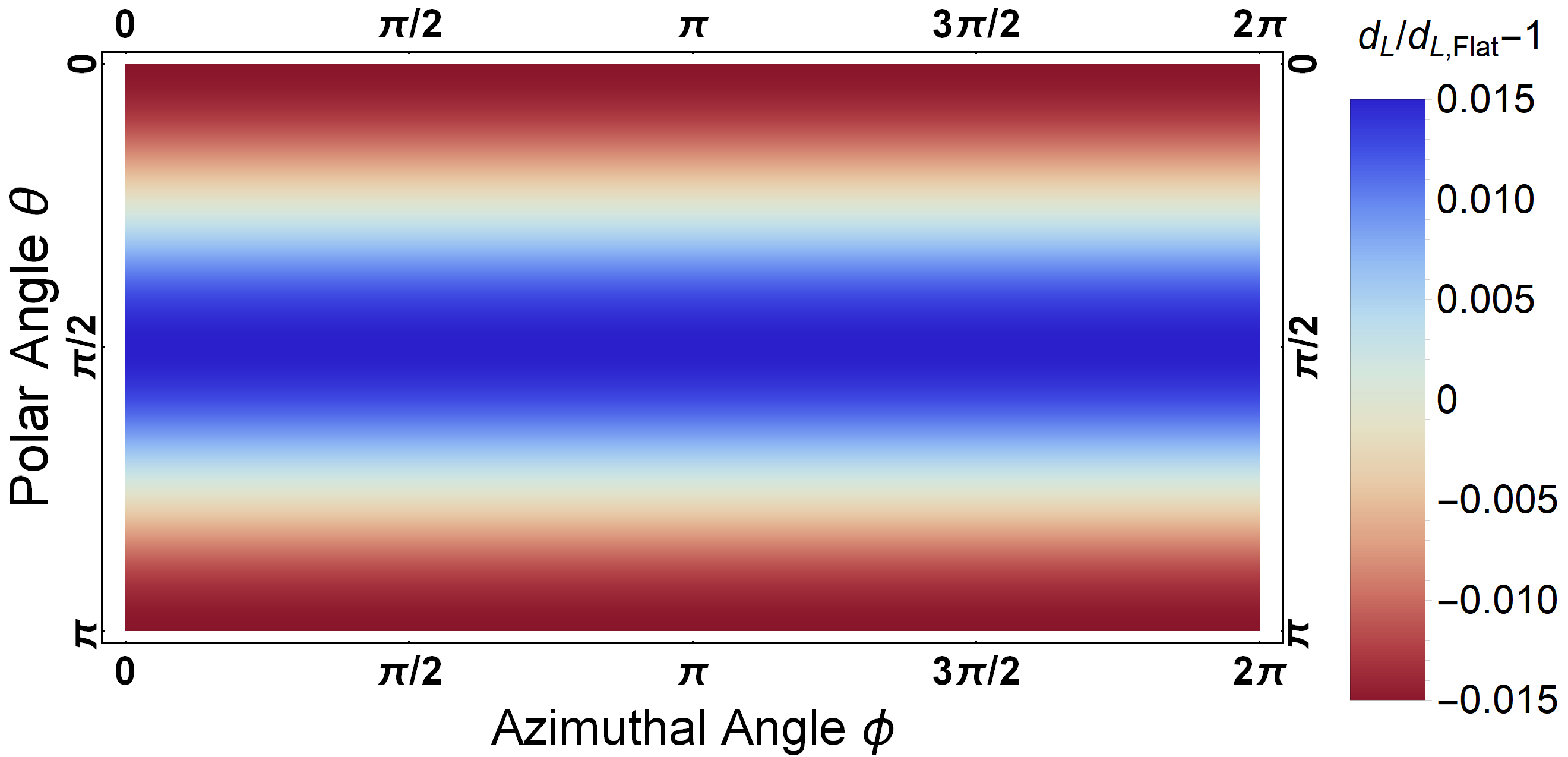}
\caption{The $(\lambda/L)^2$ modes for the Nil geometry.}
\label{figure 20}
\end{minipage}
\begin{minipage}[l]{0.5 \linewidth}
\includegraphics[width=1 \textwidth]{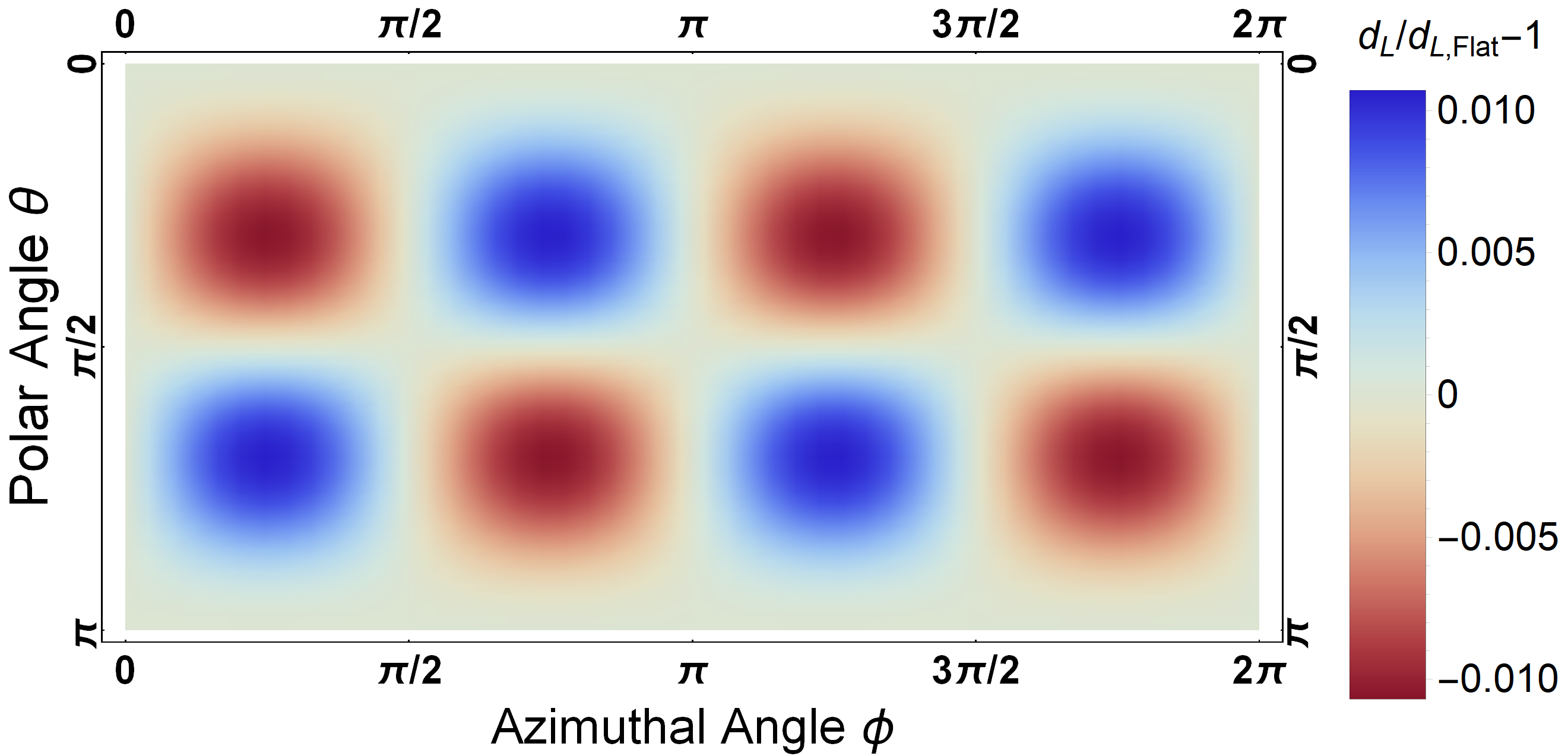}
\caption{The $(\lambda/L)^3$ modes for the Nil geometry.}
\label{figure 21}
\end{minipage}
\end{figure}

\begin{figure}[H]
\begin{minipage}[l]{0.5 \linewidth}
\includegraphics[width=1 \textwidth]{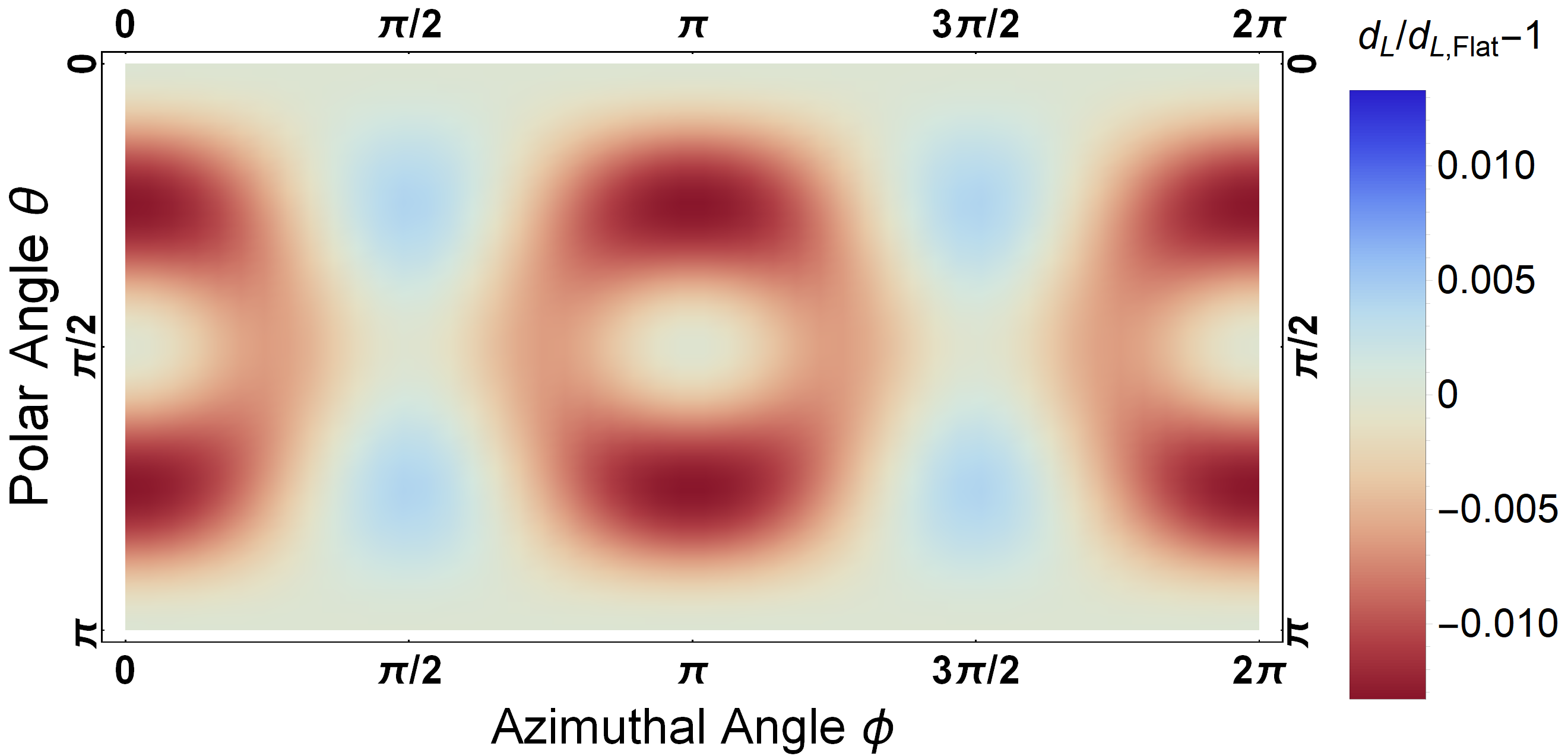}
\caption{The $(\lambda/L)^4$ modes for the Nil geometry.}
\label{figure 22}
\end{minipage}
\begin{minipage}[l]{0.5 \linewidth}
\includegraphics[width=1 \textwidth]{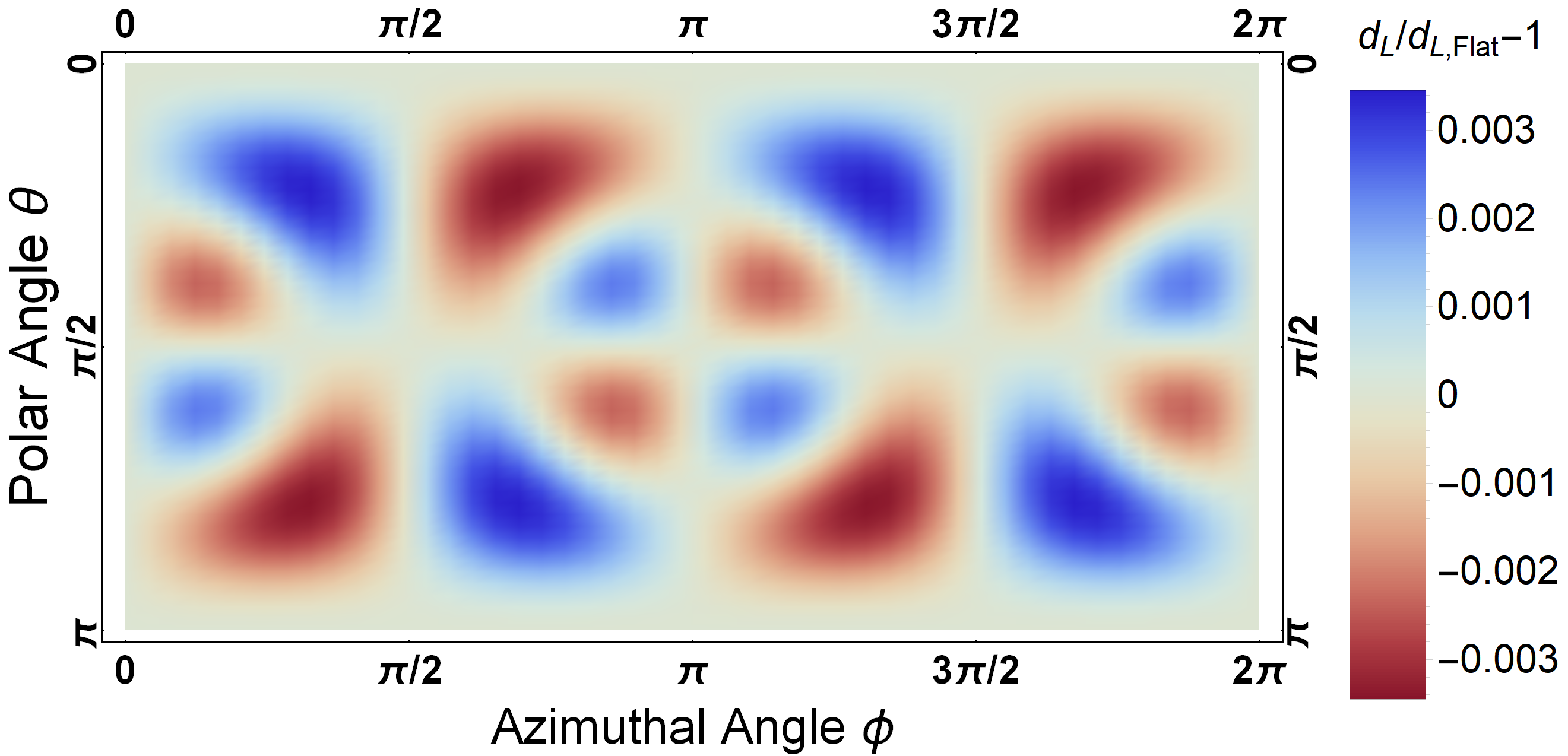}
\caption{The $(\lambda/L)^5$ modes for the Nil geometry.}
\label{figure 23}
\end{minipage}
\end{figure}

Anticipating the use of astronomical data for testing these geometries, we point at an interesting feature of these geometries when their expressions for luminosity distance are expanded in powers of $\lambda/L$. For at least the first few powers in this expansion, we observe that the series up to the $n$--th power can be expanded in spherical harmonics $Y_\ell^m$ of at most order $\ell = n$, suggesting that this is generally true.
However, we do not have a proof that this holds at higher powers. For example, the luminosity distance for the Nil geometry can be expanded as follows,
\begin{align}
\nonumber d_{L,\text{Nil}} &= \left(\frac{\sqrt{\pi}}{36} Y_0^0 - \frac{\sqrt{\pi}}{9\sqrt{5}} Y_2^0 \right)\left(\frac{\lambda}{L}\right)^2
                  + \left(\frac{i\sqrt{\pi}}{2\sqrt{210}} \left[Y_3^{2} - Y_3^{-2}\right]  \right)\left(\frac{\lambda}{L}\right)^3 \\
                 &+ \bigg(-\frac{2683\sqrt{\pi}}{43200} Y_0^0 + \frac{11\sqrt{\pi}}{945} Y_2^0 - \frac{5\sqrt{5\pi}}{168\sqrt{6}} \left[Y_2^{2} + Y_2^{-2}\right] \\
\nonumber        &+ \frac{649\sqrt\pi}{37800} Y_4^0 - \frac{5\sqrt{5\pi}}{252\sqrt{2}} \left[Y_4^{2} + Y_4^{-2}\right] + \frac{\sqrt\pi}{8\sqrt{70}} \left[Y_4^{4} + Y_4^{-4}\right] \bigg)\left(\frac{\lambda}{L}\right)^4 + \mathcal{O}\left(\frac{\lambda}{L}\right)^5.
\end{align}
A visual illustration of this feature for the Nil geometry is given in figures \ref{figure 20}-\ref{figure 23}.~\footnote{The heat maps of harmonics for other geometries can
be obtained upon request.}


\subsubsection{Violation of parity \& chiral symmetry}
\label{Sec:ParityViolationAndChirality}

Lastly, we point at two other interesting features that may be useful: all of the last three Thurston geometries violate parity, in the sense that one obtains a different metric if the geometry is mirrored through the origin; and two out of three are also chiral, in the sense that one cannot relate the geometry to its mirror image by rotations and translations. See also Table~\ref{table:MirrorReflections}. These features gain relevance in light of the recent claims that there is evidence for parity violation both in the CMB data~\cite{Minami:2020odp,Diego-Palazuelos:2022dsq,Diego-Palazuelos:2022cnh,Eskilt:2022cff} and in the LSS~\cite{Philcox:2022hkh,Creque-Sarbinowski:2023wmb,Coulton:2023oug}.

\begin{table}[!ht]
\centering
\begin{tabular}{|c|cc|ccc|ccc|c|}
\hline
                                     & Parity   & Chirality  & $x$ & $y$ & $z$ & $x$ \& $y$ & $x$ \& $z$ & $y$ \& $z$ & $x$, $y$ \& $z$ \\ \hline
$\widetilde{\text{U}(\mathds{H}^2)}$ & Violated & Chiral     & Y   & Y   & Y   & -          & -          & -          & Y               \\ \hline
Nil                                  & Violated & Chiral     & Y   & Y   & Y   & -          & -          & -          & Y               \\ \hline
Solv                                 & Violated & Not Chiral & -   & -   & Y   & -          & Y          & Y          & Y               \\ \hline
Rest                                 & Obeyed   & Not Chiral & -   & -   & -   & -          & -          & -          & -               \\ \hline
\end{tabular}
\caption{This table shows which of the last Thurston geometries violate parity and whether they are chiral or not. In addition, it is stated whether reflection through a given axis yields a different result (Y) or the same metric (-).}
\label{table:MirrorReflections}
\end{table}

\noindent This is fairly easy to derive from the metrics of these geometries, which will again present in a simple form below, using $L = \sqrt{-\kappa}$ for notational convenience.
\begin{align}
{\rm d}\Sigma_{\widetilde{\text{U}(\mathds{H}^2)}}^2
 &= {\rm d}x^2 +\cosh(2x/L) {\rm d}y^2  + {\rm d}z^2 + 2  \sinh(x/L){\rm d}y {\rm d}z \\
{\rm d}\Sigma_{\text{Nil}}^2
 &= {\rm d}x^2 + \left( 1 + \tfrac{x^2}{L^2}\right)  {\rm d}y^2+ {\rm d}z^2 - \tfrac{2x}{L} {\rm d}y {\rm d}z \\
{\rm d}\Sigma_{\text{Solv}}^2      &={\rm e}^{2z/L}{\rm d}x^2 +{\rm e}^h{-2z/L} {\rm d}y^2 + {\rm d}z^2
\end{align}
The diagonal components of the $\widetilde{\text{U}(\mathds{H}^2)}$ geometry are mirror-even for reflections along cardinal directions, but the off-diagonal component, $2 {\rm d}y {\rm d}z \sinh(x/L)$, is mirror-odd. Hence mirroring this geometry through the origin, which is really mirroring in all three cardinal directions, will flip the sign of this metric and yield a different geometry. Since any rotation can be expressed as a pair of two reflections (typically with respect to two different hyperplanes), it follows that no rotation can undo the effect of a single reflection. Hence this geometry violates parity and is chiral in the sense described above.

The reasoning is similar for the Nil geometry, whose diagonal elements are insensitive to reflections along cardinal directions, but the off-diagonal term $2x {\rm d}y {\rm d}z / L$ picks up a minus sign if any such reflection is performed. Using the same line of reasoning, Nil is also parity-violating and chiral.

The Solv geometry behaves slightly differently, as only reflection along the $z$-direction changes the metric by effectively exchanging the coordinates $x$ and $y$. This change can then simply be undone by rotating by 90 degrees in the $(x,y)$-plane to return the metric to its original form. It follows that this geometry violates parity, but is not chiral in the sense described above. \\

\noindent This violation of parity is also a visible feature in our plots for luminosity distance. To see this, we can decompose $d_L(\hat{P},\lambda)$ into a parity-even and a parity-odd component using the following expressions,
\begin{align}
\label{eq:ParityOdd}  d_L(\hat{P},\lambda)_\textit{odd}  &= \frac{1}{2} \left( d_L(\hat{P},\lambda) - d_L(-\hat{P},\lambda) \right) = \frac{1}{2} \left( d_L(\hat{P},\lambda) - d_L(\hat{P},-\lambda) \right) \\
\label{eq:ParityEven} d_L(\hat{P},\lambda)_\textit{even} &= \frac{1}{2} \left( d_L(\hat{P},\lambda) + d_L(-\hat{P},\lambda) \right) = \frac{1}{2} \left( d_L(\hat{P},\lambda) + d_L(\hat{P},-\lambda) \right) \ .
\end{align}
These expressions are plotted in Figures~\ref{LumDist-UH2-Parity-Even}--\ref{LumDist-Solv-Parity-Odd} below. Note that equations~\eqref{eq:ParityOdd} and \eqref{eq:ParityEven} imply that even (odd) powers in $\lambda/L$ must likewise be parity-even (parity-odd), which can be seen by contrasting Figures~\ref{LumDist-Nil-Parity-Even} and \ref{LumDist-Nil-Parity-Odd} for the Nil geometry with Figures~\ref{figure 20}--\ref{figure 23} above.

\begin{figure}[H]
\begin{minipage}[l]{0.48 \linewidth}
\includegraphics[width=1 \textwidth]{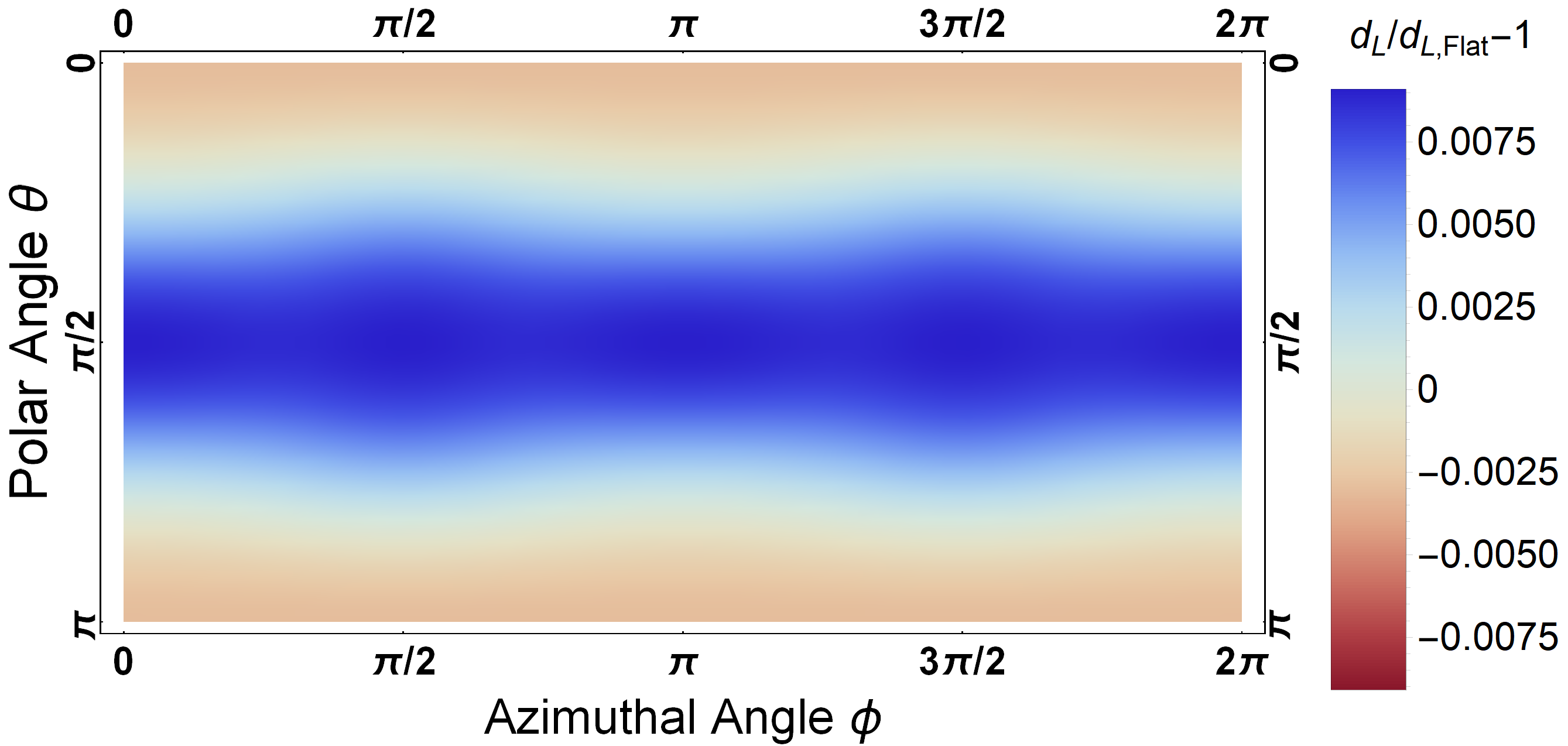}
\caption{Parity-even component of $d_L$ for $\widetilde{\text{U}(\mathds{H}^2)}$.}
\label{LumDist-UH2-Parity-Even}
\end{minipage}
\begin{minipage}[l]{0.02 \linewidth}
\end{minipage}
\begin{minipage}[l]{0.48 \linewidth}
\includegraphics[width=1 \textwidth]{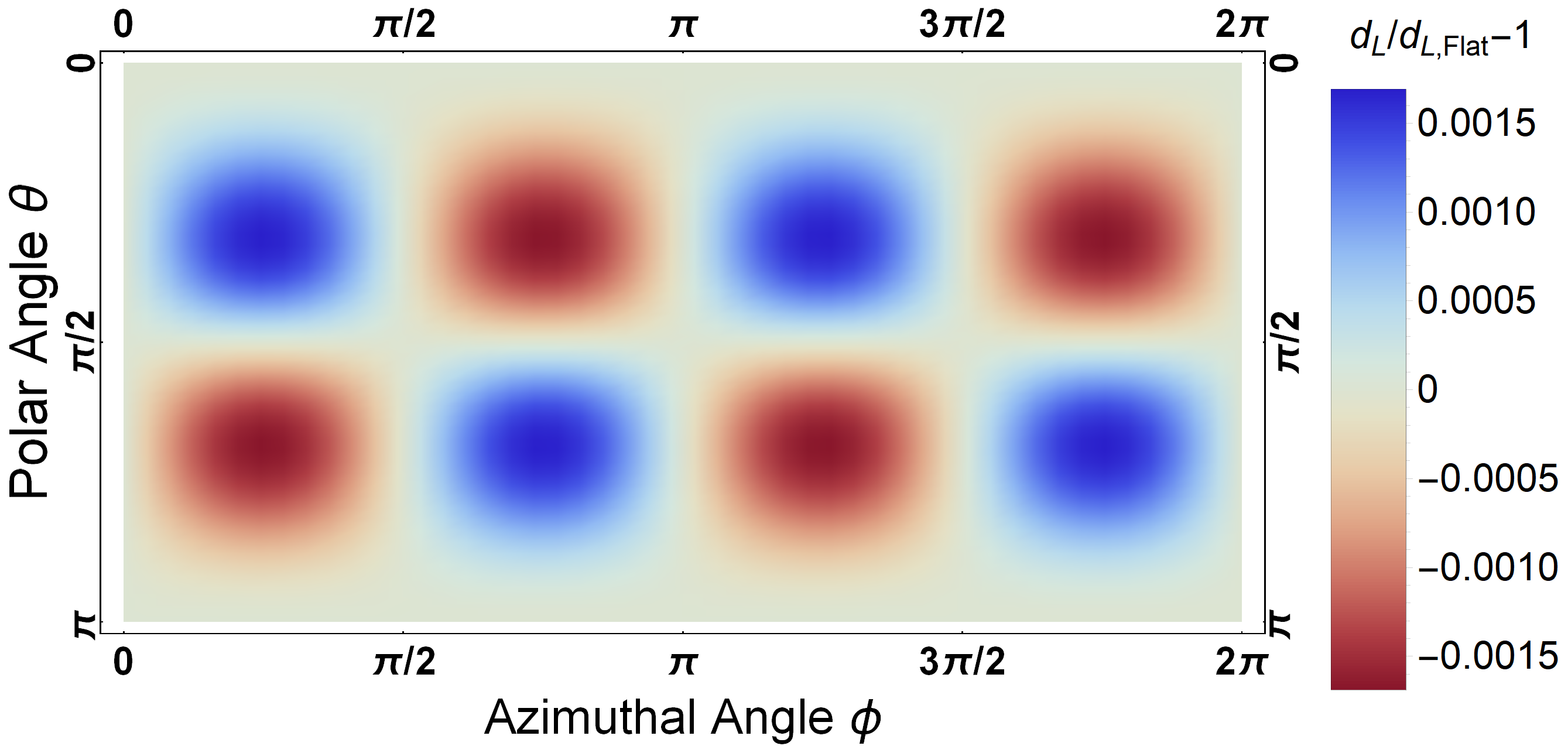}
\caption{Parity-odd component of $d_L$ for $\widetilde{\text{U}(\mathds{H}^2)}$.}
\label{LumDist-UH2-Parity-Odd}
\end{minipage}
\end{figure}

\begin{figure}[H]
\begin{minipage}[l]{0.48 \linewidth}
\includegraphics[width=1 \textwidth]{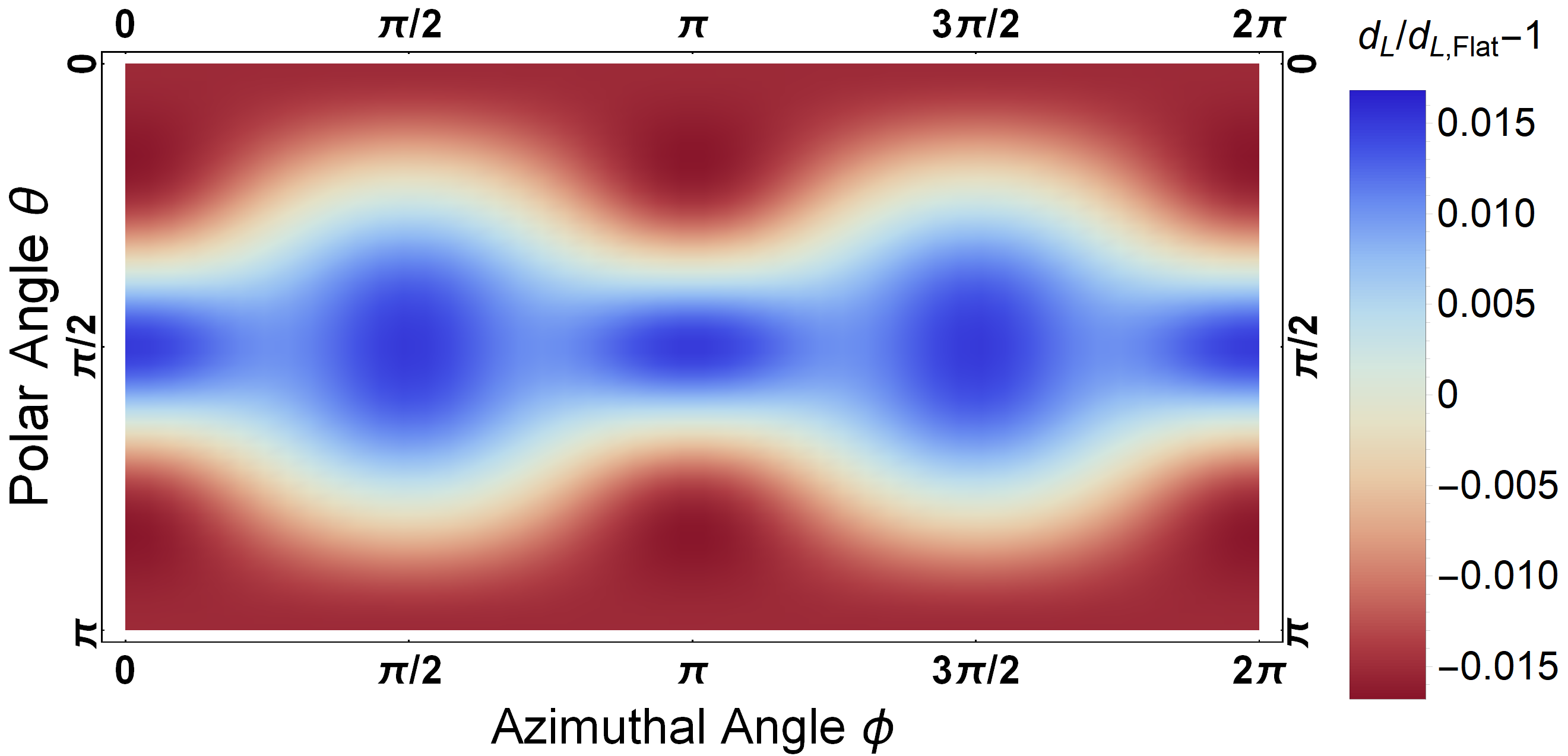}
\caption{Parity-even component of $d_L$ for Nil.}
\label{LumDist-Nil-Parity-Even}
\end{minipage}
\begin{minipage}[l]{0.02 \linewidth}
\end{minipage}
\begin{minipage}[l]{0.48 \linewidth}
\includegraphics[width=1 \textwidth]{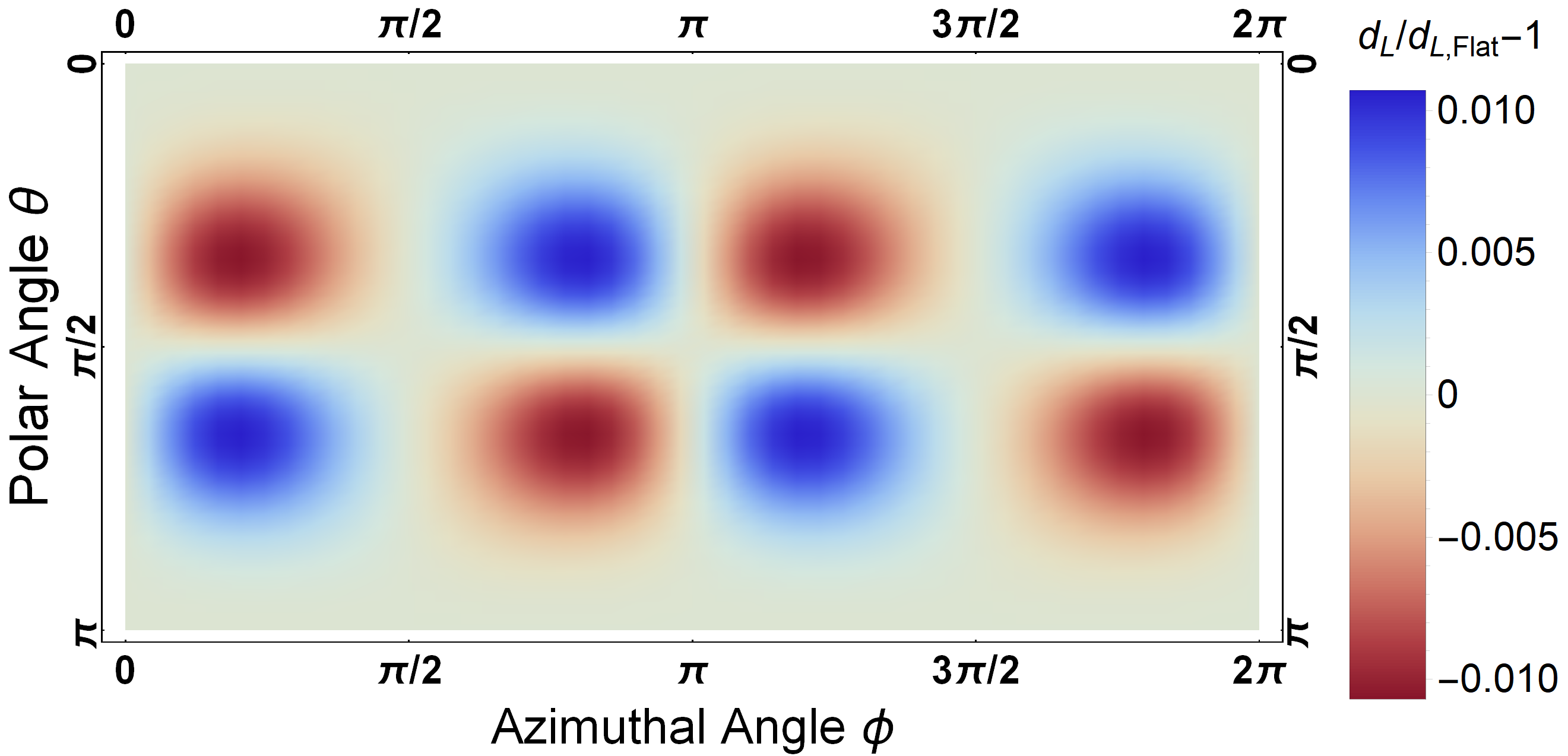}
\caption{Parity-odd component of $d_L$ for Nil.}
\label{LumDist-Nil-Parity-Odd}
\end{minipage}
\end{figure}

\begin{figure}[H]
\begin{minipage}[l]{0.48 \linewidth}
\includegraphics[width=1 \textwidth]{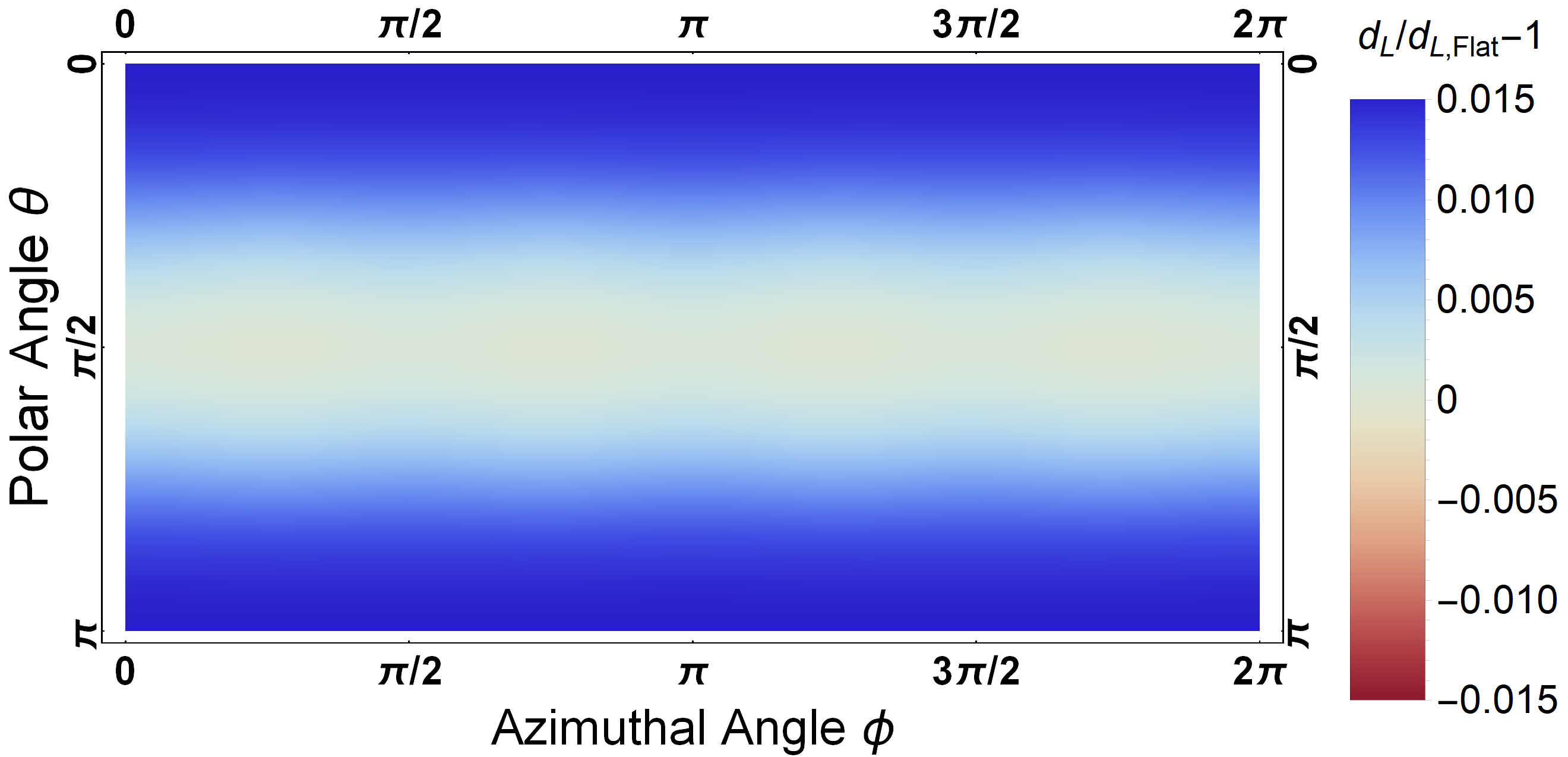}
\caption{Parity-even component of $d_L$ for Solv.}
\label{LumDist-Solv-Parity-Even}
\end{minipage}
\begin{minipage}[l]{0.02 \linewidth}
\end{minipage}
\begin{minipage}[l]{0.48 \linewidth}
\includegraphics[width=1 \textwidth]{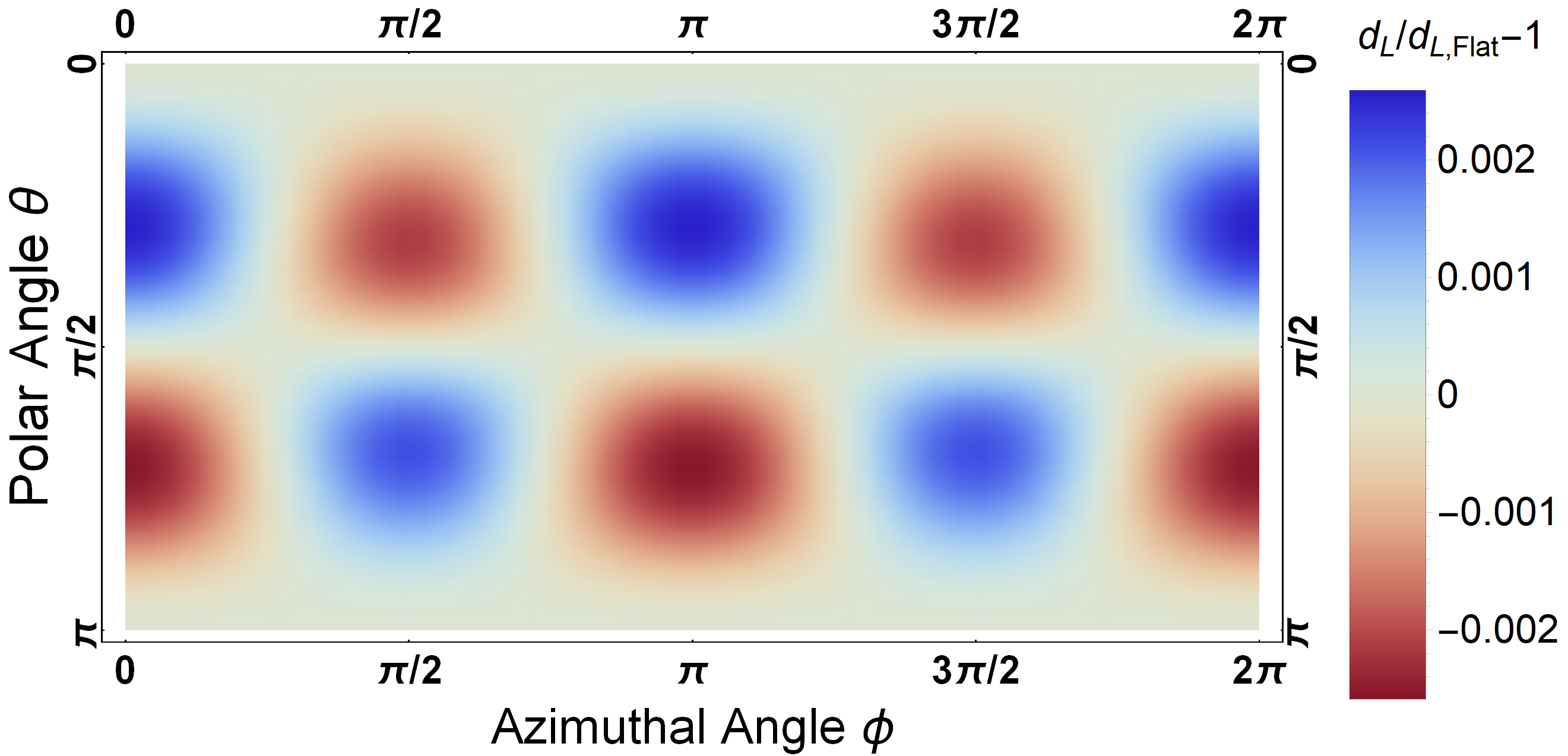}
\caption{Parity-odd component of $d_L$ for Solv.}
\label{LumDist-Solv-Parity-Odd}
\end{minipage}
\end{figure}

Note that all of the reflections marked with $Y$ in Table~\ref{table:MirrorReflections} act through the same mechanism -- which is either flipping the sign on the off-diagonal term, or effectively exchanging the $x$- and $y$-coordinate. This means that reflection-even and reflection-odd luminosity distances defined in an equivalent manner with respect to any of the reflections marked with $Y$ will yield the same result as the expressions in equations~\eqref{eq:ParityOdd} and \eqref{eq:ParityEven}.


\section{Anisotropic Scale Factors}
\label{sec:AnisotropicScaleFactors}

In Section \ref{sec:BackgroundEvolution}, we found that solving the Friedmann equations for anisotropic geometries required the introduction of a rather peculiar fluid with a nonvanishing shear tensor $\pi^i_{\hspace{1ex}j}$ and an $\propto a^{-2}$ scaling property. In this section we will ask the question whether we can get rid of this shear tensor by dropping the assumption of isotropy for the scale factor $a$. In particular, we are interested in figuring out whether this approach is compatible with a perfect fluid solution and in studying the time dependence of anisotropies in the expansion. \\

We will make a slight alteration to the metric equation \eqref{eq:ThustonCosmo2} by splitting the scale factor $a$ into three components, $A_1$, $A_2$, and $A_3$ that are {\it a priori} independent.
\begin{equation}
{\rm d}s^2 = -{\rm d}t^2 + \hspace{1ex} \gamma_{ab,\text{Thurston}} \hspace{1ex} \begin{pmatrix}
A_1(t) & 0 & 0 \\
0 & A_2(t) & 0 \\
0 & 0 & A_3(t)
\end{pmatrix}^a_{\hspace{1ex}i} \begin{pmatrix}
A_1(t) & 0 & 0 \\
0 & A_2(t) & 0 \\
0 & 0 & A_3(t)
\end{pmatrix}^b_{\hspace{1ex}j} {\rm d}x^i {\rm d}x^j
         \,,\quad
\label{Bianchi Type-I metric}
\end{equation}
The Friedmann equations in an expanding universe in the metric~(\ref{Bianchi Type-I metric}) ({\it cf. e.g.} Ref.~\cite{2067292}) depend on the Thurston geometry under consideration, but will admit a generic perfect fluid solution after some work. By splitting $a$, we introduce non-zero terms in the off-diagonal components of the Einstein tensor that depend on (derivatives of) $A_1$, $A_2$, and $A_3$. If we insist on working in a perfect fluid solution, we must require that these off-diagonal terms vanish. This means that we must solve one or more differential equations that constrain $A_1$, $A_2$, and $A_3$.

For instance, in the Solv geometry, $G^{0}_{\hspace{1ex}3}$ picks up terms dependent on $A_1$ and $A_2$, which must vanish if we wish to obtain a perfect fluid solution. This means that we must require
\begin{equation}
G^{0}_{\hspace{1ex}3,\text{Solv}} = \frac{\sqrt{-\kappa}}{A_3^2} \left( \frac{\dot{A_2}}{A_2} - \frac{\dot{A_1}}{A_1} \right) = 0.
\end{equation}
One can easily satisfy this equation by setting $A_1$ proportional to $A_2$;
their relative (constant) factor $A_2/A_1$ can be absorbed by a suitable rescaling of
the $y$-coordinate in the metric. The constraints that follow from applying this procedure to all geometries are shown in Table~\ref{table:AnisotropicExpansionParameters}.

\begin{table}[!ht]
\centering
\begin{tabular}{|c|l|c|}
\hline
\textbf{Space-time}                             & \textbf{Constraints}  & $A_{\rm dom}$ \\ \hline
$\mathbf{\mathds{R}^3}$                         & $\kappa = 0$          & n/a       \\
$\mathbf{\mathds{H}^3} / \mathbf{S^3}$          & $A_1=A_2=A_3$         & $a$       \\
$\mathbf{\mathds{R} \times \mathds{H}^2 / S^2}$ & $A_1=A_2$             & $A_1$     \\
$\widetilde{\text{U}(\mathds{H}^2)}$            & $A_1=A_2=A_3$         & $a$       \\
\textbf{Nil }                                   & $A_2=A_3$             & $A_1$     \\
\textbf{Solv}                                   & $A_1=A_2$             & $A_3$     \\ \hline
\end{tabular}
\caption{Anisotropic scale factor constraints and the dominant contribution to the Einstein field equations.}
\label{table:AnisotropicExpansionParameters}
\end{table}

\noindent The $\widetilde{\text{U}(\mathds{H}^2)}$ and Nil geometries are again somewhat exceptional here, as $G^3_{\hspace{1ex}2} = x \sqrt{-\kappa}\kappa/A_1(t)^2$
cannot be made to vanish by equating scale factors.
 However,
as discussed in Section \ref{sec:BackgroundEvolution},
this term is suppressed by $x/L\ll 1$ with respect to
the leading curvature contributions, and can therefore be neglected
at the leading order in $\kappa$.

What is of additional note about $\widetilde{\text{U}(\mathds{H}^2)}$ is that the constraint equations force the scale factors for all three directions to be the same. This means that it is not possible for this geometry to get rid of the special $\propto a^{-2}$  terms in the Einstein Tensor by introducing anisotropic expansion while leaving the underlying geometric structure static. This means we don't know how a universe with this geometry evolves,
as spatial sections evolve dynamically; but for sufficiently large $L$ we know that anisotropies remain small and for the periods of time and scales at which we can measure, this effect is likely to be small. We will not spend any additional time on this and leave it as an open problem to devise a peculiar fluid with the appropriate scaling properties or to understand how this geometry can be supported dynamically. \\

\noindent With these constraints satisfied, the diagonal elements of the Einstein field equations take the following form:
\begin{eqnarray}
\hspace{0em} ^0_{\hspace{1ex}0}\hspace{1em}\text{equation:} \hspace{-0em}
\qquad
 \frac{\dot A_1}{A_1}\frac{\dot A_2}{A_2} + \frac{\dot A_2}{A_2}\frac{\dot A_3}{A_3} +\frac{\dot A_3}{A_3}\frac{\dot A_1}{A_1}
 &=& 8 \pi G \rho + \Lambda  + \frac{K^{(0)}\kappa}{A_{\rm dom}^2}
\label{Friedmann equation 00}\\
\hspace{0em} ^1_{\hspace{1ex}1}\hspace{1em}\text{equation:} \hspace{-0em}
\qquad\qquad\,
\frac{\ddot A_2}{A_2}+\frac{\ddot A_3}{A_3} + \frac{\dot A_2}{A_2}\frac{\dot A_3}{A_3}
&=& 8 \pi G(-{\cal P}) + \Lambda + \frac{K^{(1)}\kappa}{A_{\rm dom}^2}
\label{Friedmann equation 11}\\
\hspace{0em} ^2_{\hspace{1ex}2}\hspace{1em}\text{equation:} \hspace{-0em}
\qquad\qquad\;
\frac{\ddot A_3}{A_3}+\frac{\ddot A_1}{A_1} + \frac{\dot A_3}{A_3}\frac{\dot A_1}{A_1}
 &=& 8 \pi G(-{\cal P}) + \Lambda + \frac{K^{(2)}\kappa}{A_{\rm dom}^2}
\label{Friedmann equation 22}\\
\hspace{0em} ^3_{\hspace{1ex}3}\hspace{1em}\text{equation:} \hspace{-0em}
\qquad\qquad\;
\frac{\ddot A_1}{A_1}+\frac{\ddot A_2}{A_2} + \frac{\dot A_1}{A_1}\frac{\dot A_2}{A_2}
 &=& 8 \pi G(-{\cal P}) + \Lambda + \frac{K^{(3)}\kappa}{A_{\rm dom}^2}
\label{Friedmann equation 33}
         \,,\quad
\end{eqnarray}
where ${\cal P}$ denotes the (isotropic) pressure and $K^{(0)}$, $K^{(1)}$, $K^{(2)}$ and $K^{(3)}$ are the curvature coupling parameters from Section \ref{sec:BackgroundEvolution}. Here $A_{\rm dom}$ is one of the three anisotropic scale factors, which contributes predominantly to the Einstein field equations; these can be obtained by writing out these field equations and reading off the appropriate factor (see Table~\ref{table:AnisotropicExpansionParameters} above). \\

It is instructive to introduce the Universe's volume scale,
$V=A_1A_2A_3$, which shows the growth of some referential volume, in terms of which one can define the average expansion rate $H(t)$ and the average scale factor $a(t)$ as,
\begin{equation}
H(t)\equiv \frac{\dot{a}}{a}= \frac{1}{3}\left(\frac{\dot A_1}{A_1}+\frac{\dot A_2}{A_2} +\frac{\dot A_3}{A_3}\right)
\,,\qquad
a(t) \equiv V^{1/3} = [A_1A_2A_3]^{1/3}
\,.
\label{expansion rate scale factor}
\end{equation}
By pairwise subtracting equations~(\ref{Friedmann equation 11}--\ref{Friedmann equation 33}) one easily obtains,
\begin{align}
{^{2}_{\hspace{1ex}2}}-{^{1}_{\hspace{1ex}1}} \hspace{1em} \rightarrow \hspace{1em} &\frac{{\rm d}}{{\rm d}t}\left[ \frac{\dot A_1}{A_2}-\frac{\dot A_2}{A_2}\right] + 3H\left[ \frac{\dot A_1}{A_2}-\frac{\dot A_2}{A_2}\right] = \kappa \left[ \frac{K^{(2)} - K^{(1)}}{A_{\rm dom}^2} \right]
\label{Friedmann equation 11b}\\
{^{3}_{\hspace{1ex}3}}-{^{2}_{\hspace{1ex}2}} \hspace{1em} \rightarrow \hspace{1em} &\frac{{\rm d}}{{\rm d}t}\left[ \frac{\dot A_2}{A_2}-\frac{\dot A_3}{A_3}\right] + 3H\left[ \frac{\dot A_2}{A_2}-\frac{\dot A_3}{A_3}\right] = \kappa \left[ \frac{K^{(3)} - K^{(2)}}{A_{\rm dom}^2} \right]
\label{Friedmann equation 22b}\\
{^{1}_{\hspace{1ex}1}}-{^{3}_{\hspace{1ex}3}} \hspace{1em} \rightarrow \hspace{1em} &\frac{{\rm d}}{{\rm d}t}\left[ \frac{\dot A_3}{A_3}-\frac{\dot A_1}{A_2}\right] + 3H\left[ \frac{\dot A_3}{A_3}-\frac{\dot A_1}{A_2}\right] = \kappa \left[ \frac{K^{(1)} - K^{(3)}}{A_{\rm dom}^2} \right]
\label{Friedmann equation 33b}
         \,.\quad
\end{align}
It is convenient to introduce differential expansion rates,
\begin{equation}
\Delta H_{12} = \frac{\dot A_1}{A_1}-\frac{\dot A_2}{A_2} \,,\qquad
\Delta H_{23} = \frac{\dot A_2}{A_2}-\frac{\dot A_3}{A_3} \,,\qquad
\Delta H_{31} = \frac{\dot A_3}{A_3}-\frac{\dot A_1}{A_1} \,,\qquad
\label{differential expansion rates}
\end{equation}
in terms of which equations~(\ref{Friedmann equation 11b}--\ref{Friedmann equation 33b})
become,
\begin{eqnarray}
\frac{1}{a^3}\frac{{\rm d}}{{\rm d}t}\left[a^3\Delta H_{12}(t)\right]
\!\!&=&\!\! \kappa \left[ \frac{K^{(2)} - K^{(1)}}{A_{\rm dom}^2} \right]
\label{Friedmann equation 11c}\\
\frac{1}{a^3}\frac{{\rm d}}{{\rm d}t}\left[a^3\Delta H_{23}(t)\right]
\!\!&=&\!\! \kappa \left[ \frac{K^{(3)} - K^{(2)}}{A_{\rm dom}^2} \right]
\label{Friedmann equation 22c}\\
\frac{1}{a^3}\frac{{\rm d}}{{\rm d}t}\left[a^3\Delta H_{31}(t)\right]
\!\!&=&\!\! \kappa \left[ \frac{K^{(1)} - K^{(3)}}{A_{\rm dom}^2} \right]
\label{Friedmann equation 33c}
         \,.\quad
\end{eqnarray}
It is now clear that the anisotropic expansion is generated by $\kappa$, {\it i.e.} when
$\kappa=0$ there exists isotropic Universe solutions. Therefore, when $\kappa$ is small,
one can study anisotropies in powers of $\kappa$. With this in mind,
equations~(\ref{Friedmann equation 11c}--\ref{Friedmann equation 33c}) can be simplified to,
\begin{eqnarray}
\frac{1}{a^3}\frac{{\rm d}}{{\rm d}t}\left[a^3\Delta H_{12}(t)\right]
\!\!&=&\!\! \frac{\kappa}{a^2}\left[K^{(2)}-K^{(1)}\right] +{\cal O}(\kappa^2)
\label{Friedmann equation 11d}\\
\frac{1}{a^3}\frac{{\rm d}}{{\rm d}t}\left[a^3\Delta H_{23}(t)\right]
\!\!&=&\!\! \frac{\kappa}{a^2}\left[K^{(3)}-K^{(2)}\right]+{\cal O}(\kappa^2)
\label{Friedmann equation 22d}\\
\frac{1}{a^3}\frac{{\rm d}}{{\rm d}t}\left[a^3\Delta H_{31}(t)\right]
\!\!&=&\!\! \frac{\kappa}{a^2}\left[K^{(1)}-K^{(3)}\right] +{\cal O}(\kappa^2)
\label{Friedmann equation 33d}
         \,.\quad
\end{eqnarray}
Note that $A_{\rm dom}$ drops out from these equations as, any deviation from $a$ is absorbed into the ${\cal O}(\kappa^2)$ terms.
This means that the results from hereon are generically applicable to every Thurston geometry.
Equations~(\ref{Friedmann equation 11d}--\ref{Friedmann equation 33d}) can be integrated upon introducing the following time variable, ${\rm d}\tau = a{\rm d}t$,
\begin{eqnarray}
\Delta H_{12}(t)
\!\!&=&\!\! \frac{\kappa\tau}{a^3}\left[K^{(2)}-K^{(1)}\right] +{\cal O}(\kappa^2)
\label{Friedmann equation 11e}\\
\Delta H_{23}(t)
\!\!&=&\!\! \frac{\kappa\tau}{a^3}\left[K^{(3)}-K^{(2)}\right]+{\cal O}(\kappa^2)
\label{Friedmann equation 22e}\\
\Delta H_{31}(t)
\!\!&=&\!\! \frac{\kappa\tau}{a^3}\left[K^{(1)}-K^{(3)}\right] +{\cal O}(\kappa^2)
\label{Friedmann equation 33e}
         \,,\quad
\end{eqnarray}
where
\begin{equation}
\tau(t) = \int^t a(t') {\rm d}t'
\,,
\label{tau: def}
\end{equation}
and we assumed that the initial anisotropies (incorporated in the integration constants) are
negligibly small. Notice that the individual expansion rates can be expressed in terms of
the average expansion rate $H(t)$ and differential expansion rates,
\begin{eqnarray}
 H_1 \equiv \frac{\dot A_1}{A_1} \!\!&=&\!\! H + \frac{\Delta H_{12}-\Delta H_{31}}{3}
\label{expansion rate A}\\
 H_2 \equiv \frac{\dot A_2}{A_2} \!\!&=&\!\! H + \frac{\Delta H_{23}-\Delta H_{12}}{3}
\label{expansion rate B}\\
 H_3 \equiv \frac{\dot A_3}{A_3} \!\!&=&\!\! H + \frac{\Delta H_{31}-\Delta H_{23}}{3}
\label{expansion rate C}
         \,.\quad
\end{eqnarray}
Inserting this into the Friedmann equation~(\ref{Friedmann equation 00}) gives,
\begin{eqnarray}
 \!\!&&\!\!
 3H^2 + \frac{1}{9} \begin{bmatrix}
      (\Delta H_{12}\!-\!\Delta H_{31})(\Delta H_{23}\!-\!\Delta H_{12}) \\
+ (\Delta H_{23}\!-\!\Delta H_{12})(\Delta H_{31}\!-\!\Delta H_{23}) \\
+ (\Delta H_{31}\!-\!\Delta H_{23})(\Delta H_{12}\!-\!\Delta H_{31})
\end{bmatrix}%
= 8\pi G \rho  + \Lambda + \frac{ \kappa K^{(0)}}{a^2}
\label{Friedmann equation 00b}
         \,.\quad
\end{eqnarray}
%
%
Inserting into this the solutions~(\ref{Friedmann equation 11e}--\ref{Friedmann equation 33e})
and moving these terms to the right hand side, results in,
\begin{eqnarray}
\label{Friedmann equation 00c}
H^2  \!\!&=&\!\!
\frac{8\pi G}{3} \rho  + \frac{\Lambda }{3}+ \frac{ \kappa K^{(0)}}{3a^2}
- \frac{\kappa^2\tau^2}{27a^6} %
\begin{bmatrix}
(K^{(1)}\!+\!K^{(2)}\!-\!2K^{(3)})(K^{(2)}\!+\!K^{(3)}\!-\!2K^{(1)}) \\
+ (K^{(2)}\!+\!K^{(3)}\!-\!2K^{(1)})(K^{(3)}\!+\!K^{(1)}\!-\!2K^{(2)}) \\
+ (K^{(3)}\!+\!K^{(1)}\!-\!2K^{(2)})(K^{(1)}\!+\!K^{(2)}\!-\!2K^{(3)})
\end{bmatrix} \\
\nonumber &=&\!\!
\frac{8\pi G}{3} \rho  + \frac{\Lambda }{3}+ \frac{ \kappa K^{(0)}}{3a^2}
- \frac{\kappa^2\tau^2}{27a^6} \times 3 \big[K^{(1)} K^{(2)} + K^{(2)} K^{(3)} + K^{(3)} K^{(1)} - (K^{(1)})^2 - (K^{(2)})^2 - (K^{(3)})^2 \big]
         \,,\qquad\;\;
\end{eqnarray}
where the last term is of the second order in $\kappa$, and it represents
the backreaction of the anisotropic expansion on the average (isotropic) expansion rate.
The next step is to determine how the individual scale factors deviate from the isotropic expansion.
Upon inserting equations~(\ref{Friedmann equation 11e}--\ref{Friedmann equation 33e}) and
into (\ref{expansion rate A}--\ref{expansion rate C}) one obtains,
\begin{eqnarray}
\frac{{\rm d}}{{\rm d}t}\ln\left(\frac{A_1(t)}{a(t)}\right)
                 \!\!&=&\!\!\frac{\kappa \tau(t)  }{3a^3(t)}  \left(K^{(2)}+K^{(3)}-2K^{(1)}\right)
\label{expansion rate Ac}\\
\frac{{\rm d}}{{\rm d}t}\ln\left(\frac{A_2(t)}{a(t)}\right)
                 \!\!&=&\!\!\frac{\kappa \tau(t)  }{3a^3(t)}  \left(K^{(3)}+K^{(1)}-2K^{(2)}\right)
\label{expansion rate Bc}\\
\frac{{\rm d}}{{\rm d}t}\ln\left(\frac{A_3(t)}{a(t)}\right)
                 \!\!&=&\!\!\frac{\kappa \tau(t)  }{3a^3(t)} \left(K^{(1)}+K^{(2)}-2K^{(3)}\right)
\label{expansion rate Cc}
         \,.\quad
\end{eqnarray}
By introducing a compact notation, $\Delta K_{(1)} = (K^{(2)}+K^{(3)}-2K^{(1)})/K^{(0)}$
(plus cyclical permutation for $K_{(2)}$ and $K_{(3)}$),
these equations can be written as one equation for $A_i(t)= A_1(t)$, $A_2(t)$, or $A_3(t)$ as,
\begin{equation}
\ln\left(\frac{A_i(t)}{a(t)}\right)
                 =\int^t \frac{\kappa \tau(t') \Delta K_{(i)} K^{(0)}}{3a(t')^3}\,{\rm d}t'
\,.
\label{expansion rate alpha/a}
\end{equation}
%

%
%

\subsection{Growth of anisotropies in an epoch with matter and
cosmological constant}
\label{sec:AnisotropicScalingCosmoMatter}

It is well-known that relatively recently (at a redshift $z=z_{\rm DE}\simeq 0.7$)
the Universe entered a dark energy dominated
epoch, during which it exhibits an accelerated expansion. It is
therefore essential to include dark energy when studying the dynamics of anisotropies.

\begin{figure}[H]
\begin{minipage}[l]{0.48 \linewidth}
\includegraphics[width=1 \textwidth]{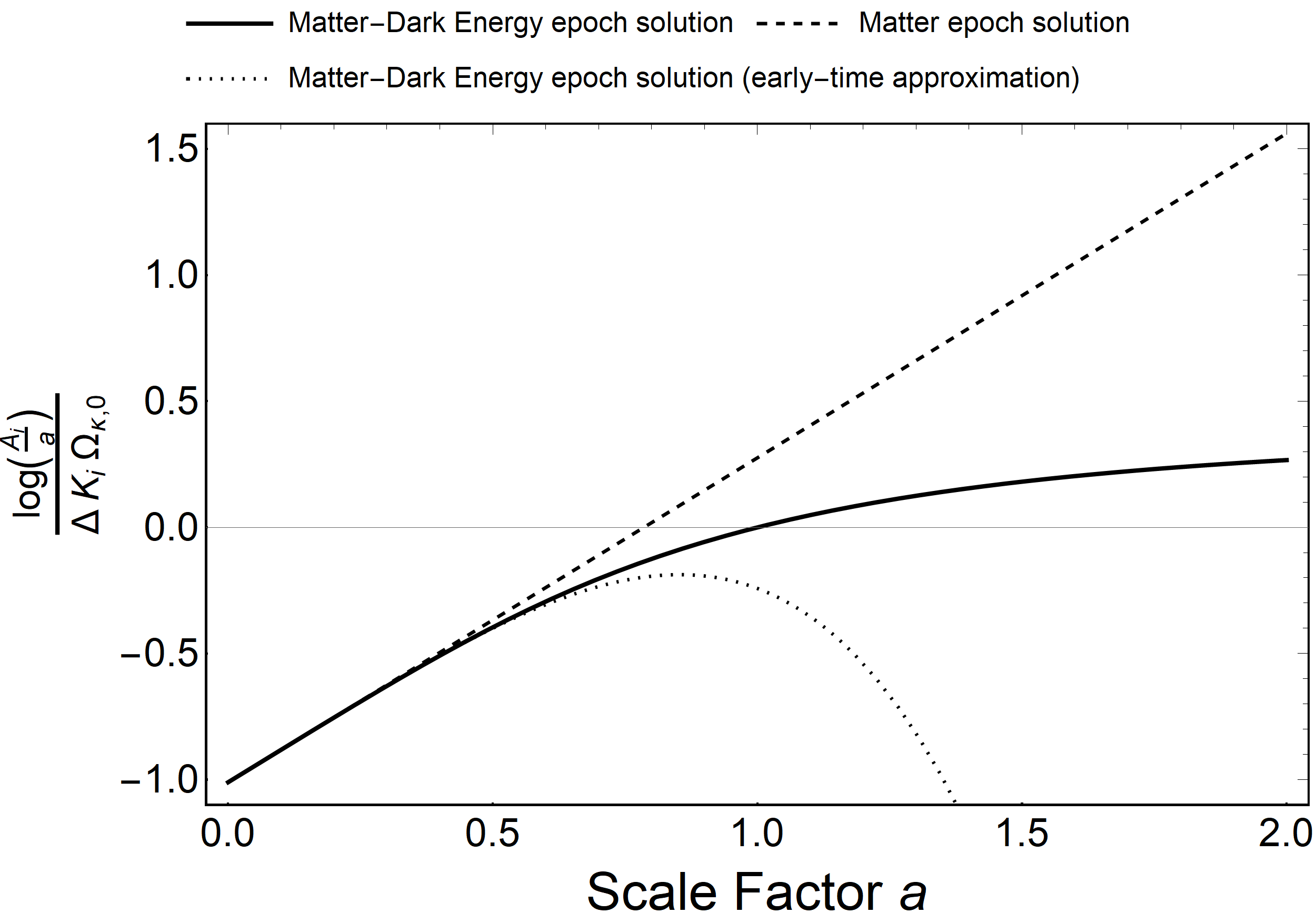}
\caption{The evolution of anisotropy in equation~\eqref{eq:LogRatioAnisotropiesFourTerms}. The integration constant is chosen such that the ratio vanishes today (when $a=1$). Also shown are the matter epoch solution (dashed) and the small-$a$, early-time expansion from equation~\eqref{expansion anisotropies: matter era} (dotted). We see that the evolution of $\ln(A_i/a)$ is linear in $a$ in the matter-dominated epoch and slows down in the dark energy-dominated epoch.}
\label{figure 24}
\end{minipage}
\hskip 0.5cm
\begin{minipage}[l]{0.48 \linewidth}
\includegraphics[width=1 \textwidth]{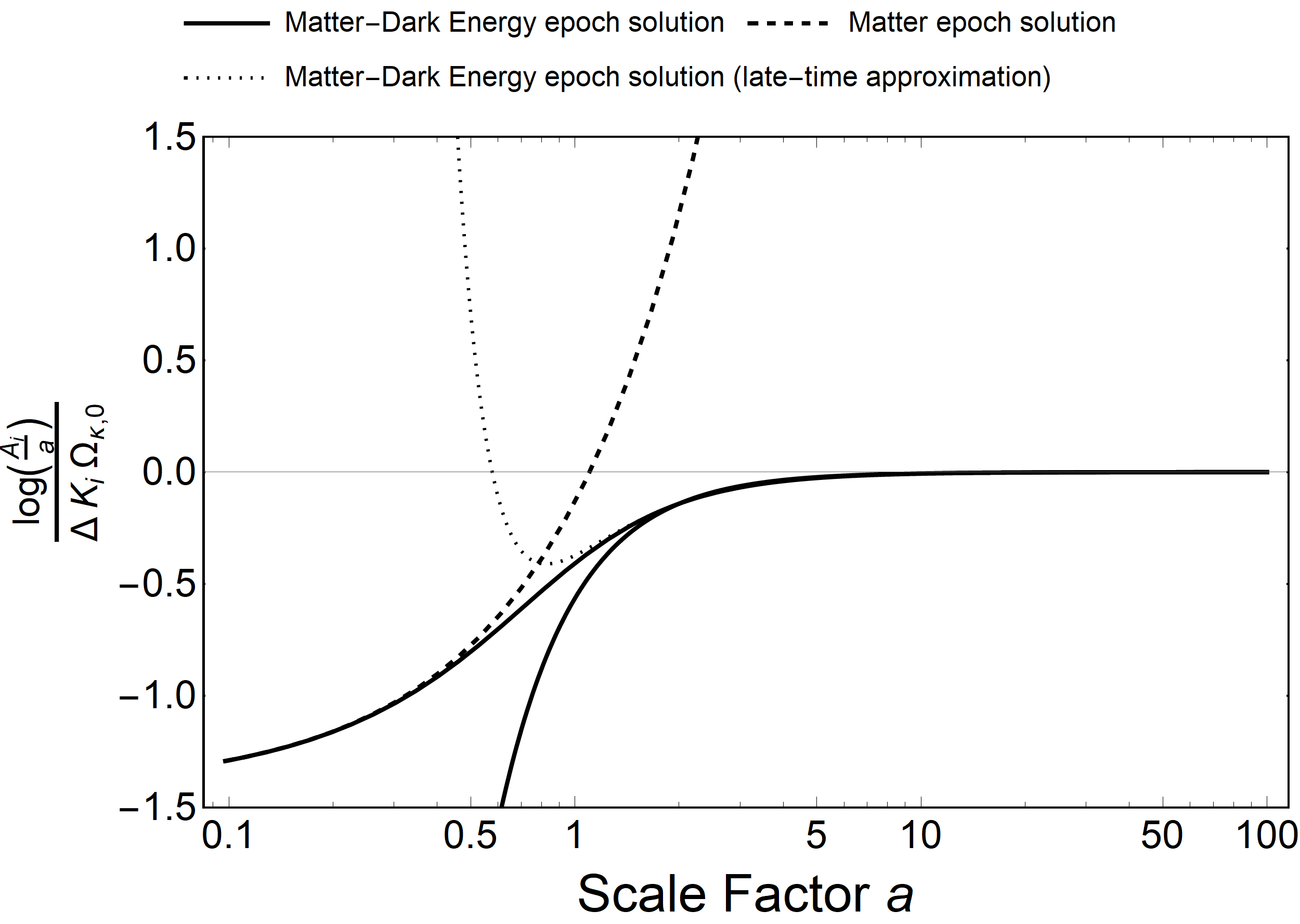}
\caption{The same graph as on the left, but with a log scale on the horizontal axis to show late-time behaviour. The integration constant is chosen such that the ratio vanishes as $a\rightarrow\infty$. Also shown are the matter epoch solution (dashed) and the large-$a$, late-time expansion from equation~\eqref{asymptotic expansion anisotropies} (dotted). Whereas in the matter-dominated era anisotropies diverge over time, in the dark energy epoch anisotropies asymptote to a constant value.}
\label{figure 25}
\end{minipage}
\end{figure}

In this section we study the growth of anisotropies in an era dominated by nonrelativistic
matter ($\rho_{\rm m}(t) = \rho_{\rm m,0}/a^3$)
 and cosmological constant $\Lambda$, which we take to represent dark energy.
This is an excellent approximation
for the observed Universe from the decoupling at $z_*\simeq 1091$ up to today ($z=0$).
Neglecting for now the curvature contribution $\propto \kappa$ (which we know is small),
the Friedmann equation~(\ref{eq:FriedmannI})
({\it cf.} also the anisotropic Friedmann equation~(\ref{Friedmann equation 00c}))
 simplifies to,
\begin{equation}
H^2 = \frac{\dot a^2}{a^2}
 = \frac{8\pi G}{3}\frac{\rho_{\rm m, 0}}{a^3}
  + \frac{\Lambda}{3}
\,,
\label{Appx E: isotropic Friedmann}
\end{equation}
whose solution is,
\begin{equation}
a(t) =  a_\text{eq} \sinh^\frac23\left(\frac{\sqrt{3\Lambda}}{2}t\right)
\,,\qquad
 a_\text{eq} = \left(\frac{8\pi G \rho_{\rm m,0}}{\Lambda}\right)^\frac13
           = \left( \frac{\Omega_\text{m,0}}{\Omega_{\Lambda,0}}\right)^{1/3}
\,,
\label{Appx E: isotropic Friedmann solution}
\end{equation}
where $\Omega_{\rm m,0}=\rho_{\rm m,0}/H_0^2$,
$\Omega_{\Lambda,0}=\Lambda/(3H_0^2)$,
 $t$ is cosmological time, with $t=0$ corresponding to
the Big Bang singularity.
The first step towards undestanding the dynamics of anisotropies dictated by
Eq.~(\ref{expansion rate alpha/a}) is to evaluate the time variable $\tau(t')$
defined in Eq.~(\ref{tau: def}),
which for the problem at hand reduces to the following integral,
%
\begin{eqnarray}
\tau(t)= \int^t a(\tilde t\,) {\rm d}\tilde t
     = \sqrt{\frac{3}{\Lambda}}\frac{1}{ (a_\text{eq})^\frac32 }
     \int^a
     \frac{\tilde a^\frac32}{\sqrt{1+\left(\frac{\tilde a}{ a_\text{eq}}\right)^3}}
         {\rm d}\tilde a
     = \frac{2}{\sqrt{3\Lambda}}  a_\text{eq}
     \int^{(a/ a_\text{eq})^{3/2}}
     \frac{\tilde x^\frac23}{\sqrt{1+\tilde x^2}}
         {\rm d}\tilde x
\label{Appx E: integral a dt}
         \,,\quad
\end{eqnarray}
where $\tilde x = (\tilde a/ a_\text{eq})^\frac32$ and we made use of,
\begin{equation}
\frac{{\rm d}a}{{\rm d}t} = \sqrt{\frac{\Lambda}{3}}  a_\text{eq}
    \frac{\cosh\left(\frac{\sqrt{3\Lambda}}{2}t\right)}
    {\sinh^\frac13\left(\frac{\sqrt{3\Lambda}}{2}t\right)}
    = \sqrt{\frac{\Lambda}{3}} \frac{ (a_\text{eq})^\frac32 }{a(t)^\frac12}
         \sqrt{1+\left(\frac{a}{ a_\text{eq}}\right)^3}
\,.
\label{Appx E: da dt}
\end{equation}
Next, by expanding
$(1+\tilde x^2)^{-\frac12}
= \sum_{n=0}^\infty \left(\frac{1}{2}\right)_n(-\tilde x^2)^n/n!$, where
$(z)_n=\Gamma(z+n)/\Gamma(z)$ denotes the Pochhammer symbol,
and exchanging the sum and the integral,
the integral in equation~(\ref{Appx E: integral a dt}) can be evaluated,
%
\begin{eqnarray}
\tau(t)&=& \int^t a(\tilde t\,) {\rm d}\tilde t
      = \frac{2 a_\text{eq}}{\sqrt{3\Lambda}}
     \sum_{n=0}^\infty \frac{\left(\frac{1}{2}\right)_n(-1)^n}{n!}
     \int^{x=(a/ a_\text{eq})^{3/2}}  \tilde x^{\frac23+2n}
         {\rm d}\tilde x
         = \frac{ a_\text{eq}}{\sqrt{3\Lambda}}
     \sum_{n=0}^\infty \frac{\left(\frac{1}{2}\right)_n(-1)^n}{n!}
  \frac{x^{\frac53+2n}}{\frac56+n}
\nonumber\\
    &=& \frac{ a_\text{eq}}{\sqrt{3\Lambda}} x^\frac53
    \frac{\Gamma\left(\frac{5}{6}\right)}
          {\Gamma\left(\frac{11}{6}\right)}
     \sum_{n=0}^\infty
     \frac{\left(\frac{1}{2}\right)_n\left(\frac{5}{6}\right)_n}
       {\left(\frac{11}{6}\right)_n}
  \frac{(-x^2)^n}{n!}
    = \frac{ a_\text{eq}}{\sqrt{3\Lambda}}
    \frac{6}{5} \left(\frac{a}{ a_\text{eq}}\right)^\frac52
          \times_{2}\!F_1\left(\frac{1}{2},\frac{5}{6};
          \frac{11}{6};- \left(\frac{a}{ a_\text{eq}}\right)^3\right)
         \,,\quad
\label{Appx E: integral a dt 2}
\end{eqnarray}
where $_{2}F_1(\alpha,\beta;\gamma;z)$ denotes
Gauss' hypergeometric function.
Upon inserting this result into Eqs.~(\ref{expansion rate Ac}--\ref{expansion rate Cc})
one obtains,
\begin{eqnarray}
\ln\left[\frac{A_1(t)}{a(t)}\right]  \!\!&=&\!\!
    \big(K^{(2)}\!+\!K^{(3)}\!-\!2K^{(1)}\big)
    \,\frac{2\kappa}{5\Lambda a_\text{eq}^3}
         \int^{a(t)}\!\!\!  \frac{{\rm d}a'}
         {\sqrt{1+\left(\frac{a'}{ a_\text{eq}}\right)^3}}
          \times_{2}\!F_1\left(\frac{1}{2},\frac{5}{6};
\frac{11}{6};-\left(\frac{a'}{ a_\text{eq}}\right)^3\right)
\,
\qquad\;\,
\label{Appx E: anisotropies A1} \\
\ln\left[\frac{A_2(t)}{a(t)}\right]  \!\!&=&\!\!
    \big(K^{(3)}\!+\!K^{(1)}\!-\!2K^{(2)}\big)
    \,\frac{2\kappa}{5\Lambda a_\text{eq}^3}
         \int^{a(t)}\!\!\!  \frac{{\rm d}a'}
         {\sqrt{1+\left(\frac{a'}{ a_\text{eq}}\right)^3}}
          \times_{2}\!F_1\left(\frac{1}{2},\frac{5}{6};
\frac{11}{6};-\left(\frac{a'}{ a_\text{eq}}\right)^3\right)
\,
\qquad\;\,
\label{Appx E: anisotropies A2} \\
\ln\left[\frac{A_3(t)}{a(t)}\right]  \!\!&=&\!\!
    \big(K^{(1)}\!+\!K^{(2)}\!-\!2K^{(3)}\big)
    \,\frac{2\kappa}{5\Lambda a_\text{eq}^3}
         \int^{a(t)}\!\!\!  \frac{{\rm d}a'}
         {\sqrt{1+\left(\frac{a'}{ a_\text{eq}}\right)^3}}
          \times_{2}\!F_1\left(\frac{1}{2},\frac{5}{6};
\frac{11}{6};-\left(\frac{a'}{ a_\text{eq}}\right)^3\right)
\,.
\qquad\;\,
\label{Appx E: anisotropies A3}
\end{eqnarray}
By making a few substitutions,
and using the compact notation from Eq.~(\ref{expansion rate alpha/a}),
Eqs.~(\ref{Appx E: anisotropies A1}-\ref{Appx E: anisotropies A3}) can be rewritten as
\begin{align}
\ln\left[\frac{A_i(t)}{a(t)}\right]
    &= \Delta K_{(i)} \Omega_{\kappa,0}
    \times
    \,\frac{2}{5\Omega_{\rm m,0}}
    \int^a da' \;\frac{{}_{2}F_1\left(\frac{1}{2},\frac{5}{6};
\frac{11}{6};-a'^3\frac{\Omega_{\Lambda,0}}{\Omega_{\rm m,0}}\right)}{\sqrt{1+a'^3\frac{\Omega_{\Lambda,0}}{\Omega_{\rm m,0}} }}
\,.
\label{eq:LogRatioAnisotropiesFourTerms}
\end{align}
Using the values of 
$\Omega_{\Lambda,0}$ and $\Omega_{\rm m,0}$ from Table~\ref{table:Planck2018data} we can put constraints on (the growth of) this ratio by constraining each term in kind.
In figures~\ref{figure 24} and~\ref{figure 25}
we plot the evolution of  anisotropies $\ln(A_i/a)$
(in units of $\Delta K_{(i)} \Omega_{\kappa,0}$) from early in the matter-dominated era up to
roughly today (left panel) and up to a distant future (right panel).
We see that $\ln(A_i)$ grows linearly with the scale factor in the matter-dominated, but the growth
slows down in the dark energy-dominated epoch, asymptoting to a constant in a distant future.
\vskip -0.2cm
\vskip -0.2cm

In order to understand whether the growth of anisotropies~(\ref{Appx E: anisotropies A1}--\ref{Appx E: anisotropies A3})
can be neglected in the analysis presented in earlier sections, note first
that in predominantly matter era ($z\gg z_{\rm DE}\simeq 0.7$)
the argument of the hypergeometric function in Eq.~(\ref{eq:LogRatioAnisotropiesFourTerms})  is small, justifying its Taylor expansion ({\it cf.} Eq.~(\ref{Appx E: integral a dt 2})).
Keeping the two leading terms, the integral in~(\ref{eq:LogRatioAnisotropiesFourTerms}) evaluates to,
\begin{equation}
\ln\left[\frac{A_i(t)}{a(t)}\right]
= \Delta K_{(i)} \Omega_{\kappa,0} \times
    \,\frac{2}{5\Omega_{\rm m,0}} a
    \left(1-\frac{2}{11}\frac{\Omega_{\Lambda,0}}{\Omega_{\rm m,0}} a^3\right)
     +  {\cal O}\big(a^{7}\big)
\,.
\label{expansion anisotropies: matter era}
\end{equation}
Here we have used gauge freedom to fix anisotropy in the scale factors at zero,
{\it ergo} $A_1 = A_2 = A_3$, at initial time by a suitable (global) rescaling of the spatial coordinates. This is analogous to the gauge freedom to set $a(t_0) = 1$ at present time in isotropic FLRW space-time by a global rescaling of the radial coordinate. This expansion
is valid for times earlier than today, {\it i.e.} when $a=1/(1+z)<1$ ($z>0$). \\

\noindent To get an expansion valid at late times ($a\gg 1$), one can use Eq.~(9.132.2) in
Ref.~\cite{Gradshteyn:2014}~\footnote{According to Eq.~(9.132.2) in
Ref.~\cite{Gradshteyn:2014},
\begin{equation}
{}_2F_1\left(\frac12,\frac56;\frac{11}{6};z\right) = \frac{5}{2}\big(\!-z\big)^{-\frac12}
\times_2\!F_1\left(\frac12,-\frac13;\frac{2}{3};\frac1z\right)
+\frac{\Gamma\big(\frac{11}{6}\big)\Gamma\big(-\frac{1}{3}\big)}{\sqrt{\pi}}
\big(\!-z\big)^{-\frac56}
\,,\qquad \left(z= -\frac{\Omega_{\Lambda,0}}{\Omega_{\rm m,0}} {a'}^3\right)
\,.
\end{equation}
} to obtain,
\begin{eqnarray}
\ln\left[\frac{A_i(t)}{a(t)}\right]
\!\! &=&\!\!
\ln\left[\frac{A_i(t)}{a(t)}\right]_{\infty}
-\Delta K_{(i)} \Omega_{\kappa,0}
    \times
    \,\frac{1}{2\Omega_{\Lambda,0}}\frac{1}{a^2}
\left[1-\frac{2}{3}\frac{\Gamma\big(\frac56\big)\Gamma\big(\frac23\big)}{\sqrt{\pi}}
\left(\frac{\Omega_{\rm m,0}}{\Omega_{\Lambda,0}}\right)^\frac13\frac{1}{a}
  -\frac{1}{10}\frac{\Omega_{\rm m,0}}{\Omega_{\Lambda,0}}\frac{1}{a^3}
\right] \!+\! {\cal O}\big(a^{-6}\big)
\,,\qquad\qquad
 \label{asymptotic expansion anisotropies}
\end{eqnarray}
where we made use of,
\begin{equation}
\frac{{}_2F_1\left(\frac12,-\frac13;\frac{2}{3};-a'^3\frac{\Omega_{\Lambda,0}}{\Omega_{\rm m,0}} \right)}
{\sqrt{1+a'^3\frac{\Omega_{\Lambda,0}}{\Omega_{\rm m,0}} }}
\simeq\frac{5}{2}\frac{\Omega_{\rm m,0}}{\Omega_{\Lambda,0}}  \frac{
1-\frac14\frac{1}{a'^3}\frac{\Omega_{\rm m,0}}{\Omega_{\Lambda,0}}
-\frac{\Gamma\big(\frac{5}{6}\big)\Gamma\big(\frac{2}{3}\big)}{\sqrt{\pi}}
\Big(\!a'^3\frac{\Omega_{\Lambda,0}}{\Omega_{\rm m,0}}\Big)^{-\frac13}}
{a'^3}
+{\cal O}\big(a^{-7}\big)
\,.
\end{equation}
Here the constant $ \ln\left[\frac{A_i(t)}{a(t)}\right]_{\infty}$ can be evaluated numerically
(or read off from the graph in Figure \ref{figure 25}). This constant has no relevance for the dynamics of anisotropies,
as it can also be subtracted by a suitable global rescaling of the spatial coordinates. This implies that, in a cosmological constant-dominated
era (such as inflation), anisotropies in the scale factors
$\ln\left[\frac{A_i(t)}{a(t)}\right]$ decay rapidly (exponentially fast) in time as
$\propto 1/a^2\propto {\rm e}^{-2H_{\Lambda}t}$, where
$ H_{\Lambda}^2=\Omega_{\Lambda,0}H_0^2=\Lambda/3$.

Next we comment on the implications of the above findings.
From observations we know that spatial
anisotropies today are small, or
more quantitatively, $|\Omega_{\kappa,0}|\ll 1$ and $\Delta K_{(i)}$
are at most of the order of unity (\textit{c.f.} Table \ref{table:thurstonparameters}).
Combining this with the observation that, from the beginning of matter era,
$\ln\left[\frac{A_i(t)}{a(t)}\right] /[\Delta K_{(i)} \Omega_{\kappa,0}]\simeq 1$
(which can be read off from Figure~\ref{figure 24}),
one can infer from Eq.~(\ref{expansion anisotropies: matter era})
 that the change in anisotropies $\Delta (A_i)$
from the beginning of the matter-dominated era up to today ($a=1$) can be estimated as,
\begin{equation}
\Delta \ln\left[\frac{A_i}{a}\right]
\simeq \Delta \Big(\frac{A_i}{a}\Big)
= \frac{A_i}{a} -1 \simeq \Delta K_{(i)} \Omega_{\kappa,0}  \ll 1
\,,
\label{change anisotropies: matter era}
\end{equation}
thus justifying our treatment in Sections~\ref{sec:ThurstonSpacetimes},
\ref{sec:BackgroundEvolution}, \ref{sec:DistanceMeasures}
and~\ref{sec:DistanceMeasurePlots},
where we neglected the dynamics of anisotropies.~\footnote{
\label{footnote:RelativeError}
Even though it is true that spatial anisotropies are small today and
were even smaller in the past, they do grow rapidly in matter era, $\Delta\big(A_i(t)/a(t)\big)\sim \Delta K_{(i)} \Omega_{\kappa,0} a(t)\propto 1/(1+z)$, which seems to put into question our estimates of the luminosity distance and angular diameter
distance at large redshifts. One can argue that is not a legitimate concern as follows.
The question boils down to whether in the distance measures $d_L$ and $d_A$
one can replace a directional scale factor $A_i(t)$ by the average scale factor $a(t)$.
The {\it relative error} one makes in this way
 (when estimating the effects of spatial anisotropies on  $d_L$ and $d_A$)
is at most of the order $\Delta\big(A_i(t)/a(t)\big)\sim \Delta K_{(i)} \Omega_{\kappa,0}$,
which is $\propto\kappa$, impacting the distance measures
at a higher (second) order in $\kappa$, such that this effect can be consistently neglected.
}

\subsection{The Anisotropy Problem}

With these results in mind, we can make the following remarks regarding whether
there is an {\it anisotropy problem} in the Universe, which can be formulated as follows:
\begin{itemize}
\item[]
{\it Given that anisotropies grow in the radiation and matter era,
and that the observed anisotropies are small today, one must fine-tune the initial geometry
of the Universe such to be isotropic to a very high precision.}
\end{itemize}
In order to better understand this problem, let us briefly recall the flatness problem.
In an expanding universe with $\epsilon = -\dot H/H^2 ={\rm constant}$,
$\rho_{\rm m}\propto a^{-2\epsilon}$
($\epsilon\simeq 0$ in inflation, $\epsilon\simeq 2$ in radiation
and $\epsilon=3/2$ in matter), and therefore,
\begin{equation}
\Omega_\kappa = \frac{\rho_\kappa}{\rho_{\rm m}}
\sim a^{2(\epsilon-1)}\sim {\rm e}^{2(\epsilon-1)N}
\,,\qquad \big(N=\ln(a)  \big)
\,,
\label{Omega kappa}
\end{equation}
%
meaning that $\Omega_\kappa$ decays in inflation as
$\Omega_\kappa\sim {\rm e}^{-2N_I}$,
and it grows in radiation and matter eras as
$\Omega_\kappa\sim {\rm e}^{2N_R}$ and $\Omega_\kappa\sim {\rm e}^{N_M}$,
respectively, where $N_I, N_R$ and $N_M$ denotes the number of e-foldings in the respective epoch. From this one easily infers that the flatness problem is solved if inflation
lasts at least,
\begin{equation}
{\tt Flatness\;\; problem:}\qquad N_I> (N_I)_{\rm min} = N_R + \frac12 N_M
\,.
\label{flatness problem}
\end{equation}

 Let us now consider the growth of anisotropies in an $\epsilon={\rm constant}$ epoch.
From Eq.~(\ref{Friedmann equation 00c}) we see that
anisotropies scale as, $\tau^2/a^6\sim a^{2\epsilon-4}$, which when compared
with $\rho\sim a^{-2\epsilon}$ yields,
$\Omega_{\rm anis}\sim a^{4(\epsilon-1)}\sim {\rm e}^{4(\epsilon-1)N}$
and $\Omega_{\rm anis}/\Omega_\kappa\sim a^{2(\epsilon-1)}\sim {\rm e}^{2(\epsilon-1)N}$, implying that both $\Omega_{\rm anis}=\rho_{\rm anis}/\rho$ and
$\Omega_{\rm anis}/\Omega_\kappa$ decay (grow) in accelerating (decelerating)
space-times. From this we infer in inflation,
$\Omega_{\rm anis}\sim {\rm e}^{-4N_I}$, and in radiation and matter era,
$\Omega_{\rm anis}\sim {\rm e}^{4N_R}$ and $\Omega_{\rm anis}\sim {\rm e}^{2N_M}$,
respectively. From these observations we see that the anisotropy problem is solved when
$-4N_I + 4N_R + 2N_M<0$, or equivalently,
\begin{equation}
{\tt Anisotropy\;\; problem:}\qquad N_I >(N_I)_{\rm min} = N_R + \frac12 N_M
\,,
\label{anisotropy problem}
\end{equation}
which is the same as the condition
in Eq.~(\ref{flatness problem}).
Curiously (but not surprisingly), the same condition is obtained
if one requires $\Omega_{\rm anis}/\Omega_\kappa<1$.
Notice that the result~(\ref{anisotropy problem}) could have been guessed from
Eqs.~(\ref{Friedmann equation 11e}--\ref{expansion rate C}),
according to which the relative differences in the Hubble rates scale as,
 $\Delta H/H\propto a^{2(\epsilon-1)}\sim {\rm e}^{2(\epsilon-1)N}$.

For our two exceptional cases, Nil and $\widetilde{\text{U}(\mathds{H}^2)}$ geometries, we note that the off-diagonal contributions to the energy momentum tensor in Eqs.~(\ref{shear tensor: Nil}--\ref{shear tensor: UH2}) scale as $\propto 1/a^2$. Even though we do not know the precise dynamical geometry of these space-times, it is reasonable to assume that the growth of these-off diagonal contributions will follow the same time patterns as the other geometries given this scaling property. Then, by the same logic as footnote~\ref{footnote:RelativeError} we, can trust that the relative error is small as long as equation \eqref{anisotropy problem} holds. \\

\noindent The principal result of this section is that inflation solves the anisotropy problem of the Universe precisely in the same way as it solves the flatness problem of standard Friedmann cosmologies. This implies that,
from a first-principle point of view, {\it any of the eight Thurston geometries is equally well-motivated to be the geometry of the Universe}. Which one of these is the preferred description of the Universe is a matter to be settled observationally. This finding is one of the main results of this work.


\section{Conclusion and Discussion}

In this paper we advance the hypothesis that the large scale geometry of spatial sections
of the Universe (see equations~(\ref{breakdown: R x Sigma}--\ref{breakdown: metric}))
can be described as a patchwork of one or more of the eight Thurston-Perelman
geometries that are sewn together smoothly. The first three of these eight geometries, as described in
section~\ref{sec:ThurstonSpacetimes}, are the well-known homogeneous and isotropic
Friedmann-Lema\^itre-Robertson-Walker (FLRW) geometries. Since FLRW space-times are widely
studied and well-understood, we have focused
our attention on the remaining
five geometries, which are all homogeneous, but violate spatial isotropy.
To our knowledge,
these geometries were not given
a serious consideration in the literature. \\

In section~\ref{sec:BackgroundEvolution} we have shown that these geometries give rise to a set of Friedmann equations~(\ref{eq:FriedmannI}--\ref{eq:EnergyEvolution}), compatible with usual cosmological inventory: matter, radiation and dark energy.
 However, satisfying these equations requires the introduction of a field with peculiar scaling properties to support the anisotropy of the underlying geometry, which we have assumed in sections~\ref{sec:BackgroundEvolution},
\ref{sec:DistanceMeasures} and~\ref{sec:DistanceMeasurePlots} to make the analysis simpler, and thus more pedagogical.

An important question is how one could observationally test
the large scale geometry of the Universe and, in particular,
how one could distinguish whether we live in
a spatially isotropic universe characterised
by one of the three FLRW geometries, or in a universe based on one of the five anisotropic Thurston geometries.
To answer this question, in Appendices~A, B, C and~D we have worked out how light propagates in all of the anisotropic Thurston geometries,
which are then used in section~\ref{sec:DistanceMeasures}
to calculate angular diameter distance and luminosity
distance for all of the geometries. Because of spatial anisotropy
of these geometries, we have derived general relations
which show how
angular diameter distance and luminosity distance depend
the underlying geometry. To improve clarity
we provide visual representations
for all Thurston geometries for both angular diameter distance
and luminosity distance  in Section~\ref{sec:DistanceMeasurePlots}. These figures show how distant objects would be deformed and dimmed or brightened by the presence of large-scale anisotropies.

In particular, one can infer from Figures~\ref{Lumdist-5-UH2-Individual-Scale}--\ref{Lumdist-7-Solv-Individual-Scale} that, when expanded
in powers of $\lambda/L$, where $\lambda$ is the geodesic distance and $L$ the curvature scale,
the geometries $\widetilde{U({\mathds H}_2)}$, Nil and Solv show a specific
angular dependence described by spherical harmonics $Y_{\ell}^m$ ($m\in[-\ell,\ell]$),
 which can be a useful feature when confronting these geometries against data.
Finally, in subsection~\ref{Sec:ParityViolationAndChirality} and Figures~\ref{LumDist-UH2-Parity-Even}--\ref{LumDist-Solv-Parity-Odd} we show that three of the Thurston geometries, namely $\widetilde{U({\mathds H}_2)}$
 Nil and Solv, violate parity, and that, in addition,  $\widetilde{U({\mathds H}_2)}$ and Nil violate chiral symmetry. These properties can also be used for testing these geometries against the data. Namely,
the data already show tantalizing hints for parity violation in
the Planck data~\cite{Minami:2020odp,Diego-Palazuelos:2022dsq,Diego-Palazuelos:2022cnh,Eskilt:2022cff}
and in the LSS four-point functions~\cite{Philcox:2022hkh,Creque-Sarbinowski:2023wmb,Coulton:2023oug}.

The validity and limitations of the assumption of spatial isotropy are discussed in detail in section~\ref{sec:AnisotropicScaleFactors}. In particular, we have shown that in the matter era most of the Thurston geometries~\footnote{The exceptions
are $\widetilde{U({\mathds H}_2)}$ and Nil. Even though we
do not know the precise dynamics of these geometries,
we give a simple argument in favour
of long time stability of these geometries.} can be supported by
a standard, spatially isotropic cosmological (perfect) fluid,
provided one introduces anisotropic scale factors, which allow
for different expansion rates in different spatial directions.
Even though anisotropies in the scale factors grow rapidly in the matter era
(see Eqs.~(\ref{eq:LogRatioAnisotropiesFourTerms}--\ref{asymptotic expansion anisotropies})
and Figures~\ref{figure 24} and~\ref{figure 25}),
one can show that the Universe's geometry can remain stable over large periods of time
(many e-foldings),
and moreover that the corrections in the distance measures induced by
the dynamics of anisotropies are of a higher order in curvature $\kappa$, {\it cf.} Eq.~(\ref{Friedmann equation 00c}), and therefore can be consistently neglected. Furthermore,
even though
Thurston geometries pose an anisotropy problem in standard Big Bang cosmologies,
it can be solved by a sufficiently long period of cosmic inflation
(see Eq.~(\ref{anisotropy problem})) in a way analogous to the flatness problem of
standard FLRW space-times.
\\

The next natural step is to make use of the results of this paper to investigate whether the current data contain evidence to single out any one of the Thurston geometries as a preferred candidate, and make forecasts for the upcoming observations. Moreover, the methods developed in the recent study~\cite{Vedder:2022spt} can aid such investigations.

Further theoretical work could also be done to study the effects of anisotropy on the polarization of light. We have seen in subsection~\ref{Sec:ParityViolationAndChirality} that both parity and chiral symmetry can be violated in the Thurston geometries. One expects that photons with different polarizations will be affected differently in such geometries. Finding such a difference may open up additional avenues for probing curvature anisotropies. \\

Lastly, it is worth remarking that Thurston's classification schema is related to the well-known Bianchi spaces, as shown among others in \cite{Fagundes:1991uy,Fischer:2006}. However a complete discussion of this correspondence is beyond the scope of this paper.

\section*{Acknowledgements}

This work is part of the Delta ITP consortium, a program of the Netherlands Organisation
for Scientific Research (NWO) that is funded by the Dutch Ministry of Education, Culture
and Science (OCW).
 \\


\newpage

\section*{Appendices}

\appendix
\label{Appendix}

\section{Geodesics of the $\mathds{R}\times\mathds{H}^2$ and $\mathds{R}\times\mathbf{S}^2$ geometries}

The fourth and fifth Thurston geometries are the first anisotropic spaces of Thurston's classification scheme, hence we will begin our treatment of geodesics here. We start from spatial sections of metric \eqref{eq:FLRW2Dmetric0} presented in Section \ref{sec:ThurstonSpacetimes},
\begin{equation}
\label{eq:FLRW2Dmetric}
{\rm d}\Sigma_3^2 =  {\rm d}z^2 + {\rm d}\chi^2 + S_\kappa^2(\chi) {\rm d}\phi^2.
\end{equation}
Since this metric is anisotropic, we cannot rely on three-dimensional rotational symmetry to place geodesics along convenient axes. We must therefore solve the Killing equation explicitly, which becomes considerably more simple by using the isometries of the metric in \eqref{eq:FLRW2Dmetric}. If $K^i$ is a Killing vector we can construct a conserved charge, 
\begin{align}
\label{eq:KillingCharge} Q_K = K_i(x) \frac{{\rm d} x^i}{{\rm d}\lambda}
\,,
\end{align}
that is constant along geodesic trajectories. With enough Killing vectors, we can solve a set of first-order equations instead. The metric \eqref{eq:FLRW2Dmetric} has two Killing vectors, $\partial_z$ and $\partial_\phi$, which leads to two conserved quantities,
\begin{align}
P_z    &= \dot{z}\,,                       \\
L_\phi &= S_\kappa^2(\chi) \dot{\phi}
\,.
\end{align}
We can use these conserved charges together with the requirement that the geodesics are space-like,
\begin{align}
\label{eq:Epsilon}
 \epsilon = - g_{ij} \frac{{\rm d}x^i}{{\rm d}\lambda}\frac{{\rm d}x^j}{{\rm d}\lambda} = - 1,
\end{align}
to write a general (implicit) expression for geodesics of the metric \eqref{eq:FLRW2Dmetric} as,
\begin{align}
                             z(\lambda) &= P_z \lambda + z_0 \\
                    \dot{\chi}(\lambda) &= \pm \sqrt{1 - P_z^2 - L^2_\phi / S_\kappa^2(\chi(\lambda))} \\
\label{eq:2DLPHI}   \dot{\phi}(\lambda) &= L_\phi / S_\kappa^2(\chi(\lambda)).
\end{align}

\vspace{.5em}
\noindent\textbf\textbf{\bf Radial Geodesics.}

\noindent Setting $L_\phi = 0$ restricts us to just the spatial radial geodesics and by further requiring that that $x^i(\lambda=0) = \vec{0}$, they will start the the origin of our coordinate system. These geodesics are then given by
\begin{align}
   z(\lambda) &= P_z    \lambda  \\
\chi(\lambda) &= P_\chi \lambda \\
\phi(\lambda) &= \phi_0
\,,
\end{align}
under the constraint that $P_z^2$ $+$ $P_\chi^2 = 1$. This strongly motivates a change of coordinates $z = \rho \cos(\theta)$, $\chi = \rho \sin(\theta)$, under which the metric \eqref{eq:FLRW2Dmetric} changes to,
\begin{equation}
\label{eq:FLRW2DmetricRadial}
 {\rm d}\Sigma_3^2 = {\rm d}\rho^2 + \rho^2 {\rm d}\theta^2 + S_\kappa^2(\rho \sin(\theta) ) {\rm d}\phi^2
\,.
\end{equation}
In these coordinates, the geodesics look like
\begin{align}
  \rho(\lambda) &= \lambda \\
\theta(\lambda) &= \theta_0 \\
  \phi(\lambda) &= \phi_0
  \,.
\end{align}
Hence (radial) proper distance to the origin is simply given by $\ell_\text{rad} = \lambda_f \equiv \rho_0$.

\section{Geodesics of the $\widetilde{\text{U}(\mathds{H}^2)}$ geometry}
\noindent We start with the metric that we derived in Section \ref{sec:ThurstonSpacetimes},
\begin{equation}
{\rm d}\Sigma_3^2 = {\rm d}x^2 + {\rm d}y^2 \cosh^2(x/L) + \big({\rm d}y \sinh^2(x/L) + {\rm d}z\big)^2
= {\rm d}x^2 + {\rm d}y^2 \cosh(2x/L) + {\rm d}z^2 + 2 {\rm d}y {\rm d}z \sinh(x/L),
\end{equation}

\noindent where we have written $L = 1/\sqrt{-\kappa}$ for notational convenience. Nil has two obvious Killing vectors, $\partial_y$ and $\partial_z$, which leads to two conserved quantities and therefore to two first-order equations.
\begin{align}
\label{eq:UH2Q1} Q_1 &= \dot{y}\cosh(2x/L) + \dot{z}\sinh(x/L) & &\longrightarrow & \dot{y} &= \frac{  Q_1            - Q_2 \sinh( x/L)}{\cosh^2(x/L)}. &&& & &&& & \\
\label{eq:UH2Q2} Q_2 &= \dot{y}\sinh( x/L) + \dot{z}           & &\longrightarrow & \dot{z} &= \frac{- Q_1 \sinh(x/L) + Q_2 \cosh(2x/L)}{\cosh^2(x/L)}.  &&& & &&& &
\end{align}
\noindent We can again use the space-like nature of the geodesics so solve for $x$.
\begin{align}
\label{eq:UH2lineelement}
1 = g_{ij} \frac{{\rm d}x^i}{{\rm d}\lambda} \frac{{\rm d}x^j}{{\rm d}\lambda}
= \dot{x}^2 + \dot{y}^2 \cosh^2(x/L) + \Big(\dot{y} \sinh(x/L) + \dot{z}\Big)^2\,.
\end{align}
Plugging in conserved charges from \eqref{eq:UH2Q1} and \eqref{eq:UH2Q2} gives,
\begin{align}
\label{eq:UH2xequation}
\dot{x} = \pm \sqrt{1 - Q_2^2 - \frac{\Big(Q_1 - Q_2 \sinh( x/L)\Big)^2}{\cosh^2(x/L)}}.
\end{align}
At the origin $(x,y,z) = \vec{0}$ the first-order equations reduce to,
\begin{align}
\dot{x} &= \pm \sqrt{1 - Q_1^2 - Q_2^2} \\
\dot{y} &= Q_1 \\
\dot{z} &= Q_2.
\end{align}
This means we can reformulate the conserved charges for geodesics crossing the origin in terms of initial momenta, $P_x = \pm \sqrt{1 - Q_1^2 - Q_2^2}$, $P_y = Q_1$ and $P_z = Q_2$, subject to the constraint that $P_x^2 + P_y^2 + P_z^2 = 1$. Since $P_x$ must be real-valued, this tells us that $Q_1^2 + Q_2^2 \leq 1$. \\

\noindent We can now rewrite equation \eqref{eq:UH2xequation} as an integral equation by using separation of variables,
\begin{align}
\int {\rm d}\lambda = \int {\rm d}x \left[1 - Q_2^2 - \frac{\Big(Q_1 - Q_2 \sinh( x/L)\Big)^2}{\cosh^2(x/L)}\right]^{-\frac{1}{2}}.
\end{align}
We now substitute $w = \sinh(x/L) $, so that ${\rm d}w = \tfrac{{\rm d}x}{L} \cosh(x/L) = \tfrac{{\rm d}x}{L} \sqrt{1 + w^2}$. With this substitution we can rewrite the integral equation as,
\begin{align}
\int {\rm d}\lambda &= L \int \frac{{\rm d}w}{\sqrt{1 + w^2}} \left[1 - Q_2^2 - \tfrac{\big(Q_1 - Q_2 w \big)^2}{1 + w^2} \right]^{-\frac{1}{2}} \\
\label{eq:UH2xintegral}
\int {\rm d}\lambda  &= L \int \frac{{\rm d}w}{\sqrt{
            \smash[b]{\underbrace{\Big(1-2 Q_2^2\Big        )}_A} w^2 +
            \smash[b]{\underbrace{\Big(2 Q_1 Q_2\Big        )}_B} w +
            \smash[b]{\underbrace{\Big(1 - Q_1^2 - Q_2^2\Big)}_C}
            }}.
\end{align}

\vspace{1em}

\noindent We have now reduced solving the geodesic equation to solving the following integral:
\begin{align}
\label{eq:UH2xintegralPolynomial}
\mathcal{I} = \int \frac{{\rm d}w}{\sqrt{A w^2 + B w + C}}.
\end{align}
Since we are interested in real-valued solutions, we will restrict ourselves to the region where $p \equiv A w^2 + B w + C > 0$. Note that $A$ and $B$ take values in the interval $[-1,1]$ depending on $Q_1$ and $Q_2$ while $C = P_x^2$ is restricted to $[0,1]$. This means we must proceed carefully as both $A$ and the discriminant $\Delta \equiv B^2 - 4 A C$ have the potential to take positive and negative values. \\

\begin{figure}[!ht]
\centering
\includegraphics[width=0.8 \textwidth]{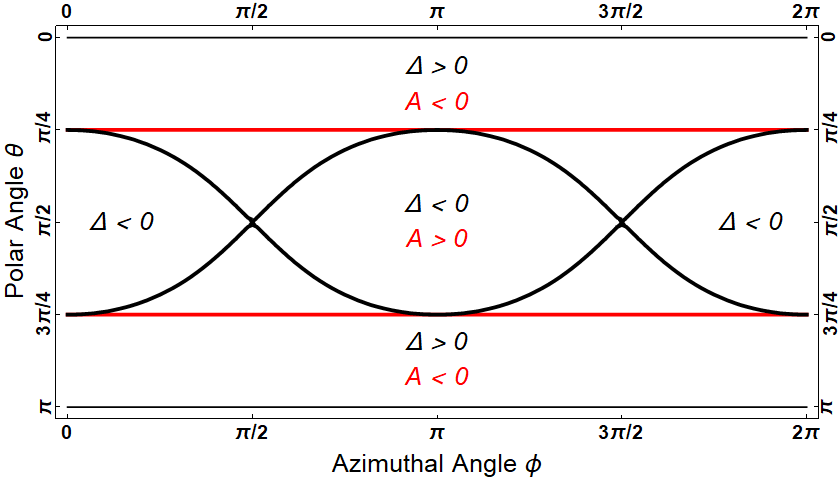}
\caption{An overview of the regions}
\label{fig:UH2regions}
\end{figure}

\noindent To more easily study the different regions, we parameterise the initial velocities as
\begin{align}
P_x &= \sin(\theta) \cos(\phi) \\
P_y &= \sin(\theta) \sin(\phi) \\
P_z &= \cos(\theta).
\end{align}
We can then plot the sign of $A$ and $\Delta$ on the $(\phi,\theta)$ plane in Figure \ref{fig:UH2regions} below to visualise the relevant regions.

\vspace{0.1cm}
\noindent\textbf{Case 1: $\mathbf{A > 0}$.}

\noindent The first big region we consider is the region where $A$ is positive. In this case, we can complete the square and substitute $u = w + B/2A$ to get,
\begin{align}
\label{eq:UH2Case1}
\mathcal{I} = \frac{1}{\sqrt{A}} \int \frac{{\rm d}w}{\sqrt{\left(w + \frac{B}{2A}\right)^2 - \frac{\Delta}{4A^2}}}
= \frac{1}{\sqrt{A}} \int \frac{{\rm d}u}{\sqrt{u^2 - \frac{\Delta}{4A^2}}}\,.
\end{align}
If sign$(\Delta) = \pm 1$ then we can define $a = \sqrt{\mp \Delta} / 2 A$ and write equation \eqref{eq:UH2Case1} as
\begin{align}
\mathcal{I} = \begin{dcases}
\frac{1}{\sqrt{A}} \int \frac{{\rm d}u}{\sqrt{u^2 + a^2}}
= \frac{1}{\sqrt{A}} \ln\left|\frac{u + \sqrt{u^2 + a^2}}{a}\right| = \frac{1}{\sqrt{A}} \ln\left|\frac{2 A w + B + 2 \sqrt{A}\sqrt{p}}{\sqrt{-\Delta}}\right|\,, & \text{if } \Delta < 0
\\
\frac{1}{\sqrt{A}} \int \frac{{\rm d}u}{\sqrt{u^2 - a^2}} = \frac{1}{\sqrt{A}} \ln\left|\frac{u + \sqrt{u^2 - a^2}}{a}\right| = \frac{1}{\sqrt{A}} \ln\left|\frac{2 A w + B + 2 \sqrt{A}\sqrt{p}}{\sqrt{ \Delta}}\right|\,, & \text{if } \Delta > 0
\,.
\end{dcases}
\end{align}
Let's investigate the sign of the expression inside the absolute value brackets. Since $A > 0$ and $p > 0$ by assumption, we can only have an overall minus sign inside the logarithm if $|2Aw+B| > 2\sqrt{A}\sqrt{A w^2 + B w +C}$ and $2 A w + B < 0$. The first condition is equivalent to $\Delta > 0$ (square both sides and rearrange), and the second one to $w < -B/2A$. This means that we never get an overall minus sign in the $\Delta < 0$ case and only sometimes for the $\Delta > 0$ case. \\

Since the first condition is met for the $\Delta > 0$ case by assumption, we know that the polynomial $p$ has two distinct real-valued roots, $w_\pm = -B/2A \pm\sqrt{\Delta}/2A$. Since $A > 0$ we know that $p$ takes non-positive values for $w \in [w_-,w_+]$ and so $\mathcal{I}$ will only have real-valued solutions for $w \in (-\infty,w_-) \cup (w_+,\infty)$. Since $2 A w + B$ is negative if $w < -B/2A$, this means that the argument of the logarithm in the $\Delta > 0$ case is negative when $w < w_-$ and positive when $w > w_+$. \\

\begin{figure}[!ht]
\centering
\includegraphics[width=0.4 \textwidth]{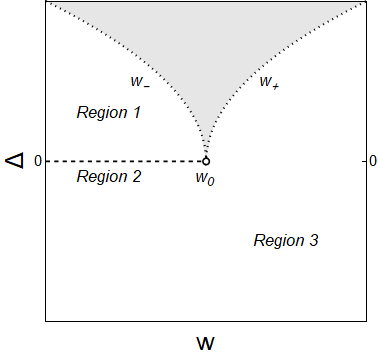}
\caption{An overview of the domain of $\mathcal{I}$ in the case where $A>0$. The three solutions above have their domain on the three regions specified, the first being above the dashed line, the second being on the dashed line and the third being on the other side of the dashed line. $p = 0$ on the dotted lines and at (0,0), and the grey area indicates that $p < 0$; there are no real-valued solutions to $\mathcal{I}$ in these regions.}
\label{fig:UH2regions2}
\end{figure}

\noindent In the third case, $\Delta = 0$,  $p$ has exactly one root $w_0 = -B / 2A$ and we can rewrite equation \eqref{eq:UH2Case1} as
\begin{align}
\mathcal{I} = \frac{1}{\sqrt{A}} \int \frac{{\rm d}u}{\sqrt{u^2}} = \frac{\text{sign}(u)}{\sqrt{A}} \ln|u| = \begin{dcases}
-\frac{1}{\sqrt{A}} \ln\left(-w - B/2A \right)\,, & \text{if } \Delta = 0 \text{ and } w < w_0\\
+\frac{1}{\sqrt{A}} \ln\left(+w + B/2A \right)\,, & \text{if } \Delta = 0 \text{ and } w > w_0
\,.
\end{dcases}
\end{align}
So far we have left the integration constant implicit. We can pick and choose convenient values for this integration constant to write a complete solution to equation \eqref{eq:UH2xintegralPolynomial} when $A>0$:
\begin{align}
\mathcal{I} = \begin{dcases}
+\frac{1}{\sqrt{A}} \ln\left(- 2Aw - B - 2\sqrt{A}\sqrt{A w^2 - B w + C}\right)\,, & \text{if } \Delta > 0 \text{ and } w < w_- \\
-\frac{1}{\sqrt{A}} \ln\left(- 2Aw - B - 2\sqrt{A}\sqrt{A w^2 + B w + C}\right)\,, & \text{if } \Delta = 0 \text{ and } w < w_0 \\
+\frac{1}{\sqrt{A}} \ln\left(+ 2Aw + B + 2\sqrt{A}\sqrt{A w^2 + B w + C}\right)\,, & \text{otherwise}
\,.
\end{dcases}
\end{align}
Figure~\ref{fig:UH2regions2} below visualises the different regions of this solution.

\noindent We now plug this result into equation \eqref{eq:UH2xintegralPolynomial} to get an equation for the geodesic,
\begin{align}
\begin{cases}
\chi {\rm e}^{+\sqrt{A}\frac{\lambda}{L}} = - 2Aw - B - 2\sqrt{A}\sqrt{A w^2 - B w + C}\,, & \text{if } \Delta > 0 \text{ and } w < w_- \\
\tfrac{1}{\chi}{\rm e}^{-\sqrt{A}\frac{\lambda}{L}} = - 2Aw - B - 2\sqrt{A}\sqrt{A w^2 + B w + C}\,, & \text{if } \Delta = 0 \text{ and } w < w_0 \\
\chi  {\rm e}^{+\sqrt{A}\frac{\lambda}{L}} = + 2Aw + B + 2\sqrt{A}\sqrt{A w^2 + B w + C}\,, & \text{otherwise}\,.
\end{cases}
\end{align}
Here $\chi > 0$ is an integration constant coming from the ${\rm d}\lambda$-integral. We can straightforwardly invert these relations to get $w$ as a function of $\lambda$,
\begin{align}
w(\lambda) = \begin{dcases}
+\frac{B^2 - 4 A C - \chi^2}{4 A \chi} \sinh\left(\sqrt{A} \frac{\lambda}{L} \right) - \frac{B^2 - 4 A C + \chi^2} {4 A \chi} \cosh\left(\sqrt{A} \frac{\lambda}{L} \right) - \frac{B}{2A}\,, & \text{if } \Delta > 0 \text{ and } w < w_- \\
-\frac{B^2 - 4 A C - \chi^2}{4 A \chi} \sinh\left(\sqrt{A} \frac{\lambda}{L} \right) - \frac{B^2 - 4 A C + \chi^2} {4 A \chi} \cosh\left(\sqrt{A} \frac{\lambda}{L} \right) - \frac{B}{2A}\,, & \text{if } \Delta = 0 \text{ and } w < w_0 \\
-\frac{B^2 - 4 A C - \chi^2}{4 A \chi} \sinh\left(\sqrt{A} \frac{\lambda}{L} \right) + \frac{B^2 - 4 A C + \chi^2} {4 A \chi} \cosh\left(\sqrt{A} \frac{\lambda}{L} \right) - \frac{B}{2A}\,, & \text{otherwise}\,.
\end{dcases}
\end{align}
The requirement that these geodesics are radial translates to $x(0) = w(0) = 0$, which fixes $\chi$.
\begin{align}
\chi = \begin{dcases}
-B \pm 2 \sqrt{A}\sqrt{C}\,, & \text{if } \Delta > 0 \text{ and } w < w_- \\
-B \pm 2 \sqrt{A}\sqrt{C}\,, & \text{if } \Delta = 0 \text{ and } w < w_0 \\
+B \pm 2 \sqrt{A}\sqrt{C}\,, & \text{otherwise}\,.
\end{dcases}
\end{align}
If we plug these values for $\chi$ into the expressions for $w(\lambda)$ they collapse to a single expression that is valid for all cases:
\begin{align}
w(\lambda) &= \pm \frac{\sqrt{C}}{\sqrt{A}       } \sinh\left(\sqrt{A       }\frac{\lambda}{L} \right) + \frac{B}{2A}             \left[ \cosh\left(\sqrt{A       }\frac{\lambda}{L} \right) - 1 \right] \\
w(\lambda) &=     \frac{P_x     }{\sqrt{1-2P_z^2}} \sinh\left(\sqrt{1-2P_z^2}\frac{\lambda}{L} \right) + \frac{P_y P_z}{1-2P_z^2} \left[ \cosh\left(\sqrt{1-2P_z^2}\frac{\lambda}{L} \right) - 1 \right] \\
x(\lambda) &= L \sinh^{-1}\left[\frac{P_x     }{\sqrt{1-2P_z^2}} \sinh\left(\sqrt{1-2P_z^2}\frac{\lambda}{L} \right) + \frac{P_y P_z}{1-2P_z^2} \left[ \cosh\left(\sqrt{1-2P_z^2}\frac{\lambda}{L} \right) - 1 \right]\right].
\end{align}
This expression is a solution to \eqref{eq:UH2lineelement} and has the correct large-$L$ behaviour, $\lim_{L\rightarrow\infty} x(\lambda) = P_x \lambda$.

\vspace{1em}
\noindent\textbf{Case 2: $\mathbf{A = 0}$.}

\noindent If $A=0$ then $\Delta = B^2$ and we only need to consider the $\Delta =0$ and $\Delta >0$ scenarios. If $\Delta > 0$ then $p$ reduces to a first-order polynomial $B w + C$ and if $\Delta = 0$ then $p$ reduces to a constant $C$. We can then simply write,
\begin{align}
\lambda-\lambda_0 = \begin{dcases}
L \frac{2\sqrt{B w + C}}{B} \,, & \text{if } \Delta > 0 \\
L \frac{w}{\sqrt{C}}\,,         & \text{if } \Delta = 0\,.
\end{dcases}
\end{align}
We can invert these relationships and require that $x(0) = 0$ to get
\begin{align}
w(\lambda) &= \pm\sqrt{C}\frac{\lambda}{L}              + \frac{B}{4}\frac{\lambda^2}{L^2} \\
w(\lambda) &= P_x\frac{\lambda}{L}                      + \frac{P_y P_z}{2}\frac{\lambda^2}{L^2} \\
x(\lambda) &= L \sinh^{-1}\left[P_x \frac{\lambda}{L}   + \frac{P_y P_z}{2}\frac{\lambda^2}{L^2} \right].
\end{align}
This expression again covers both the $\Delta = 0$ and $\Delta > 0$ cases and solves \eqref{eq:UH2lineelement}.

\vspace{1em}
\noindent\textbf{Case 3: $\mathbf{A < 0}$.}

\noindent When $A <0$ we know that $\Delta = B^2 - 4 A C \geq 0$ since $C$ is non-negative. If $\Delta = 0$ then $P_x =0 $, $P_y = 0$ and $P_z = \pm 1$ so that $\vec{x} = (0,0,P_z \lambda)$ is a solution to the geodesic equation. \\

\noindent If $\Delta > 0$ then we proceed similarly to the $A>0$ case. We now only find real-valued solutions for $\mathcal{I}$ if $\Delta > 0$ and $w \in (w_-, w_+)$. This means we can again complete the square and substitute $u = w + B/2A$ and $a = \sqrt{\Delta}/2A$ to get
\begin{align}
\label{eq:UH2Case3}
\mathcal{I} = \frac{1}{\sqrt{-A}} \int \frac{{\rm d}w}{\sqrt{- \left(w + \frac{B}{2A}\right)^2 + \frac{\Delta}{4A^2}}} = \frac{1}{\sqrt{-A}} \int \frac{{\rm d}u}{\sqrt{-u^2 + a^2}}\,.
\end{align}
This is a standard integral and we can therefore find a complete solution for \eqref{eq:UH2xintegral} in the $A<0$ case as
\begin{align}
\lambda-\lambda_0 = -\frac{L}{\sqrt{-A}} \arcsin\left(\frac{u}{a}\right) = -\frac{L}{\sqrt{-A}} \arcsin\left(\frac{2Aw +B}{\sqrt{\Delta}}\right)\,,  &\quad  \text{where } |2Aw+B| < \sqrt{\Delta}\,.
\end{align}
We again invert these relationships and require that $x(0) = 0$ to get
\begin{align}
w(\lambda) &= \frac{\sqrt{\Delta}}{2(-A)}   \sin\left( \sqrt{-A}\frac{\lambda}{L} - \arcsin\left(\frac{B}{\sqrt{\Delta}}\right) \right) - \frac{B}{2A} \\
w(\lambda) &= \frac{\sqrt{P_y^2 P_z^2 + P_x^2(2P_z^2 -1)}}{2 P_z^2 -1}   \sin\left[ \sqrt{2 P_z^2 -1}\frac{\lambda}{L} - \arcsin\left(\frac{P_y P_z}{\sqrt{P_y^2 P_z^2 + P_x^2(2P_z^2 -1)}}\right) \right] - \frac{P_y P_z}{2 P_z^2 -1} \\
x(\lambda) &= L \sinh^{-1} \left[ w(\lambda) \right]
\,.
\end{align}

\vspace{1em}
\noindent\textbf{Final geodesics.}

\noindent We can now write the geodesics for the $\widetilde{\text{U}(\mathds{H}^2)}$ geometry in a compact form.
\begin{align}
x(\lambda) &= L \sinh^{-1} \left[w(\lambda)\right] \\
y(\lambda) &= \int {\rm d}\lambda \; \frac{- P_z w + P_y}{1+w^2} \\
z(\lambda) &= \int {\rm d}\lambda \; \frac{2 P_z w^2 - P_y w + P_z}{1+w^2}
\,,
\end{align}
where $w$ is determined by the value of $P_z$,
\begin{align}
\!\!\!
w(\lambda) =
\begin{cases}
\frac{P_x}{\sqrt{1-2P_z^2}} \sinh\left(\sqrt{1-2P_z^2}\frac{\lambda}{L} \right) + \frac{P_y P_z}{1-2P_z^2} \left[ \cosh\left(\sqrt{1-2P_z^2}\frac{\lambda}{L} \right) - 1 \right]\,,                                           & \text{if } P_z^2 < 1/2 \\
P_x\frac{\lambda}{L} + \frac{P_y P_z}{2}\frac{\lambda^2}{L^2}\,,                                                                                                                                                                    & \text{if } P_z^2 = 1/2 \\
\frac{\sqrt{P_y^2 P_z^2 + P_x^2(2P_z^2 -1)}}{2 P_z^2 -1}   \sin\!\left[ \sqrt{2 P_z^2 -1}\frac{\lambda}{L} - \arcsin\left(\frac{P_y P_z}{\sqrt{P_y^2 P_z^2 + P_x^2(2P_z^2 -1)}}\right) \right] - \frac{P_y P_z}{2 P_z^2 -1}\,,  & \text{if } 1/2 < P_z^2 < 1
,
\end{cases}
\end{align}
and we have an additional constraint that $P_x^2 + P_y^2 + P_z^2 = 1$. The fact that these expressions are not smooth with respect to the initial velocity $\hat{P}$ should not worry us. This feature exists within many other dynamical systems, c.f. elliptical, parabolic and hyperbolic trajectories on orbital mechanics. \\

\begin{figure}
    \centering
    \begin{minipage}[t]{0.31\textwidth}
        \centering
        \includegraphics[width=0.95\textwidth]{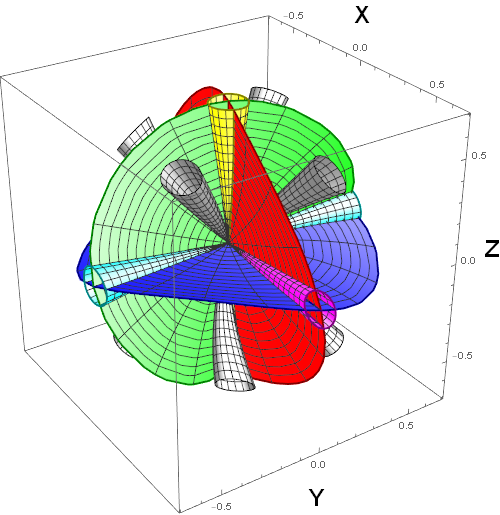}
        \caption{{\it Geodesics of the $\widetilde{\text{U}(\mathds{H}^2)}$ geometry.} Note that the cyan cone is stretched horizontally, the magenta cone is stretched more harshly vertically, while the yellow cone is largely unaffected.}
        \label{fig:UH2Geodesics1}
    \end{minipage}\hfill
    \begin{minipage}[t]{0.31\textwidth}
        \centering
        \includegraphics[width=0.95\textwidth]{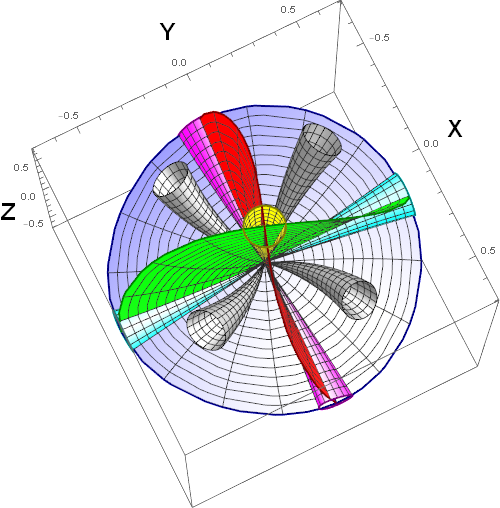}
        \caption{The same visualisation, viewed from a higher angle to show a clockwise twisting effect of geodesics about the $z$-axis of the geometry; a similar counter-clockwise twist can be observed around the $y$-axis in the first plot.}
        \label{fig:UH2Geodesics2}
    \end{minipage}\hfill
    \begin{minipage}[t]{0.31\textwidth}
        \centering
        \includegraphics[width=0.95\textwidth]{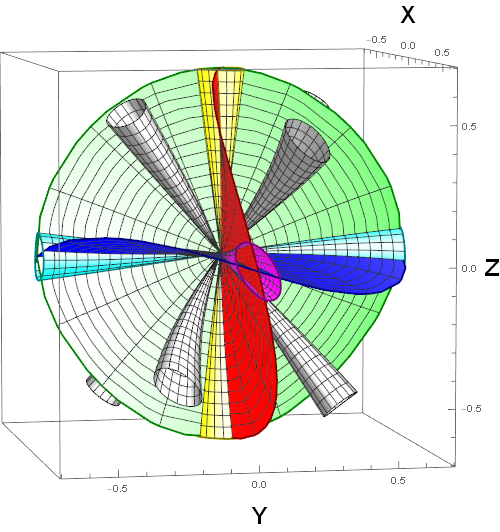}
        \caption{However, around $x$-axis of the geometry geodesics are skewed towards a diagonal direction rather than twisted.}
        \label{fig:UH2Geodesics3}
    \end{minipage}
\end{figure}

To visualise these geodesics, Figures \ref{fig:UH2Geodesics1}, \ref{fig:UH2Geodesics2} and \ref{fig:UH2Geodesics3} show the trajectory of the above geodesics along selected directions emanating from the origin. The curved red, green and blue surfaces traced by the set of geodesics with an initial direction orthogonal to the $x$-, $y$- and $z$-direction respectively. The magenta, cyan and yellow surfaces are traced by sets of geodesics with an initial direction in a small cone around the $x$-, $y$- and $z$-direction respectively. The grey surfaces are similar, but are cones around initial directions with $|P_x| = |P_y| = |P_z|$.  To give a clearer visualisation of the effects of curvature, we have increased its effects by drawing the surfaces from distance $\lambda = 0$ to distance $\lambda = 5 \eta_*$.


\section{Geodesics of the Nil geometry}

The Nil case starts along the same general lines from spatial sections of the Nil-metric \eqref{eq:Nilmetric0},
\begin{align}
{\rm d}\Sigma_3^2 = {\rm d}x^2 + \left(1+ \frac{x^2}{L^2} \right) {\rm d}y^2 + {\rm d}z^2 - \frac{2x}{L} \ {\rm d}y {\rm d}z\,.
\end{align}
Note that we have written $L = 1/\sqrt{-\kappa}$ for notational convenience. Nil has two obvious Killing vectors, $\partial_y$ and $\partial_z$, which lead two conserved quantities and therefore to two first-order equations,
\begin{align}
\label{eq:NilQ1} Q_1 &= \dot{y}\left(1+\frac{x^2}{L^2}\right) - x \dot{z} / L  & &\longrightarrow & \dot{y} &= Q_1 + Q_2 x / L. &&& & &&& & \\
\label{eq:NilQ2} Q_2 &= \dot{z} - x \dot{y} / L                                & &\longrightarrow & \dot{z} &= Q_2 \left( 1 + \frac{x^2}{L^2}\right) + Q_1 x / L\,.  &&& & &&& &
\end{align}
Unfortunately, the Nil equivalent of \eqref{eq:Epsilon} does not allow for separation of variables and we must solve the geodesic equation for $x$ directly,
\begin{align}
\ddot{x}(\lambda) + \frac{Q_2^2}{L^2}\left(x(\lambda) + \frac{Q_1 L}{Q_2}\right) = 0
\,.
\end{align}
This is the equation for a (shifted) simple harmonic oscillator with angular frequency $\omega = Q_2/L$ and it is solved by,
\begin{align}
x(\lambda) + \frac{Q_2 L}{Q_1} = C_1 \cos\left(\omega \lambda \right) + C_2 \sin\left(\omega \lambda \right)
\,.
\end{align}
From this result it is a straightforward, if tedious, exercise to derive a full solution for $y$ and $z$. We will omit the details of this calculation and skip directly to the solution in a convenient form,
\begin{align}
x(\lambda) &= x_0 + \frac{P_x}{\omega} \sin( \omega \lambda ) - \frac{P_y}{\omega} \Big[ 1 - \cos ( \omega \lambda ) \Big] \\
y(\lambda) &= y_0 + \frac{P_y}{\omega} \sin( \omega \lambda ) + \frac{P_x}{\omega} \Big[ 1 - \cos ( \omega \lambda ) \Big] \\
z(\lambda) &= z_0 + L \omega \lambda + \frac{x_0 P_y}{L \omega} \sin(\omega\lambda)
            + \frac{P_x^2}{4 L \omega^2} \Big[ \sin(2\omega\lambda) - 2\omega\lambda \Big]
            + \frac{P_y^2}{4 L \omega^2} \Big[ 2\omega\lambda - 4 \sin(\omega\lambda) + \sin(2\omega\lambda)\Big] \\
           &+ \frac{2 P_x}{L\omega^2}\Big[x_0 \omega + P_y \cos(\omega\lambda)\Big] \sin^2\left(\frac{\omega\lambda}{2}\right)
\,,
\end{align}
where we have chosen the $Q$s, $C$s and the constants arising from integrating \eqref{eq:NilQ1} and \eqref{eq:NilQ2} so that $x^i(0) = x^i_0$ and $\dot{x}^i(0) = P_i$,
and  the angular frequency is defined as,
$\omega = \dfrac{L P_z - P_y x_0}{L^2}$.

\vspace{1em}
\noindent\textbf{Radial geodesics.}

\noindent To make these geodesics radial, we set $x_0 = y_0 = z_0 = 0$. The angular frequency now simplifies to $\omega = P_z/L$ and we may write exact expressions for the geodesics as
\begin{align}
\label{eq:Nilx} x(\lambda) &= \frac{L P_x}{P_z} \sin\left( P_z \frac{\lambda}{L}\right) - \frac{L P_y}{P_z} \left[ 1 - \cos \left( P_z \frac{\lambda}{L}\right) \right] \\
\label{eq:Nily} y(\lambda) &= \frac{L P_y}{P_z} \sin\left( P_z \frac{\lambda}{L}\right) + \frac{L P_x}{P_z} \left[ 1 - \cos \left( P_z \frac{\lambda}{L}\right) \right] \\
\label{eq:Nilz} z(\lambda) &= P_z \lambda
                            + \frac{P_x^2}{4 P_z^2} \left[ 2 P_z \lambda -  L \sin\left(\frac{2 P_z \lambda}{L}\right) \right]
                            + \frac{P_y^2}{4 P_z^2} \left[ 2 P_z \lambda - 4L \sin\left(\frac{ P_z \lambda}{L}\right) + L \sin\left(\frac{2 P_z \lambda}{L}\right) \right] \\
                           &+ \frac{2 L P_x P_y}{P_z^2} \cos\left(\frac{P_z \lambda}{L}\right)\sin^2\left(\frac{P_z \lambda}{2L}\right)
\,.
\end{align}
Given these expressions, it is productive to reconsider equation \eqref{eq:Epsilon}, which now reads
\begin{align}
\label{eq:NilEpsilon}  1 = \epsilon = P_x^2 + P_y^2 + P_z^2.
\end{align}
In effect, this tells us that if we choose initial velocity vector $\vec{P} = (P_x,P_y,P_z)$ to be of unit length, then $\lambda$ parameterises geodesic distance along the curve. Hence, proper (radial) distance is again given by  $\ell_\text{rad} = \lambda_f \equiv \rho_0$. \\

\begin{figure}
    \centering
    \begin{minipage}[t]{0.31\textwidth}
        \centering
        \includegraphics[width=0.95\textwidth]{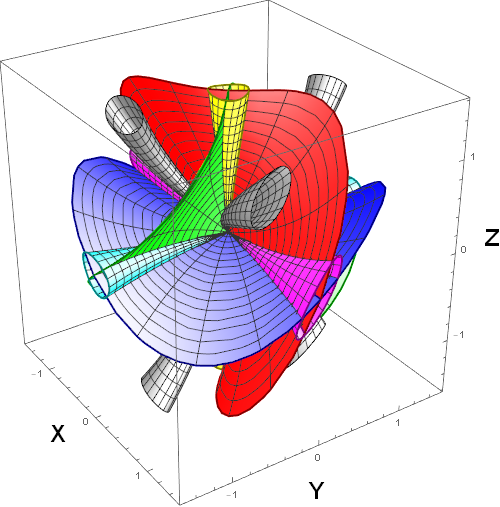}
        \caption{{\it Geodesics of the Nil geometry.} Like the $\widetilde{\text{U}(\mathds{H}^2)}$ geometry, the cyan cone is stretched horizontally, the magenta cone is stretched more harshly vertically, while the yellow cone is largely unaffected.}
        \label{fig:NilGeodesics1}
    \end{minipage}\hfill
    \begin{minipage}[t]{0.31\textwidth}
        \centering
        \includegraphics[width=0.95\textwidth]{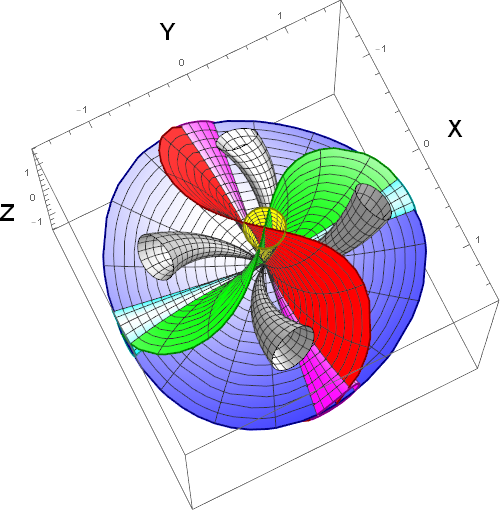}
        \caption{The same visualisation, viewed from a higher angle to show a counter-clockwise twisting effect of geodesics about the $z$-axis of the geometry; a similar twist can be observed around the $y$-axis in the first plot.}
        \label{fig:NilGeodesics2}
    \end{minipage}\hfill
    \begin{minipage}[t]{0.31\textwidth}
        \centering
        \includegraphics[width=0.95\textwidth]{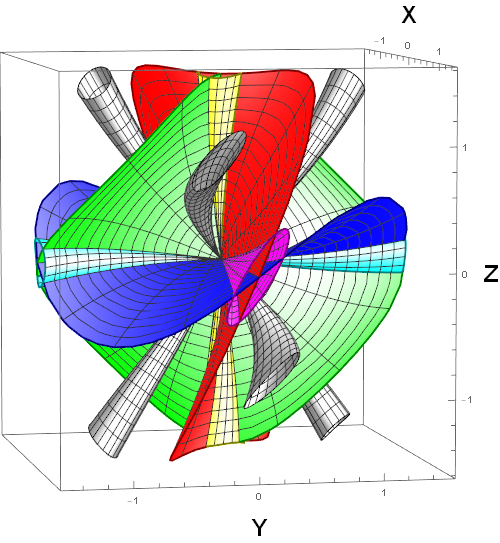}
        \caption{However, around $x$-axis of the geometry geodesics are skewed towards a diagonal direction rather than twisted. Note that the diagonal points the other way when compared to $\widetilde{\text{U}(\mathds{H}^2)}$.}
        \label{fig:NilGeodesics3}
    \end{minipage}
\end{figure}

To visualise these geodesics we have again plotted select geodesics of this geometry in Figures \ref{fig:NilGeodesics1}, \ref{fig:NilGeodesics2} and \ref{fig:NilGeodesics3}. For a description of the surfaces, please refer to the previous appendix. As in Appendix~C, we have augmented the effects of curvature for a clearer visualisation. For this geometry it was sufficient to draw the surfaces from distance $\lambda = 0$ to distance $\lambda = 2.5 \eta_*$.

\section{Geodesics of the Solv geometry}

We again start from metric introduced in \eqref{eq:Solvmetric0} and look at spatial sections,
\begin{equation}
\label{eq:Solvmetric}
{\rm d}\Sigma_3^2 = {\rm e}^{2z/L} \ {\rm d}x^2 + {\rm e}^{-2z/L} \ {\rm d}y^2 + {\rm d}z^2
\,.
\end{equation}
This metric admits three Killing vectors, the first is $K_1 = \partial_x$, the second $K_2 = \partial_y$ and the third $K_3 = -\tfrac{x}{L} \partial_x  \  +  \  \tfrac{y}{L} \partial_y \ + \ \partial_z$. These lead to the following first-order equations,
\begin{eqnarray}
\dot{x} &=& P_x {\rm e}^{-2z/L}
\label{Solv: dot x}\\
\dot{y} &=& P_y {\rm e}^{ 2z/L}
\label{Solv: dot y}\\
\dot{z} &=& P_z + P_x \frac{x}{L} - P_y \frac{y}{L}
\,.
\label{Solv: dot z}
\end{eqnarray}
Unlike in the previous cases, it is hard to solve this system exactly~\footnote{
Upon inserting
equations~(\ref{Solv: dot x}--\ref{Solv: dot y})
into equations~(\ref{eq:Solvmetric}) one obtains,
\begin{equation}
1  = \frac{P_x^2}{{\rm e}^{2z/L}} +\frac{P_y^2}{{\rm e}^{-2z/L}} + \dot z^2
\;\Longrightarrow
\int\frac{{\rm d}z}{\sqrt{1-{\rm e}^{-2z/L}P_x^2 - {\rm e}^{2z/L}P_y^2}}  = \lambda
\,.
\end{equation}
A convenient variable is $w=e^{z/L}$, in terms of which
this equation simplifies to,
\begin{equation}
\int\frac{{\rm d}w}{\sqrt{w^2-P_x^2 - w^4P_y^2}}  = \frac{\lambda}{L}
\,.
\end{equation}
This is an elliptic integral, and so its solution can be expressed in terms of
elliptic functions. Instead of studying these general solutions,
we opt for
a much simpler expansion in powers of $1/L$, with the results given
in equations~(\ref{Eq D10}--\ref{Eq D12}). \\
}
and so we will  opt for a perturbative approach.

\vskip 0.2cm
\noindent\textbf{Radial geodesics.}

\noindent We can write $\vec{x} = \sum_{i=0}^n \vec{x}_{i} L^{-i}$ and expand the system above to a desired power $L^n$. To linear order, for example, this would look like
\begin{align}
\dot{x} &= P_x \big(1 - 2z/L\big)  \\
\dot{y} &= P_y \big(1 + 2z/L\big) \\
\dot{z} &= P_z + P_x \frac{x}{L} - P_y \frac{y}{L}.
\end{align}
We can then require $x^i(0) = 0$ and equate terms of equal power of $L$ to solve the system. This gives the following result,
\begin{align}
x(\lambda) &= \lambda P_x \left( 1 - P_z \frac{\lambda}{L} - \frac{1}{3}(1 - 2 P_y^2 + 3 P_z^2) \frac{\lambda^2}{L^2} + ...  \right)
\label{Eq D10}
\\
y(\lambda) &= \lambda P_y \left( 1 + P_z \frac{\lambda}{L} + \frac{1}{3}(1 - 2 P_y^2 + 3 P_z^2) \frac{\lambda^2}{L^2} + ...  \right)
\label{Eq D11} \\
z(\lambda) &= P_z \lambda + \frac{P_x^2 - P_y^2}{2} \frac{\lambda}{L} + \frac{P_z^3 - P_z}{3} \frac{\lambda^2}{L^2} + ...
\,.
\label{Eq D12}
\end{align}
We can evaluate the constraint equation \eqref{eq:Epsilon} at $\lambda = 0$, which tells us that $P_x^2 + P_y^2 + P_z^2 = 1$ holds as before. Hence (radial) proper distance to the origin is again given by $\ell_\text{rad} = \lambda_f \equiv \rho_0$. \\

\begin{figure}
    \centering
    \begin{minipage}[t]{0.31\textwidth}
        \centering
        \includegraphics[width=0.95\textwidth]{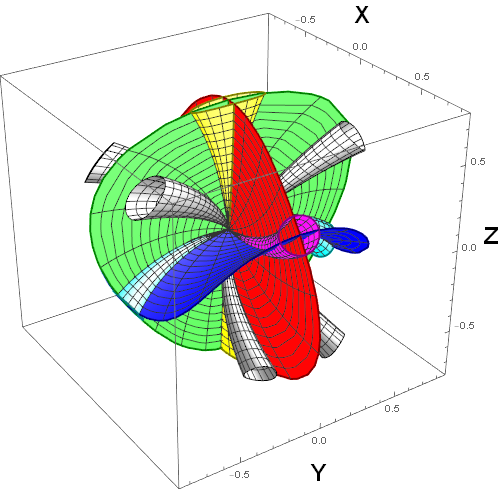}
        \caption{{\it Geodesics of the Solv geometry.} The geometry is characterized by geodesics being pushed up- or downward respectively depending on whether $|x|$ is larger or smaller than $|y|$.}
        \label{fig:SolvGeodesics1}
    \end{minipage}\hfill
    \begin{minipage}[t]{0.31\textwidth}
        \centering
        \includegraphics[width=0.95\textwidth]{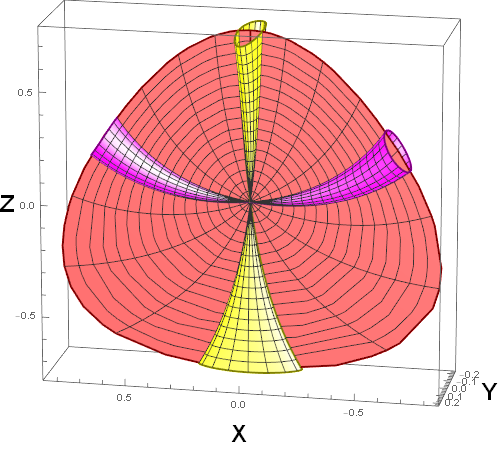}
        \caption{The same visualisation, viewed along the $y$-direction and with some elements removed to show more clearly how geodesics along the $x$-$z$ plane are deflected upwards along a distinctive guitar pick shape.}
        \label{fig:SolvGeodesics2}
    \end{minipage}\hfill
    \begin{minipage}[t]{0.31\textwidth}
        \centering
        \includegraphics[width=0.95\textwidth]{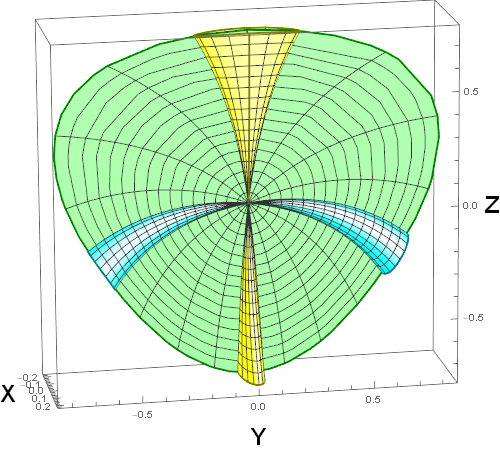}
        \caption{The same visualisation, viewed along the $x$-direction and with some elements removed to show more clearly how geodesics along the $y$-$z$ plane are deflected downwards.}
        \label{fig:SolvGeodesics3}
    \end{minipage}
\end{figure}

To visualise these geodesics we have again plotted select geodesics of this geometry in Figures \ref{fig:SolvGeodesics1}, \ref{fig:SolvGeodesics2} and \ref{fig:SolvGeodesics3}. For a description of the surfaces, please refer to the previous appendix. As before, for a clearer visualisation we have augmented the effects of curvature by drawing the surfaces
from distance $\lambda = 0$ to distance $\lambda = 2.5 \eta_*$.

\begin{thebibliography}{99}

\bibitem{Milne:1933}
E.~A.~Milne, ``World-Structure and the Expansion of the Universe.'' Zeitschrift fur Astrophysik {\bf 6} (1933) 1.

\bibitem{deOliveira-Costa:2003utu}
A.~de Oliveira-Costa, M.~Tegmark, M.~Zaldarriaga and A.~Hamilton,
``The Significance of the largest scale CMB fluctuations in WMAP,''
Phys. Rev. D \textbf{69} (2004), 063516
doi:10.1103/PhysRevD.69.063516
[arXiv:astro-ph/0307282 [astro-ph]].

\bibitem{Abdalla:2022yfr}
E.~Abdalla, G.~Franco Abell\'an, A.~Aboubrahim, A.~Agnello, O.~Akarsu, Y.~Akrami, G.~Alestas, D.~Aloni, L.~Amendola and L.~A.~Anchordoqui, \textit{et al.}
``Cosmology intertwined: A review of the particle physics, astrophysics, and cosmology associated with the cosmological tensions and anomalies,''
JHEAp \textbf{34} (2022), 49-211
doi:10.1016/j.jheap.2022.04.002
[arXiv:2203.06142 [astro-ph.CO]].

\bibitem{Planck:2015gmu}
P.~A.~R.~Ade \textit{et al.} [Planck],
``Planck 2015 results - XVIII. Background geometry and topology of the Universe,''
Astron. Astrophys. \textbf{594} (2016), A18
doi:10.1051/0004-6361/201525829
[arXiv:1502.01593 [astro-ph.CO]].

\bibitem{Planck:2013okc}
P.~A.~R.~Ade \textit{et al.} [Planck],
``Planck 2013 results. XXVI. Background geometry and topology of the Universe,''
Astron. Astrophys. \textbf{571} (2014), A26
doi:10.1051/0004-6361/201321546
[arXiv:1303.5086 [astro-ph.CO]].


\bibitem{COMPACT:2022gbl}
Y.~Akrami \textit{et al.} [COMPACT],
``The Search for the Topology of the Universe Has Just Begun,''
[arXiv:2210.11426 [astro-ph.CO]].

\bibitem{Land:2004bs}
K.~Land and J.~Magueijo,
``Cubic anomalies in WMAP,''
Mon. Not. Roy. Astron. Soc. \textbf{357} (2005), 994-1002
doi:10.1111/j.1365-2966.2005.08707.x
[arXiv:astro-ph/0405519 [astro-ph]].

\bibitem{Land:2005ad}
K.~Land and J.~Magueijo,
``The Axis of evil,''
Phys. Rev. Lett. \textbf{95} (2005), 071301
doi:10.1103/PhysRevLett.95.071301
[arXiv:astro-ph/0502237 [astro-ph]].

\bibitem{Land:2006bn}
K.~Land and J.~Magueijo,
``The Axis of Evil revisited,''
Mon. Not. Roy. Astron. Soc. \textbf{378} (2007), 153-158
doi:10.1111/j.1365-2966.2007.11749.x
[arXiv:astro-ph/0611518 [astro-ph]].

\bibitem{Copi:2013jna}
C.~J.~Copi, D.~Huterer, D.~J.~Schwarz and G.~D.~Starkman,
``Large-scale alignments from WMAP and Planck,''
Mon. Not. Roy. Astron. Soc. \textbf{449} (2015) no.4, 3458-3470
doi:10.1093/mnras/stv501
[arXiv:1311.4562 [astro-ph.CO]].

\bibitem{Planck:2019evm}
Y.~Akrami \textit{et al.} [Planck],
``Planck 2018 results. VII. Isotropy and Statistics of the CMB,''
Astron. Astrophys. \textbf{641} (2020), A7
doi:10.1051/0004-6361/201935201
[arXiv:1906.02552 [astro-ph.CO]].

\bibitem{Planck:2015igc}
P.~A.~R.~Ade \textit{et al.} [Planck],
``Planck 2015 results. XVI. Isotropy and statistics of the CMB,''
Astron. Astrophys. \textbf{594} (2016), A16
doi:10.1051/0004-6361/201526681
[arXiv:1506.07135 [astro-ph.CO]].

\bibitem{Schwarz:2015cma}
D.~J.~Schwarz, C.~J.~Copi, D.~Huterer and G.~D.~Starkman,
``CMB Anomalies after Planck,''
Class. Quant. Grav. \textbf{33} (2016) no.18, 184001
doi:10.1088/0264-9381/33/18/184001
[arXiv:1510.07929 [astro-ph.CO]].

\bibitem{BennettHalpernHinshaw:2003}
C.~L.~Bennett, M.~Halpern, G.~Hinshaw, et al.,
``First-Year Wilkinson Microwave Anisotropy Probe (WMAP) Observations: Preliminary Maps and Basic Results,''
Astrophys. J. Supp. \textbf{148} (2003) 1 doi:10.1086/377253


\bibitem{Eriksen:2003db}
H.~K.~Eriksen, F.~K.~Hansen, A.~J.~Banday, K.~M.~Gorski and P.~B.~Lilje,
``Asymmetries in the Cosmic Microwave Background anisotropy field,''
Astrophys. J. \textbf{605} (2004), 14-20
[erratum: Astrophys. J. \textbf{609} (2004), 1198]
doi:10.1086/382267
[arXiv:astro-ph/0307507 [astro-ph]].


\bibitem{Vielva:2003et}
P.~Vielva, E.~Martinez-Gonzalez, R.~B.~Barreiro, J.~L.~Sanz and L.~Cayon,
``Detection of non-Gaussianity in the WMAP 1 - year data using spherical wavelets,''
Astrophys. J. \textbf{609} (2004), 22-34
doi:10.1086/421007
[arXiv:astro-ph/0310273 [astro-ph]].

\bibitem{Cruz:2004ce}
M.~Cruz, E.~Martinez-Gonzalez, P.~Vielva and L.~Cayon,
``Detection of a non-gaussian spot in wmap,''
Mon. Not. Roy. Astron. Soc. \textbf{356} (2005), 29-40
doi:10.1111/j.1365-2966.2004.08419.x
[arXiv:astro-ph/0405341 [astro-ph]].

\bibitem{Jones:2023ncn}
J.~Jones, C.~J.~Copi, G.~D.~Starkman and Y.~Akrami,
[arXiv:2310.12859 [astro-ph.CO]].

\bibitem{Peebles:2022akh}
P.~J.~E.~Peebles,
``Anomalies in Physical Cosmology,''
[arXiv:2208.05018 [astro-ph.CO]].

\bibitem{Jain:2003sg}
P.~Jain, G.~Narain and S.~Sarala,
``Large scale alignment of optical polarizations from distant QSOs using coordinate invariant statistics,''
Mon. Not. Roy. Astron. Soc. \textbf{347} (2004), 394
doi:10.1111/j.1365-2966.2004.07169.x
[arXiv:astro-ph/0301530 [astro-ph]].

\bibitem{Marinello:2016aay}
G.~E.~Marinello, R.~G.~Clowes, L.~E.~Campusano, G.~M.~Williger, I.~K.~S\"ochting and M.~J.~Graham,
``Compatibility of the Large Quasar Groups with the Concordance Cosmological Model,''
Mon. Not. Roy. Astron. Soc. \textbf{461} (2016) no.3, 2267-2281
doi:10.1093/mnras/stw1513
[arXiv:1603.03260 [astro-ph.CO]].

\bibitem{Aluri:2022hzs}
P.~K.~Aluri, P.~Cea, P.~Chingangbam, M.~C.~Chu, R.~G.~Clowes, D.~Hutsem\'ekers, J.~P.~Kochappan, A.~Krasi\'nski, A.~M.~Lopez and L.~Liu, \textit{et al.}
``Is the Observable Universe Consistent with the Cosmological Principle?,''
[arXiv:2207.05765 [astro-ph.CO]].

\bibitem{Secrest:2022uvx}
N.~Secrest, S.~von Hausegger, M.~Rameez, R.~Mohayaee and S.~Sarkar,
``A Challenge to the Standard Cosmological Model,''
[arXiv:2206.05624 [astro-ph.CO]].


\bibitem{Perivolaropoulos:2021jda}
L.~Perivolaropoulos and F.~Skara,
``Challenges for $\Lambda$CDM: An update,''
New Astron. Rev. \textbf{95} (2022), 101659
doi:10.1016/j.newar.2022.101659
[arXiv:2105.05208 [astro-ph.CO]].


\bibitem{Planck:2018vyg}
N.~Aghanim \textit{et al.} [Planck],
``Planck 2018 results. VI. Cosmological parameters,''
Astron. Astrophys. \textbf{641} (2020), A6
[erratum: Astron. Astrophys. \textbf{652} (2021), C4]
doi:10.1051/0004-6361/201833910
[arXiv:1807.06209 [astro-ph.CO]].

\bibitem{ACTPol:2016kmo}
T.~Louis \textit{et al.} [ACTPol],
``The Atacama Cosmology Telescope: Two-Season ACTPol Spectra and Parameters,''
JCAP \textbf{06} (2017), 031
doi:10.1088/1475-7516/2017/06/031
[arXiv:1610.02360 [astro-ph.CO]].

\bibitem{ACT}
ACT website: act.princeton.edu (2023).

\bibitem{EUCLID}
EUCLID collaboration website: https://www.cosmos.esa.int/web/euclid (2023).

\bibitem{Euclid:2021icp}
R.~Scaramella \textit{et al.} [Euclid],
``Euclid preparation - I. The Euclid Wide Survey,''
Astron. Astrophys. \textbf{662} (2022), A112
doi:10.1051/0004-6361/202141938
[arXiv:2108.01201 [astro-ph.CO]].

\bibitem{SKAO}
SKAO collaboration website: https://www.skao.int/en/about-us/skao (2023).


\bibitem{Leonard:2016evk}
C.~D.~Leonard, P.~Bull and R.~Allison,
``Spatial curvature endgame: Reaching the limit of curvature determination,''
Phys. Rev. D \textbf{94} (2016) no.2, 023502
doi:10.1103/PhysRevD.94.023502
[arXiv:1604.01410 [astro-ph.CO]].

\bibitem{DiDio:2016ykq}
E.~Di Dio, F.~Montanari, A.~Raccanelli, R.~Durrer, M.~Kamionkowski and J.~Lesgourgues,
``Curvature constraints from Large Scale Structure,''
JCAP \textbf{06} (2016), 013
doi:10.1088/1475-7516/2016/06/013
[arXiv:1603.09073 [astro-ph.CO]].

\bibitem{Cornish:1997rp}
N.~J.~Cornish, D.~N.~Spergel and G.~D.~Starkman,
``Measuring the topology of the universe,''
Proc. Nat. Acad. Sci. \textbf{95} (1998), 82
doi:10.1073/pnas.95.1.82
[arXiv:astro-ph/9708083 [astro-ph]].

\bibitem{Luminet:1999qh}
J.~P.~Luminet and B.~F.~Roukema,
``Topology of the universe: Theory and observation,''
[arXiv:astro-ph/9901364 [astro-ph]].

\bibitem{Sandhu:2016gbz}
J.~Sandhu,
``Cosmic Topology,''
[arXiv:1612.04157 [astro-ph.CO]].

\bibitem{COMPACT:2022nsu}
P.~Petersen \textit{et al.} [COMPACT],
``Cosmic topology. Part~I. Limits on orientable Euclidean manifolds from circle searches,''
JCAP \textbf{01} (2023), 030
doi:10.1088/1475-7516/2023/01/030
[arXiv:2211.02603 [astro-ph.CO]].

\bibitem{Starobinsky:1993yx}
A.~A.~Starobinsky,
``New restrictions on spatial topology of the universe from microwave background temperature fluctuations,''
JETP Lett. \textbf{57} (1993), 622-625
[arXiv:gr-qc/9305019 [gr-qc]].

\bibitem{deOlivieraCosta:1994eb}
A.~de Oliviera Costa and G.~F.~Smoot,
``Constraints on the topology of the universe from the 2-year COBE data,''
Astrophys. J. \textbf{448} (1995), 477
doi:10.1086/175977
[arXiv:astro-ph/9412003 [astro-ph]].

\bibitem{Cornish:2003db}
N.~J.~Cornish, D.~N.~Spergel, G.~D.~Starkman and E.~Komatsu,
``Constraining the topology of the universe,''
Phys. Rev. Lett. \textbf{92} (2004), 201302
doi:10.1103/PhysRevLett.92.201302
[arXiv:astro-ph/0310233 [astro-ph]].

\bibitem{Vaudrevange:2012da}
P.~M.~Vaudrevange, G.~D.~Starkman, N.~J.~Cornish and D.~N.~Spergel,
``Constraints on the Topology of the Universe: Extension to General Geometries,''
Phys. Rev. D \textbf{86} (2012), 083526
doi:10.1103/PhysRevD.86.083526
[arXiv:1206.2939 [astro-ph.CO]].

\bibitem{Luminet:2003dx}
J.~P.~Luminet, J.~Weeks, A.~Riazuelo, R.~Lehoucq and J.~P.~Uzan,
``Dodecahedral space topology as an explanation for weak wide - angle temperature correlations in the cosmic microwave background,''
Nature \textbf{425} (2003), 593
doi:10.1038/nature01944
[arXiv:astro-ph/0310253 [astro-ph]].

\bibitem{Bernui:2018wef}
A.~Bernui, C.~P.~Novaes, T.~S.~Pereira and G.~D.~Starkman,
``Topology and the suppression of CMB large-angle correlations,''
[arXiv:1809.05924 [astro-ph.CO]].

\bibitem{Thurston}
William P.~Thurston,
``Three dimensional Manifolds, Kleinian groups and hyperbolic geometry,''
Bull. Am. Math. Soc. (N.S.) 6, 357 (1982).

\bibitem{Perelman:2003uq}
G.~Perelman,
``Finite extinction time for the solutions to the Ricci flow on certain three-manifolds,''
[arXiv:math/0307245 [math.DG]].

\bibitem{Perelman:2006un}
G.~Perelman,
``The Entropy formula for the Ricci flow and its geometric applications,''
[arXiv:math/0211159 [math.DG]].

\bibitem{Perelman:2006up}
G.~Perelman,
``Ricci flow with surgery on three-manifolds,''
[arXiv:math/0303109 [math.DG]].

\bibitem{Fagundes:1991uy}
H.~V.~Fagundes,
``Closed spaces in cosmology,''
Gen. Rel. Grav. \textbf{24} (1992), 199
doi:10.1007/BF00756787
[arXiv:0812.4103 [gr-qc]].

\bibitem{Coquereaux:2014sva}
R.~Coquereaux,
``The history of the universe is an elliptic curve,''
Class. Quant. Grav. \textbf{32} (2015) no.11, 115013
[erratum: Class. Quant. Grav. \textbf{33} (2016) no.15, 159601]
doi:10.1088/0264-9381/32/11/115013
[arXiv:1411.2192 [gr-qc]].

\bibitem{Coquereaux:1981ya}
R.~Coquereaux and A.~Grossmann,
``Analytic Discussion of Spatially Closed Friedmann Universes With Cosmological Constant and Radiation Pressure,''
Annals Phys. \textbf{143} (1982), 296
doi:10.1016/0003-4916(82)90030-6

\bibitem{Minami:2020odp}
Y.~Minami and E.~Komatsu,
Phys. Rev. Lett. \textbf{125} (2020) no.22, 221301
doi:10.1103/PhysRevLett.125.221301
[arXiv:2011.11254 [astro-ph.CO]].

\bibitem{Diego-Palazuelos:2022dsq}
P.~Diego-Palazuelos, J.~R.~Eskilt, Y.~Minami, M.~Tristram, R.~M.~Sullivan, A.~J.~Banday, R.~B.~Barreiro, H.~K.~Eriksen, K.~M.~G\'orski and R.~Keskitalo, \textit{et al.}
``Cosmic Birefringence from the Planck Data Release 4,''
Phys. Rev. Lett. \textbf{128} (2022) no.9, 091302
doi:10.1103/PhysRevLett.128.091302
[arXiv:2201.07682 [astro-ph.CO]].

\bibitem{Diego-Palazuelos:2022cnh}
P.~Diego-Palazuelos, E.~Mart\'\i{}nez-Gonz\'alez, P.~Vielva, R.~B.~Barreiro, M.~Tristram, E.~de la Hoz, J.~R.~Eskilt, Y.~Minami, R.~M.~Sullivan and A.~J.~Banday, \textit{et al.}
``Robustness of cosmic birefringence measurement against Galactic foreground emission and instrumental systematics,''
JCAP \textbf{01} (2023), 044
doi:10.1088/1475-7516/2023/01/044
[arXiv:2210.07655 [astro-ph.CO]].

\bibitem{Eskilt:2022cff}
J.~R.~Eskilt and E.~Komatsu,
``Improved constraints on cosmic birefringence from the WMAP and Planck cosmic microwave background polarization data,''
Phys. Rev. D \textbf{106} (2022) no.6, 063503
doi:10.1103/PhysRevD.106.063503
[arXiv:2205.13962 [astro-ph.CO]].

\bibitem{Philcox:2022hkh}
O.~H.~E.~Philcox,
``Probing parity violation with the four-point correlation function of BOSS galaxies,''
Phys. Rev. D \textbf{106} (2022) no.6, 063501
doi:10.1103/PhysRevD.106.063501
[arXiv:2206.04227 [astro-ph.CO]].

\bibitem{Creque-Sarbinowski:2023wmb}
C.~Creque-Sarbinowski, S.~Alexander, M.~Kamionkowski and O.~Philcox,
``Parity-Violating Trispectrum from Chern-Simons Gravity,''
[arXiv:2303.04815 [astro-ph.CO]].

\bibitem{Coulton:2023oug}
W.~R.~Coulton, O.~H.~E.~Philcox and F.~Villaescusa-Navarro,
``Signatures of a Parity-Violating Universe,''
[arXiv:2306.11782 [astro-ph.CO]].

\bibitem{2067292}
M.~Koussour, H.~Filali and M.~Bennai,
``Two Minimally Interacting Fluids: Matter and Holographic Dark Energy in Bianchi Type-I Universe,''
doi:10.2139/ssrn.4028697

\bibitem{Gradshteyn:2014}
 Izrail Solomonovich Gradshteyn and Iosif Moiseevich Ryzhik,
``Table of integrals, series, and products'', Academic press (2014)

\bibitem{Vedder:2022spt}
C.~J.~G.~Vedder, E.~Belgacem, N.~E.~Chisari and T.~Prokopec,
``Fluctuating Dark Energy and the Luminosity Distance,''
[arXiv:2209.00440 [astro-ph.CO]].

\bibitem{Fischer:2006}
A.~E.~Fischer and V.~Moncrief
``Hamiltonian Reduction of Einstein's Equations''
doi:10.1016/B0-12-512666-2/00498-3
\end{thebibliography}
\end{document}